\documentclass[pra,twocolumn,eqsecnum,superscriptaddress,showpacs,amsmath,amstex,amssymb,citeautoscript]{revtex4-1}

\usepackage{amsmath,amssymb,bm,bbm,mathrsfs}

\usepackage[T1]{fontenc}
\usepackage[utf8]{inputenc}
\usepackage{lipsum}
\usepackage{amsmath}
\usepackage{amssymb}
\usepackage{bbm}
\usepackage{braket}
\usepackage{xcolor}
\usepackage{pifont}
\usepackage[mathscr]{euscript}
\usepackage[shortlabels]{enumitem}
\usepackage[justification=justified]{subcaption}
\usepackage{graphicx}
\usepackage{lipsum}
\allowdisplaybreaks
\usepackage{float}
\usepackage{graphicx}
\usepackage{dsfont}
\usepackage{comment}
\usepackage[colorlinks=true]{hyperref}  
\hypersetup{
    bookmarks=true,         
    unicode=false,          
    pdftoolbar=true,        
    pdfmenubar=true,        
    pdffitwindow=false,     
    pdfstartview={FitH},    
    pdftitle={Theory of zero-field superconducting diode effect in twisted trilayer graphene},    
    pdfauthor={Scammell, Li, Scheurer},     
    pdfsubject={},   
    pdfcreator={},   
    pdfproducer={}, 
    pdfkeywords={} {} {}, 
    pdfnewwindow=true,      
    colorlinks=true,       
    linkcolor=blue, 
    citecolor=blue,        
    filecolor=magenta,      
    urlcolor=blue           
}

\newcommand{\equref}[1]{Eq.~(\ref{#1})}
\newcommand{\equsref}[2]{Eqs.~(\ref{#1}) and (\ref{#2})}
\newcommand{\secref}[1]{Sec.~\ref{#1}}
\newcommand{\figref}[1]{Fig.~\ref{#1}}
\newcommand{\refcite}[1]{Ref.~\onlinecite{#1}}

\newcommand{\tableref}[1]{Table~\ref{#1}}
\newcommand{\appref}[1]{Appendix~\ref{#1}}

\newcommand{\vpe}{E}

\newcommand{\pdagger}{{\phantom{\dagger}}}
\newcommand{\diff}{\mathrm{d}}
\newcommand{\sign}{\,\text{sign}}
\renewcommand{\approx}{\simeq}

\renewcommand{\vec}[1]{\boldsymbol{#1}}

\definecolor{wrongultramarine}{rgb}{1,0.5,0}

\begin{document}

\title{Theory of zero-field superconducting diode effect in twisted trilayer graphene}

\author{Harley D. Scammell}
\affiliation{School of Physics, the University of New South Wales, Sydney, NSW, 2052, Australia}

\author{J.I.A. Li}
\affiliation{Department of Physics, Brown University, Providence, RI 02912, USA}

\author{Mathias S.~Scheurer}
\affiliation{Institute for Theoretical Physics, University of Innsbruck, Innsbruck A-6020, Austria}

\begin{abstract}
In a recent experiment [Lin \textit{et al.}, arXiv:2112.07841], the superconducting phase hosted by a heterostructure of mirror-symmetric twisted trilayer graphene and WSe$_2$  was shown to exhibit significantly different critical currents in opposite directions in the absence of external magnetic fields. 
We here develop a microscopic theory and analyze necessary conditions for this zero-field superconducting diode effect. 
Taking into account the spin-orbit coupling induced in trilayer graphene via the proximity effect, we classify the pairing instabilities and normal-state orders and derive which combinations are consistent with the observed diode effect, in particular, its field trainability. We perform explicit calculations of the diode effect in several different models, including the full continuum model for the system, and illuminate the relation between the diode effect and finite-momentum pairing. Our theory also provides a natural explanation of the observed sign change of the current asymmetry with doping, which can be related to an approximate chiral symmetry of the system, and of the enhanced transverse resistance above the superconducting transition.
Our findings not only elucidate the rich physics of trilayer graphene on WSe$_2$, but also establish a means to distinguish between various candidate interaction-induced orders in spin-orbit-coupled graphene moir\'e systems, and could therefore serve as a guide for future experiments as well.
\end{abstract}

\maketitle
\tableofcontents

\section{Introduction}

Semiconductor diodes play an essential role in modern electronics---computation, communication and sensing \cite{kitai2011principles}. The diode generates a nonreciprocity, hosting low resistance in one direction, and high resistance in the opposite. In a superconducting diode, the critical supercurrent in one direction is larger than in the opposite.  This feature has elicited fundamental theoretical and experimental studies to uncover the underlying mechanisms. To this end, recent reports of the superconducting diode effect---induced by magnetic field  \cite{Ando2020diodes, DaidoSCDiode, Yuan2021diodes, HeSCDiode, Lyu2021, bauriedl2021supercurrent,2021arXiv210800209I}, magnetic proximity \cite{shin2021magnetic,2021arXiv210800209I}, or magnetic Josephson or tunnel junctions \cite{PhysRevLett.99.067004,Buzdin2008, Szombati2016, Kopasov2021, Baumgartner2021, Diez2021magnetic, baumgartner2021effect, wu2021realization, strambini2021rectification,2021arXiv211101242H}---have emerged and attracted considerable  attention.
Having nonreciprocity in common with the semiconductor, yet boasting zero resistance, superconducting diodes have potential as building blocks for future quantum electronics.

\begin{figure}[tb]
   \centering
    \includegraphics[width=1 \linewidth]{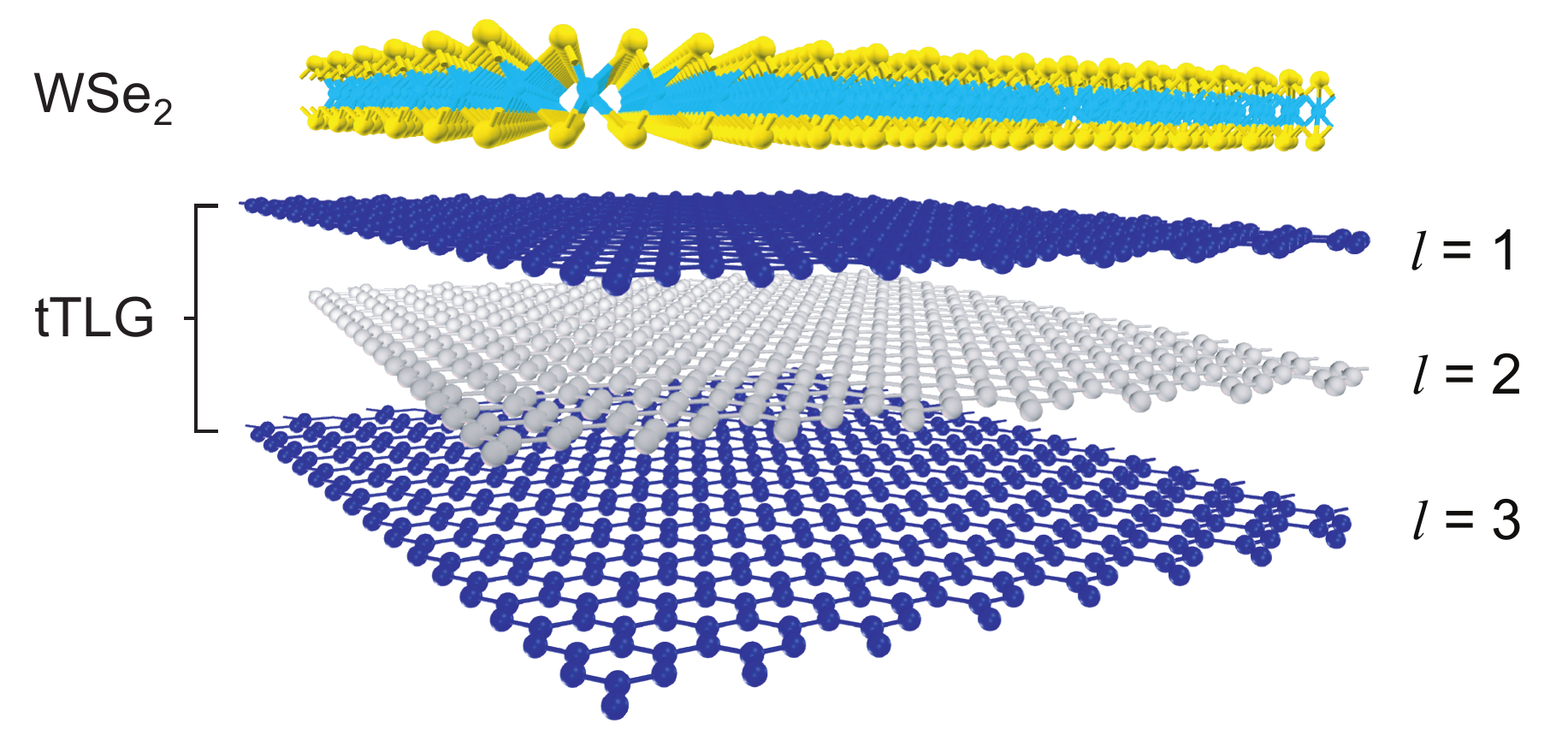}
    \caption{Schematic of the  tTLG and WSe$_2$ heterostrucure; graphene layers are labelled by $l=1,2,3$, and have alternating twist angles such that the tTLG subsystem is mirror symmetric.}
    \label{fig:overview}
\end{figure}

A recent study \cite{SCPaper} (companion to this work) considers a heterostructure consisting of twisted trilayer graphene (tTLG) \cite{Park_2021,Hao_2021,2021arXiv210312083C,2021arXiv210912127K,2021arXiv210912631T,2021arXiv210803338L} and WSe$_2$, as depicted in \figref{fig:overview}, and demonstrates a superconducting diode effect in the absence of external magnetic fields, magnetic proximity or a magnetic junction; for brevity, we here refer to this effect as the zero-field superconducting diode effect (ZFDE). In addition, several revealing features of the ZFDE were reported: (i) the diode effect, i.e., the asymmetry $\delta J_c$ of the current along opposite directions, can be trained by a small out-of-plane magnetic field, (ii) $\delta J_c$ can be reversed by doping, and (iii) the system exhibits an enhanced transverse resistance in a small temperature range above the superconducting $T_c$. ``Untraining'' the ZFDE, also suppresses this enhancement, implying a direct connection to the diode effect.

The emergence of a ZFDE in tTLG/WSe$_2$ implies a coexistence between superconductivity and spontaneously broken time-reversal and $C_{2z}$ symmetries.  Such a coexistence is a rare occurrence---superconducting states are typically restricted to systems whereby pairing occurs between time-reversed partners. Establishing an understanding of the ZFDE is therefore of fundamental concern. 
Moreover, understanding the manipulation of the ZFDE---as per the observations (i--iii)---offers potential for future technological applications. 
The reported phenomenology of superconductivity and possible related instabilities in tTLG on WSe$_2$ \cite{SCPaper} therefore offers an exciting challenge to theory, and demands examination. 

The purpose of this work is to provide a theoretical understanding of the phenomenology of the tTLG/WSe$_2$ heterostructure, with a primary focus on the ZFDE. Our analysis comprises general symmetry arguments as well as explicit model calculations, which taken together illuminate the ZFDE, the possible superconducting and normal-state instabilities of the system, the emergence of vestigial orders, and the influence of spin-orbit coupling (SOC) and external magnetic fields. It also provides important constraints on the possible origins of the ZFDE in \cite{SCPaper} which, in turn, reveal information about the many-body physics in the tTLG/WSe$_2$ heterostructure. 
Moreover, our analysis offers an explanation of the findings (i--iii). 
In particular, we consider a diode effect arising due to coexistence of superconductivity and a normal-state order, and determine the symmetry requirements---namely, which perturbation (or combination) out of SOC, strain and displacement field are sufficient---for the ZFDE and the its field training [see (i) above]. 
We find that only a small set of (four) possible normal state orders are consistent with the observed field trainabiltiy, and which further become symmetry equivalent in the limit of strong SOC. 
We provide an explanation of the doping dependence (ii) of the diode effect, which can be understood by invoking the approximate chiral symmetry of the moir\'e bands.   
And, concerning (iii), we argue how vestigial nematic order arises, and that quite generally it is expected to remain ordered above, yet in the vicinity of, the superconducting critical temperature---in which case offering an explanation of the enhanced transverse resistance reported \cite{SCPaper}. 
Additionally, we show that there is one unconventional pairing state that spontaneously breaks time-reversal symmetry and allows for a ZFDE without normal-state order.
Along the way, we illuminate the relation between finite-momentum pairing and the diode effect.

The rest of the paper is organized as follows:  We begin in \secref{ModelSymmetries} by providing a continuum noninteracting model of the tTLG setup, and present the symmetries on the model.  In  \secref{PossiblePairingStates} we discuss the possible pairing states in tTLG starting from zero SOC, and adiabatically turning it on. In  \secref{SymmetryAnalysisDiodeEffect} we provide a detailed symmetry analysis of the diode effect: determining which candidate normal-state orders can support the ZDE; finding which orders allow for magnetic field training of the diode effect; and finally presenting a means to generate the ZDE without normal state order. In \secref{ModelCalculations} we turn to explicit model calculations, presenting first the general formalism to compute the critical current, and subsequently applying it to a semi-analytic patch theory, toy models on the full MBZ, and finally to the full continuum model of tTLG with and without SOC. In \secref{DopingDependence} we turn to a curious and striking experimental feature---the doping dependence of the diode effect---and present a mechanism that explains the experimental observations thereof \cite{SCPaper}. Conclusion and outlook are provided in \secref{Outlook}

\section{Model and symmetries}\label{ModelSymmetries}
To set the stage for the analysis in the subsequent section, we will here define the models we will use for tTLG on WSe$_2$ throughout this work, and discuss its symmetries. 

\subsection{Notation and continuum model}\label{ContinuumModel} 
The heterostructure studied in \refcite{SCPaper} consists of tTLG near its magic angle and WSe$_2$, as depicted in \figref{fig:overview}. 
To describe the three layers, $l=1,2,3$, of graphene with alternating twist angle, we will employ the three-layer generalization of the commonly used continuum model of twisted-bilayer graphene \cite{dos2007graphene,bistritzer2011moire,dos2012continuum}, where the magic angle occurs at around $1.58^{\circ}$. The impact of WSe$_2$ is taken into account via the proximity-induced spin-orbit terms \cite{Gmitra2015,2021arXiv210806126N}. Starting in a real-space description, with $c^\dagger_{\vec{r};\rho,l,\eta,s}$ denoting the creating operator of an electron at position $\vec{r}\in\mathbbm{R}^2$, on sublattice $\rho$, in layer $l$, valley $\eta$, and of spin $s$, the non-interacting Hamiltonian can be written as
\begin{subequations}\begin{equation}
    H_0 = \int\diff \vec{r} \, c^\dagger_{\vec{r}}\, h(\vec{r},\vec{\nabla}) \, c^\pdagger_{\vec{r}}, \label{FullContinuumHamiltonian}
\end{equation}
where $h$ is a matrix in sublattice, layer, valley, and spin space [indices suppressed in \equref{FullContinuumHamiltonian}] and consists of the following terms
\begin{equation}
    h = \sum_{l=1}^3 h_l^{\text{g}} + h^{\text{t}} + h^{\text{D}} + h^{\text{SOC}}. \label{DifferentPartsofContHam}
\end{equation}\label{ContinuumModelDefinition}\end{subequations}
Here, $h_l^{\text{g}}$, $l=1,2,3$, are the Dirac Hamiltonians associated with each individual graphene layer $l$, twisted by angle $\theta_l = (-1)^l \theta/2$; it reads as $h_l^{\text{g}} = -iv_F e^{i \frac{\theta_l}{2} \rho_3} \vec{\rho}_\eta \vec{\nabla} e^{-i \frac{\theta_l}{2} \rho_3} $, where $\vec{\rho}_\eta = (\eta_z\rho_x,\rho_y)$ and $\rho_j$ are Pauli matrices in sublattice space. The second term in \equref{DifferentPartsofContHam} captures the tunneling between adjacent graphene layers, $(h^{\text{t}})_{\rho,l,\eta,s;\rho',l',\eta',s'}= \delta_{s,s'}\delta_{\eta,\eta'}(\delta_{l+1,l'} T_{\eta\vec{r}} + \text{H.c.})_{\rho,\rho'}$, which is modulated on the moir\'e scale \cite{bistritzer2011moire}, $T_{+,\vec{r}} = e^{-i\vec{q}_1\vec{r}}[ \mathcal{T}_1 + \mathcal{T}_2 e^{-i\vec{G}_1\vec{r}} + \mathcal{T}_3 e^{-i(\vec{G}_1+\vec{G}_2)\vec{r}}] = T_{-,\vec{r}}^*$. The momenta involved here are the momentum transfer, $\vec{q}_1 = k_{\theta} (0,-1)$, $k_\theta = 2|\text{K}_g|\sin(\theta/2)$, from the K to the K$'$ point at the corners of the moir\'e Brillouin zone (MBZ) as well as the basis vectors, $\vec{G}_1 = -\sqrt{3}k_\theta (1,\sqrt{3})^T/2$ and $\vec{G}_2 = \sqrt{3}k_\theta(1,0)^T$, of the reciprocal moir\'e lattice (RML). In this expression, $\mathcal{T}_j$ are matrices in sublattice space which only exhibit two independent real parameters---the intra ($w_0$) and intersublattice ($w_1$) hopping---and can be written as $\mathcal{T}_j = [w_0 \rho_0 + w_1 e^{i\frac{2\pi}{3}(1-j)}(\rho_x + i \rho_y)+\text{H.c.}]/2$.
Since $h^{\text{t}}$ breaks the continuous translation symmetry of the Dirac Hamiltonians $h_l^{\text{g}}$ but preserves translations on the moir\'e scale, it reconstructs the Dirac cones into moir\'e bands. The latter are derived by Fourier transformation of $c_{\vec{r};\rho,l,\eta,s}$, leading to the momentum-space operators $c_{\vec{k};\rho,l,\eta,s,\vec{G}}$ where $\vec{k} \in\text{MBZ}$ and $\vec{G}\in\text{RML}$. For convenience, \appref{a:continuum} provides the explicit form of the continuum model \equref{ContinuumModelDefinition} written in momentum space.

The third term in \equref{DifferentPartsofContHam} describes the impact of a perpendicular electric field (displacement field) $D_0$ and is given by $(h^{\text{D}})_{l,l'} = D_0 \rho_0\eta_0s_0 \delta_{l,l'} (\delta_{l,1} - \delta_{l,3})$, where we use, as above, the same symbol for Pauli matrices and the associated index, i.e., $\rho_j$, $\eta_j$, and $s_j$ are Pauli matrices in sublattice, valley, and spin space, respectively.

Finally, the last term in \equref{DifferentPartsofContHam} captures the impact of the WSe$_2$ crystal and, thus, constitutes the crucial difference between tTLG and the system that has been shown to exhibit a diode effect in \refcite{SCPaper}. It is also the part of the Hamiltonian (\ref{DifferentPartsofContHam}) that has not been discussed in previous theoretical works on tTLG \cite{KhalafKruchkov2019, CarrKruchkov2020, MoraRegnault2019}. The form of $h^{\text{SOC}}$ we use is motivated by the fact that the overlap of wavefunctions of WSe$_2$ and of the graphene layers $l=2,3$ in \figref{fig:overview} is negligibly small and, hence, the proximity effect predominately affects the graphene layer $l=1$ and, in that layer, is of the same form as for a single layer of graphene on WSe$_2$. Furthermore, to capture the relevant low-energy moir\'e bands it is sufficient to use the single-layer model expanded around each Dirac cone. This is not only true for $h_l^{\text{g}}$ but also for the impact of SOC. So we can focus on the leading, momentum-independent, terms in $h^{\text{SOC}}$ which can be written as \cite{Gmitra2015}
\begin{align}\begin{split}
    h^{\text{SOC}} =& P_1 \Bigl[ \lambda_{\text{I}} s_z \eta_z + \lambda_{\text{R}} \left(\eta_z \rho_x s_y - \rho_y s_x \right) \\ 
    & \qquad + \lambda_{\text{KM}} \eta_z \rho_z s_z + m \rho_z  \Bigr], \label{SOCTerms}
\end{split}\end{align}
where $(P_1)_{l,l'} = \delta_{l,l'}\delta_{l,1}$ projects onto the first graphene layer. The first three contributions in \equref{SOCTerms} are SOC terms as they intertwine spin, $\vec{s}$, with orbital degrees of freedom, in this case valley and sublattice. These terms  are often referred to as ``Ising'' ($\lambda_{\text{I}}$), ``Rashba'' ($\lambda_{\text{R}}$), and a ``Kane-Mele'' ($\lambda_{\text{KM}}$) SOC. The last term in \equref{SOCTerms} is a sublattice-imbalance term that can also be induced by the WSe$_2$.

Recent first-principle calculations \cite{2021arXiv210806126N} show that, in particular for the rather large twist angles between WSe$_2$ and graphene in experiment \cite{SCPaper}, $m$ and $\lambda_{\text{KM}}$ are much smaller than $\lambda_{\text{I}}$ and $\lambda_{\text{R}}$. While our analysis can be straightforwardly generalized to include both $m$ and $\lambda_{\text{KM}}$, we set $m=\lambda_{\text{KM}}=0$ from here on, to simplify the presentation of the results. Additionally, we note that to partially account for relaxation effects, we take $w_0/w_1=0.875$, with $w_1=110$ meV, these values are chosen to best agree with experiment \cite{DWPaper}. Finally, for demonstration in Section \ref{s:tTLGdiode}, we will specialize to twist angle $1.5^{\circ}$, which lies within the range of twist angles considered experimentally \cite{SCPaper}.

\subsection{Symmetries}\label{DiscussionOfSymmetries}
Having motivated and defined the model, $H_0$, we use, let us next discuss its symmetries. As the symmetries of the continuum-model description of tTLG without WSe$_2$ have already been discussed in detail in previous theoretical works \cite{Christos29543,2021PhRvB.103s5411C}, we will here focus on the modifications when $\lambda_{\text{I}}$ and/or $\lambda_{\text{R}}$ in \equref{SOCTerms} are non-zero.

\begin{table}[tb]
\begin{center}
\caption{Action of the different point symmetries $S$ on the microscopic field operators ($\psi_{\vec{k}}$) of the continuum-model description of tTLG (see \secref{ContinuumModel}) and on the operators $f_{\vec{k}}$ in \equref{LowEnergyBandstructure} for the Bloch states closest to the Fermi level. The last column shows which of strain, $\beta$, Rashba, $\lambda_{\text{R}}$, and Ising, $\lambda_{\text{I}}$, SOC break the respective symmetry. As in the main text, we use $\rho_j$, $\eta_j$, and $s_j$ to denote Pauli matrices in sublattice, valley, and spin space. Note that $\Theta$ and $\Theta_s$ are anti-unitary operators.}
\label{ActionOfSymmetries}
\begin{ruledtabular}
 \begin{tabular} {cccc} 
$S$  & $S\psi_{\vec{k};\ell,\vec{G}}S^\dagger$ & $S f_{\vec{k}} S^\dagger$ & broken by \\ \hline
SO(3)$_{s}$    & $e^{i\vec{\varphi}\cdot \vec{s}}  \psi_{\vec{k};\ell,\vec{G}}$ &  $e^{i\vec{\varphi}\cdot \vec{s}} f_{\vec{k}}$  & $\lambda_{\text{R}},\lambda_{\text{I}}$  \\ 
SO(2)$_{s}$     & $e^{i \varphi s_z}  \psi_{\vec{k};\ell,\vec{G}}$ & $e^{i \varphi s_z} f_{\vec{k}}$ & $\lambda_{\text{R}}$  \\ \hline
$C_{3z}$     & $e^{i\frac{2\pi}{3} \rho_z\eta_z}  \psi_{C_{3z}\vec{k};\ell,C_{3z}\vec{G}}$ & $f_{C_{3z}\vec{k}}$ & $\lambda_{\text{R}},\beta$  \\
$C^s_{3z}$     & $e^{i\frac{2\pi}{3} (\rho_z\eta_z + s_z)}  \psi_{C_{3z}\vec{k};\ell,C_{3z}\vec{G}}$ & $e^{i\frac{2\pi}{3}s_z}f_{C_{3z}\vec{k}}$ & $\beta$  \\ \hline
$C_{2z}$    & $\eta_x \rho_x \psi_{-\vec{k};\ell,-\vec{G}}$ & $\eta_x f_{-\vec{k}}$ & $\lambda_{\text{R}},\lambda_{\text{I}}$  \\
$C^s_{2z}$    & $s_z\eta_x \rho_x \psi_{-\vec{k};\ell,-\vec{G}}$ & $s_z\eta_x f_{-\vec{k}}$ & $\lambda_{\text{I}}$  \\
$C^{s'}_{2z}$    & $s_{x,y}\eta_x \rho_x \psi_{-\vec{k};\ell,-\vec{G}}$ & $s_{x,y}\eta_x f_{-\vec{k}}$ & $\lambda_{\text{R}}$  \\ 
$\sigma_h$    & $ (1,1,-1)_\ell \psi_{\vec{k};\ell,\vec{G}}$ & $\pm f_{\vec{k}}$ & $D_0,\lambda_{\text{R}},\lambda_{\text{I}}$  \\
$I$     & $ \eta_x\rho_x(1,1,-1)_\ell \psi_{-\vec{k};\ell,\vec{G}}$ &  $\pm\eta_x f_{-\vec{k}}$ & $D_0,\lambda_{\text{R}},\lambda_{\text{I}}$  \\ \hline
$\Theta$     & $\eta_x \psi_{-\vec{k};\ell,-\vec{G}}$ & $\eta_x f_{-\vec{k}}$  & $\lambda_{\text{R}},\lambda_{\text{I}}$  \\
$\Theta^s$    & $is_y\eta_x \psi_{-\vec{k};\ell,-\vec{G}}$ & $is_y\eta_x f_{-\vec{k}}$  & ---  \\ \hline
U(1)$_v$    & $e^{i\varphi \eta_z} \psi_{\vec{k};\ell,\vec{G}}$ & $e^{i\varphi \eta_z}  f_{\vec{k}}$  & ---
 \end{tabular}
\end{ruledtabular}
\end{center}
\end{table}

As with any SOC term, both $\lambda_{\text{I}}$ and/or $\lambda_{\text{R}}$ break the spin-rotation symmetry, SO(3)$_s$; while $\lambda_{\text{I}}$ leaves a residual spin-rotation along the $s_3$ axis as a symmetry, hence, the name ``Ising SOC'', $\lambda_{\text{R}}$ breaks SO(3)$_s$ completely. This is summarized in the first two lines of \tableref{ActionOfSymmetries}, where we also list the action of symmetries in the continuum model outlined in \secref{ContinuumModel} above. For simplicity and future reference, we use $\psi_{\vec{k};\rho,\ell,\eta,s,\vec{G}}$ to define the representation of the symmetries in \tableref{ActionOfSymmetries}; these operators are related to $c_{\vec{k};\rho,l,\eta,s,\vec{G}}$ by a unitary transformation in layer space \cite{Khalaf_2019}, $c_{\vec{r};\rho,l,\eta,s} = V_{l,\ell} \psi_{\vec{r};\rho,\ell,\eta,s}$, such that $\ell=1,2$ correspond to the mirror-even (invariant under $\sigma_h$ which interchanges the top and bottom layers of tTLG) and $\ell=3$ to the mirror-odd (odd under $\sigma_h$) sector.

Furthermore, the SOC terms intertwine symmetries in real space with symmetries in spin space: while spin-less two-fold rotational symmetry, $C_{2z}$, is broken by both $\lambda_{\text{I}}$ and $\lambda_{\text{R}}$, a certain combination of $C_{2z}$ and a rotation in spin-space, which we denote by $C_{2z}^s$ ($C_{2z}^{s'}$) in \tableref{ActionOfSymmetries}, is preserved if only $\lambda_{\text{R}}$ ($\lambda_{\text{I}}$) is non-zero. Note that there is no two-fold out-of-plane rotation symmetry left once both $\lambda_{\text{I}}$ and $\lambda_{\text{R}}$ are non-zero. Similarly, once $\lambda_{\text{R}}\neq 0$, spin-less $C_{3z}$ is not a symmetry either, while a combination with a three-fold spin-rotation, $C^s_{3z}$, will remain a symmetry for any value of $\lambda_{\text{I}}$ and $\lambda_{\text{R}}$. This symmetry is only broken if either strain is present in the samples \cite{Huder2018, Kazmierczak2021, Kerelsky2019} or if the system develops electronic nematic order \cite{Kerelsky2019, Jiang2019, Choi2019, CaoNematicity2021,rubioverdu2020universal}. We refer to \cite{PhysRevB.100.035448} and \cite{10.1088/2053-1583/abfcd6} for a microscopic description of strain and nematic order in graphene moir\'e systems and here only use a phenomenological parameter $\beta$ to describe the presence ($\beta \neq 0$) or absence ($\beta=0$) of strain in our analysis. Furthermore, also spin-less time-reversal symmetry, $\Theta$, is broken once any of the SOC terms is non-zero, but its spin-full analogue, $\Theta^s$, is always preserved by the non-interacting bandstructure. The same holds for U(1)$_v$ symmetry, see last line in \tableref{ActionOfSymmetries}: as $h$ in \equref{DifferentPartsofContHam} is diagonal in the valley index, the model conserves charge in the two valleys separately.

As is clear geometrically, see \figref{fig:overview}, the mirror symmetry $\sigma_h$ is not only broken by a displacement field but also by the presence of WSe$_2$ on only one side of tTLG. This is why both $\lambda_{\text{I}}$ and $\lambda_{\text{R}}$ [and any term in \equref{SOCTerms} for that matter] break $\sigma_h$, which leads to an admixture of mirror-odd and mirror-even bands. Since three-dimensional inversion symmetry, $I$, is simply the product of a two-fold out-of-plane rotation and $\sigma_h$, the same holds for $I$.

\subsection{Effective low-energy descriptions}\label{LowEnergyModel}
While we use the full continuum model in \equref{ContinuumModelDefinition} to compute the moir\'e bands in the vicinity of the charge neutrality point, we focus on those bands closest to the Fermi level when studying superconductivity and the diode effect below. For a given value of the filling fraction $\nu$, let us denote the creation operator of an electron in the band that is closest to the Fermi level at momentum $\vec{k} \in\text{MBZ}$, of spin species $s$, and in valley $\eta$ by $f_{\vec{k},\eta,s}$. Focusing only on these low-energy electronic degrees of freedom, $H_0$ in \equref{FullContinuumHamiltonian} can be approximated by the effective Hamiltonian,
\begin{equation}
    H_0^{\text{LE}} = \sum_{\vec{k}\in\text{MBZ}} f^\dagger_{\vec{k},\eta,s} \left(h^{\text{LE}}_{\vec{k};\eta}\right)_{s,s'}f^\pdagger_{\vec{k},\eta,s'}, \label{LowEnergyBandstructure}
\end{equation}
which has to be diagonal in $\eta$ due to U(1)$_v$ and $\Theta^s$ imposes the constraint $h^{\text{LE}}_{\vec{k};\eta} = s_2(h^{\text{LE}}_{-\vec{k};-\eta})s_2$. In \tableref{ActionOfSymmetries}, we list the representation of the symmetries discussed in \secref{DiscussionOfSymmetries} above, after appropriate gauge fixing. For instance and future reference, if $\lambda_{\text{R}}=0$, SO(2)$_s$ and $\Theta^s$ imply
\begin{equation}
h^{\text{LE}}_{\vec{k};\eta} = \xi_{\eta\cdot\vec{k}} + \zeta_{\eta\cdot\vec{k}} \,s_z \eta, \label{DispersionForm}
\end{equation}
where $\xi_{\vec{k}}$ and $\zeta_{\vec{k}}$ are smooth functions of momentum [only constrained by $\xi_{\vec{k}} = \xi_{C_{3z}\vec{k}}$ and $\zeta_{\vec{k}} = \zeta_{C_{3z}\vec{k}}$ if $\beta=0$]. Note that $\zeta_{\vec{k}}$ is odd under $\lambda_{\text{I}} \rightarrow -\lambda_{\text{I}}$ and thus vanishes if $\lambda_{\text{I}}=0$.

\section{Superconducting order parameters}\label{PossiblePairingStates}

We here discuss the possible pairing states in the system by starting from trilayer in the absence of WSe$_2$ and adiabatically following its pairing states, in particular, the admixture of further components to them, upon turning on the proximity-induced SOC terms $\lambda_{\text{I}}$ and $\lambda_{\text{R}}$ in \equref{SOCTerms}. 

For now, we will assume that the normal state, in particular, its symmetries, out of which superconductivity emerges is well described by the model in \equref{ContinuumModelDefinition} and postpone the discussion of additional symmetry-breaking particle-hole instabilities to \secref{SymmetryAnalysisDiodeEffect}. As such, the normal state exhibits time-reversal symmetry $\Theta^s$ and it is natural to focus on pairing between electrons with opposite momenta and opposite valley quantum numbers. Using the low-energy description introduced in \secref{LowEnergyModel}, the coupling between the superconducting order parameter $\Delta_{\vec{k};\eta}$ and the fermions can be written as
\begin{equation}
    \Delta H_{\text{SC}} = \sum_{\vec{k}\in\text{MBZ}} f^\dagger_{\vec{k},\eta,s} (\Delta_{\vec{k};\eta})_{s,s'} f^\dagger_{-\vec{k},-\eta,s'} + \text{H.c.}, \label{CouplingOfSCOrderParameter}
\end{equation}
which we decompose into singlet, $\Delta^s_{\vec{k};\eta}$, and triplet, $\vec{d}_{\vec{k};\eta}$, according to 
\begin{equation}
    \Delta_{\vec{k};\eta} = \left( \Delta^s_{\vec{k};\eta} + \vec{d}_{\vec{k};\eta} \cdot \vec{s} \right) i s_y. \label{FormOfSCOrderparameter}
\end{equation}
As mentioned above, we will start in the high-symmetry limit $\lambda_{\text{I}}=\lambda_{\text{R}}=0$ and $D_0=0$, where the system becomes equivalent to mirror-symmetric tTLG. First note that the low-energy Bloch states at $\vec{k}$ and in valley $\eta$ will have the same mirror-symmetry eigenvalue as the state at $-\vec{k}$ and in valley $-\eta$. Therefore, any pairing state in \equref{CouplingOfSCOrderParameter} will be even in $\sigma_h$, which is expected to be energetically most favorable \cite{2021arXiv210602063C}. Although pairing in tTLG in other representations is possible \cite{2021arXiv211011294G,PhysRevResearch.2.033062}, let us further assume that the pairing state transforms trivially under $C_{3z}$, which also avoids nodes in the gap function. This is motivated by the remarkably strong superconductivity in tTLG and the fact that it is enhanced when screening the Coulomb interaction \cite{2021arXiv210803338L}.  The SO(3)$\times C_{6h}$ point symmetry then leaves us with \cite{PhysRevResearch.2.033062} only two remaining superconducting states: first, there is the $A_g^1$ singlet, where \begin{subequations}\begin{equation}
    \psi_{\vec{k},\eta} = \chi_{\vec{k},\eta}, \quad \vec{d}_{\vec{k},\eta} = 0, 
\end{equation}
and, second, the $B_u^3$ triplet with
\begin{equation}
    \psi_{\vec{k},\eta} = 0, \quad \vec{d}_{\vec{k},\eta} = \eta \chi_{\vec{k},\eta} \widehat{\vec{d}},
\end{equation}\label{SCStartingPoint}\end{subequations}
where $\widehat{\vec{d}}$ is a three-component unit vector.
In \equref{SCStartingPoint}, the momentum dependence is parametrized with $\chi_{\vec{k},\eta}$ which is only required to obey $\chi_{\vec{k},\eta} = \chi_{-\vec{k},-\eta}$ and $\chi_{C_3\vec{k},\eta} = \chi_{\vec{k},\eta}$.
If the interactions in the system just couple the densities of electrons in the two valleys but do not exhibit an intervalley Hund's coupling, there will be an enhanced SU(2)$_+\times$SU(2)$_-$ spin symmetry. In that case, the $A_g^1$ and $B_u^3$ states will be exactly degenerate with their order parameter being paramterized by the same basis function $\chi_{\vec{k},\eta}$, as in \equref{SCStartingPoint}. While there are further interesting consequences for superconductivity in the vicinity of this point \cite{PhysRevResearch.2.033062} even without SOC, we will here focus on what happens once $\lambda_{\text{R}}$ and $\lambda_{\text{I}}$ are non-zero.

\begin{figure}
   \centering
    \includegraphics[width=0.95 \linewidth]{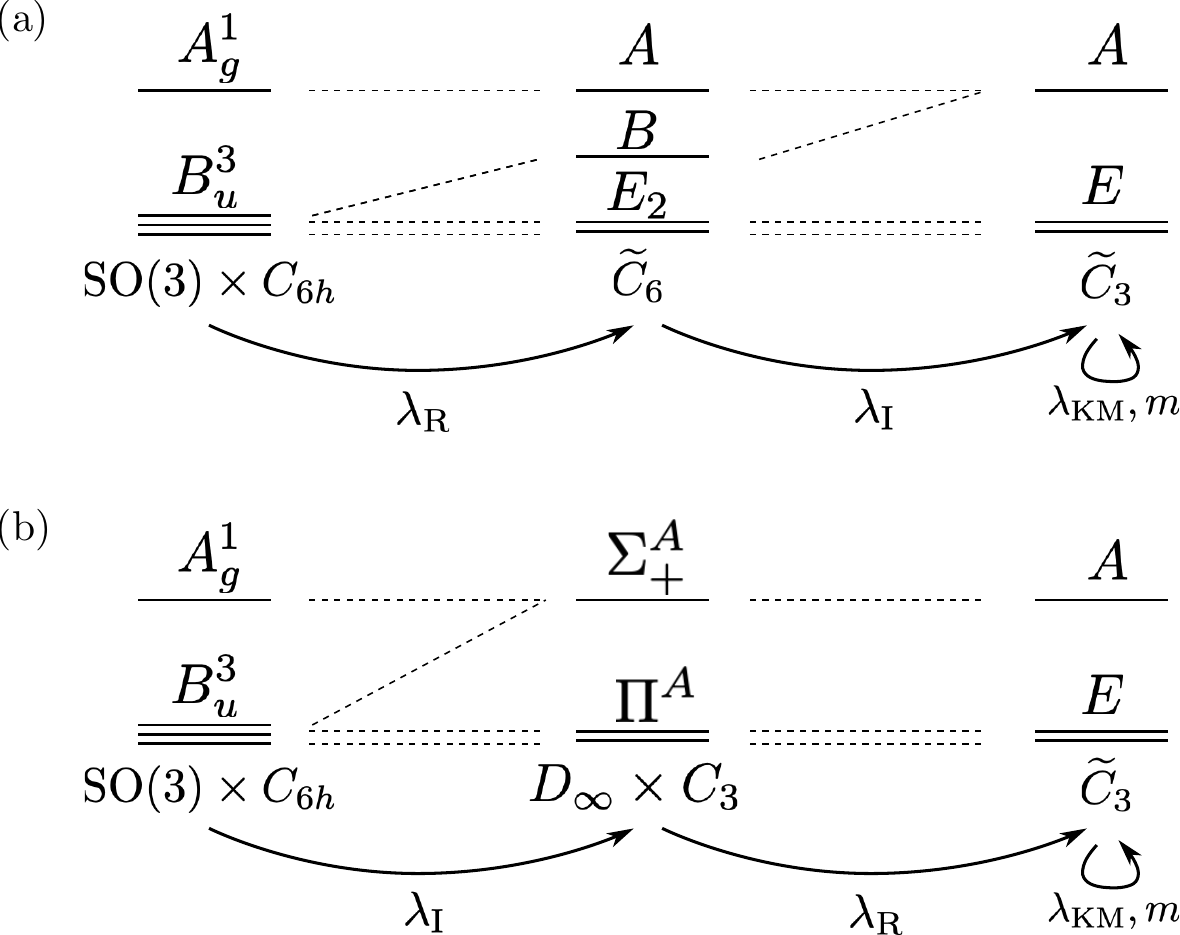}
    \caption{Summary of superconducting states, upon (a) first turning on $\lambda_{\text{R}}$ and then $\lambda_{\text{I}}$ in \equref{SOCTerms}, and vice versa (b). See \secref{PossiblePairingStates} and \appref{FirstLambdaI} for details. Note that further turning on $m$ and $\lambda_{\text{KM}}$ as perturbations at the end does not change the form of the symmetry-allowed components in the superconducting order parameter, as indicated.}
    \label{fig:states}
\end{figure}

Let us begin by discussing the case where $\lambda_{\text{R}}$ is first turned on and then $\lambda_{\text{I}}$, see Fig.~\ref{fig:states}(a).
Finite $\lambda_{\text{R}}$ reduces the point group SO(3)$\times C_{6h}$ to $\widetilde{C}_6$, where the tilde indicates that the elements are combinations of spatial rotations and appropriate spin rotations [formally, $\widetilde{C}_6$ is defined as the group generated by $C_{3z}^sC_{2z}^s$]. As can be worked out by investigation of the representations, the $A_g^1$ singlet transitions into the spin-singlet-triplet admixed $A$ state with
\begin{equation}
    \psi_{\vec{k},\eta} = \chi_{\vec{k},\eta}, \quad \vec{d}_{\vec{k},\eta} = \alpha_1 \begin{pmatrix} X_{\vec{k}} \\ Y_{\vec{k}} \\ 0 \end{pmatrix}  + \alpha_2\, \eta \begin{pmatrix} 2 X_{\vec{k}} Y_{\vec{k}} \\  X^2_{\vec{k}}-Y^2_{\vec{k}} \\ 0 \end{pmatrix},
\end{equation}
where $X_{\vec{k}}$ and $Y_{\vec{k}}$ are MBZ-periodic, real-valued functions transforming as $k_x$ and $k_y$ under $C_6$. Further, $\alpha_j\in\mathbbm{R}$ describe the, in general temperature dependent, admixture of the new component [transforming under $E_{1u}^3$ of SO(3)$\times C_{6h}$] to the pairing states induced by finite $\lambda_{\text{R}}$. 

The $B_u^3$ triplet splits into an $E_2$ doublet, which has an admixed $E_{2g}^1$ singlet component,
\begin{equation}
    \begin{pmatrix} \psi^1_{\vec{k},\eta} \\ \psi^2_{\vec{k},\eta}  \end{pmatrix} =  \alpha_1\begin{pmatrix}2X_{\vec{k}}Y_{\vec{k}}  \\ X^2_{\vec{k}}-Y^2_{\vec{k}} \end{pmatrix}, \quad \begin{pmatrix} \vec{d}^1_{\vec{k},\eta} \\ \vec{d}^2_{\vec{k},\eta} \end{pmatrix} = \eta \chi_{\vec{k},\eta} \begin{pmatrix}
    \vec{e}_x \\ \vec{e}_y
    \end{pmatrix},
\end{equation}
and a purely spin-triplet state transforming under the one-dimensional $B$ representation where
\begin{equation}
    \psi_{\vec{k},\eta} = 0, \quad \vec{d}_{\vec{k},\eta} = \eta \chi_{\vec{k},\eta} \vec{e}_z.
\end{equation}
Note that $C_{2z}^s$ prohibits any singlet component in the $B$ state, despite the presence of SOC.

Once we also turn on $\lambda_{\text{I}}$ [and $m$, $\lambda_{\text{KM}}$ in \equref{SOCTerms} for that matter] we further reduce the point symmetry to $\widetilde{C}_3$. The $E_2$ of $\widetilde{C}_3$ doublet then simply becomes the $E$ of $\widetilde{C}_3$ doublet and mixes with the previous $E_1$ state (with both singlet and triplet components). The resulting order parameter is of the form
\begin{align}\begin{split}
    \begin{pmatrix} \psi^1_{\vec{k},\eta} \\ \psi^2_{\vec{k},\eta}  \end{pmatrix} = \alpha_1\begin{pmatrix}2X_{\vec{k}}Y_{\vec{k}} \\ X^2_{\vec{k}}-Y^2_{\vec{k}} \end{pmatrix} + \alpha_2\, \eta \begin{pmatrix}X_{\vec{k}}\\ Y_{\vec{k}} \end{pmatrix}, \\ 
    \begin{pmatrix} \vec{d}^1_{\vec{k},\eta} \\ \vec{d}^2_{\vec{k},\eta} \end{pmatrix} = \eta \chi_{\vec{k},\eta} \begin{pmatrix}
    \vec{e}_x \\ \vec{e}_y
    \end{pmatrix} + \alpha_3 \vec{e}_z \begin{pmatrix}X_{\vec{k}}\\ Y_{\vec{k}} \end{pmatrix}.
\label{FormOfTheEState}\end{split}\end{align}
Furthermore, the $A$ and $B$ states merge into the singlet-triplet admixed $A$ state of $\widetilde{C}_3$ with order parameter
\begin{subequations}\begin{align}\begin{split}
    \psi_{\vec{k},\eta} &= \chi_{\vec{k},\eta}, \\ 
    \vec{d}_{\vec{k},\eta} &= \alpha_1 \begin{pmatrix} X_{\vec{k}} \\ Y_{\vec{k}} \\ 0 \end{pmatrix}  + \alpha_2\, \eta \begin{pmatrix} 2 X_{\vec{k}} Y_{\vec{k}} \\  X^2_{\vec{k}}-Y^2_{\vec{k}} \\ 0\end{pmatrix}  + \alpha_3 \eta\, \chi_{\vec{k},\eta} \vec{e}_z,
\label{ABAdmixed}\end{split}\end{align}
if the $A$ state is dominant at $\lambda_{\text{I}}=0$; if instead, the $B$ state dominates without $\lambda_{\text{I}}$, \equref{ABAdmixed} becomes
\begin{align}\begin{split}
    \psi_{\vec{k},\eta} &= \alpha_1\chi_{\vec{k},\eta}, \\ 
    \vec{d}_{\vec{k},\eta} &= \eta \chi_{\vec{k},\eta} \vec{e}_z + \alpha_2 \begin{pmatrix} X_{\vec{k}} \\ Y_{\vec{k}} \\ 0 \end{pmatrix}  + \alpha_3 \, \eta \begin{pmatrix} 2 X_{\vec{k}} Y_{\vec{k}} \\  X^2_{\vec{k}}-Y^2_{\vec{k}} \\ 0\end{pmatrix}.
\end{split}\label{ABAdmixed2}\end{align}\end{subequations}
We thus see that there are two main classes of superconducting instabilities in tTLG proximity coupled to WSe$_2$, which are associated with the irreducible representation (IR) $A$ and $E$. More specifically, by virtue of being one-dimensional, the IR $A$ is only associated with a single superconducting state, with order parameter given in \equref{ABAdmixed} or \equref{ABAdmixed2}, which preserves both $C_{3z}^s$ and $\Theta^s$; obeying all symmetries of the normal state, this state can be fully gapped and is expected to be realized if electron-phonon coupling or the fluctuation of a time-reversal-even normal state order provides the pairing glue \cite{PhysRevB.93.174509,OurMicroscopicTDBG}. 
The IR $E$ is two-dimensional and associated with two distinct pairing states: expanding the superconducting order parameter (\ref{FormOfSCOrderparameter}) as 
\begin{equation}
    \Delta_{\vec{k};\eta} = \sum_{j=1,2} c_j (\psi^j_{\vec{k},\eta} + \vec{d}_{\vec{k},\eta}^j \cdot \vec{s}), \label{EExpansionInBasisFunc}
\end{equation}
with complex coefficients $c_{1,2}$ and $\psi^j_{\vec{k},\eta}$, $\vec{d}_{\vec{k},\eta}^j $ given in \equref{FormOfTheEState}, the first state is the nematic $E_{(1,0)}$ superconductor with $c=(1,0)^T$ (and symmetry-related configurations). It breaks $C_{3z}^s$ but respects $\Theta^s$ and will have point nodes. The second state is the chiral $E_{(1,i)}$ superconductor with $c=(1,i)^T$ (and symmetry-related configurations), which preserves $C_{3z}^s$ but breaks $\Theta^s$; unless the Fermi surface crosses the $\Gamma$, K or K' point of the MBZ, it will be fully gapped. Both $E$ states can only be stabilized by an unconventional pairing mechanism based on fluctuations of a time-reversal-odd order parameter \cite{PhysRevB.93.174509}, such as spin fluctuations. As long as fluctuation corrections to mean-field \cite{PhysRevResearch.2.033062} can be neglected in the computation of the quartic terms in the GL expansion, the chiral state will always be favored over the nematic superconductor \cite{SelectionRulePaper}.

The order in which we turned on $\lambda_{\text{R}}$ and $\lambda_{\text{I}}$ in our analysis above has important consequences if $\lambda_{\text{R}}$ is sizeable but $\lambda_{\text{R}} \gg \lambda_{\text{I}}$. In that case, one should primarily think in terms of the three candidate states $A$, $B$, $E_2$ of $\widetilde{C}_6$. For instance, if $B$ is preferred, the order parameter has the form of \equref{ABAdmixed2} with $\alpha_{1,2,3}$ being small in $\lambda_{\text{I}}$. In case of $E_2$, the order parameter is given by \equref{FormOfTheEState} where $\alpha_1$ captures an order-one singlet-triplet admixture while $\alpha_{2,3}$ are small.

To understand pairing in the opposite limit, $\lambda_{\text{I}} \gg \lambda_{\text{R}}$, we have studied the evolution of pairing state when first turning on $\lambda_{\text{I}}$ before $\lambda_{\text{R}}$. The result is summarized graphically in Fig.~\ref{fig:states}(b) and the detailed form of the order parameters can be found in \appref{FirstLambdaI}.

\section{Symmetry analysis of diode effect}\label{SymmetryAnalysisDiodeEffect}
To begin our symmetry discussion of the diode effect, let us first assume that the necessary symmetry requirements are due to some additional normal-state order, i.e., interaction-induced spontaneous symmetry breaking that is already present in the normal state out of which superconductivity emerges, while no additional symmetries are spontaneously broken at the superconducting transition.

\begin{table*}[tb]
\begin{center}
\caption{Summary of candidate instabilities based on the analysis of \cite{2021arXiv210602063C}. For each state, we list a momentum-independent form of its order parameter $m_j$ in the TBG-like bands, see \equref{SimpleDefinitionOfStates}, using $\sigma_j$, $\eta_j$, and $s_j$ to denote Pauli matrices in band, valley, and spin space, respectively. We indicate how each $m_j$ transforms under the symmetries of the system, listed in \tableref{ActionOfSymmetries}. To this end, we use $A$ ($E$) to denote the trivial (complex) IR of $C_{3z}^s \cong \mathbb{Z}_3$ and $\vec{1}$ ($\vec{3}$) for the singlet (triplet) representation of the SO(3)$_s$ spin rotation group. In the column ``Hund's p.'' we indicate the Hund's partner \cite{2021arXiv210602063C,PhysRevResearch.2.033062} of each candidate order, i.e., the state it becomes degenerate with in the absence of SOC and intervalley Hund's interactions, leading to an enhanced SU(2)$_+\times$SU(2)$_-$ spin symmetry. Finally, the last two columns indicate which states become symmetry-equivalent, i.e., will start to mix, once $\lambda_{\text{I}}$ or $\lambda_{\text{R}}$ are non-zero. For convenience of the reader, this is also summarized graphically in \figref{fig:evolutionofnormalstateorders}.}
\label{ListOfOrderParameters}
\begin{ruledtabular}
 \begin{tabular} {cccccccccccc} 
type & $m_j$ & SO(3)$_s$ & $C_{2z}^s$ & $C_{2z}^{s'}$ & $C_{3z}^s$ & U(1)$_v$ & $\Theta_s$ & Hund's p.& $\lambda_{\text{I}}\neq 0$ & $\lambda_{\text{R}}\neq 0$ \\ \hline
SP$_\perp$/SP$_\parallel$ & $\sigma_0\eta_0s_z$/$\sigma_0\eta_0(s_x,s_y)$ & $\vec{3}$ & +/$-$ & $-$/$\pm$ & A/E & $0$ & $-$ & SVP & VP/SSLP$^+_{\parallel}$ & SLP$^-$/--- \\
SVP$_\perp$/SVP$_\parallel$ & $\sigma_0\eta_zs_z$/$\sigma_0\eta_z(s_x,s_y)$ & $\vec{3}$ & $-$/+ & $+$/$\mp$ & A/E & $0$ & $+$ & SP & ---/SSLP$^-_{\parallel}$ & SLP$^+$/---  \\ 
VP & $\sigma_0\eta_zs_0$ & $\vec{1}$ & $-$ & $-$ & A & $0$ & $-$ & --- & SP$_\perp$ & SSLP$^+_{\perp}$ \\  \hline
SLP$^-$ & $\sigma_y\eta_0s_0$ & $\vec{1}$ & $+$ & $+$ & A & $0$ & $-$ & --- & SSLP$^+_\perp$ & SP$_\perp$ \\ 
SLP$^+$ & $\sigma_y\eta_zs_0$ & $\vec{1}$ & $-$ & $-$ & A & $0$ & $+$ & --- & SSLP$^-_\perp$ & SVP$_\perp$ \\ 
SSLP$^-_{\perp}$/SSLP$^-_{\parallel}$ & $\sigma_y\eta_0s_z$/$\sigma_y\eta_0(s_x,s_y)$ & $\vec{3}$ & $+$/$-$ & $-$/$\pm$ & A/E & $0$ & $+$ & SSLP$^+$ & SLP$^+$/SVP$_\parallel$ & ---/--- \\
SSLP$^+_{\perp}$/SSLP$^+_{\parallel}$ & $\sigma_y\eta_zs_z$/$\sigma_y\eta_z(s_x,s_y)$ & $\vec{3}$ & $-$/$+$ & $+$/$\mp$ & A/E & $0$ & $-$ & SSLP$^-$ & SLP$^-$/SP$_{\parallel}$ & VP/--- \\ \hline
IVC$^+$ & $\sigma_0\eta_{x,y}s_0$ & $\vec{1}$ & $\pm$ & $\pm$ & A & $1$ & $+$ & SIVC$^+$ & SIVC$^-_\perp$ & SIVC$^-_{\perp}$ \\
IVC$^-$ & $\sigma_y\eta_{x,y}s_0$ & $\vec{1}$ & $\pm$ & $\pm$ & A & $1$ & $-$ & SIVC$^-$ & SIVC$^+_\perp$ &  SIVC$^+_\perp$ \\
SIVC$^+_\perp$/SIVC$^+_{\parallel}$ & $\sigma_0\eta_{x,y}s_z$/$\sigma_0\eta_{x,y}(s_x,s_y)$ & $\vec{3}$ & $\pm/\mp$ & $\mp/\pm$ & A/E & $1$ & $-$ & IVC$^+$ & IVC$^-$/--- & IVC$^-$/--- \\
SIVC$^-_\perp$/SIVC$^-_{\parallel}$ & $\sigma_y\eta_{x,y}s_z$/$\sigma_y\eta_{x,y}(s_x,s_y)$ & $\vec{3}$ & $\pm/\mp$ & $\mp/\pm$ & A/E & $1$ & $+$ & IVC$^-$ & IVC$^+$/--- & IVC$^+$/---
 \end{tabular}
 \end{ruledtabular}
\end{center}
\end{table*}

\begin{figure}[tb]
   \centering
    \includegraphics[width=\linewidth]{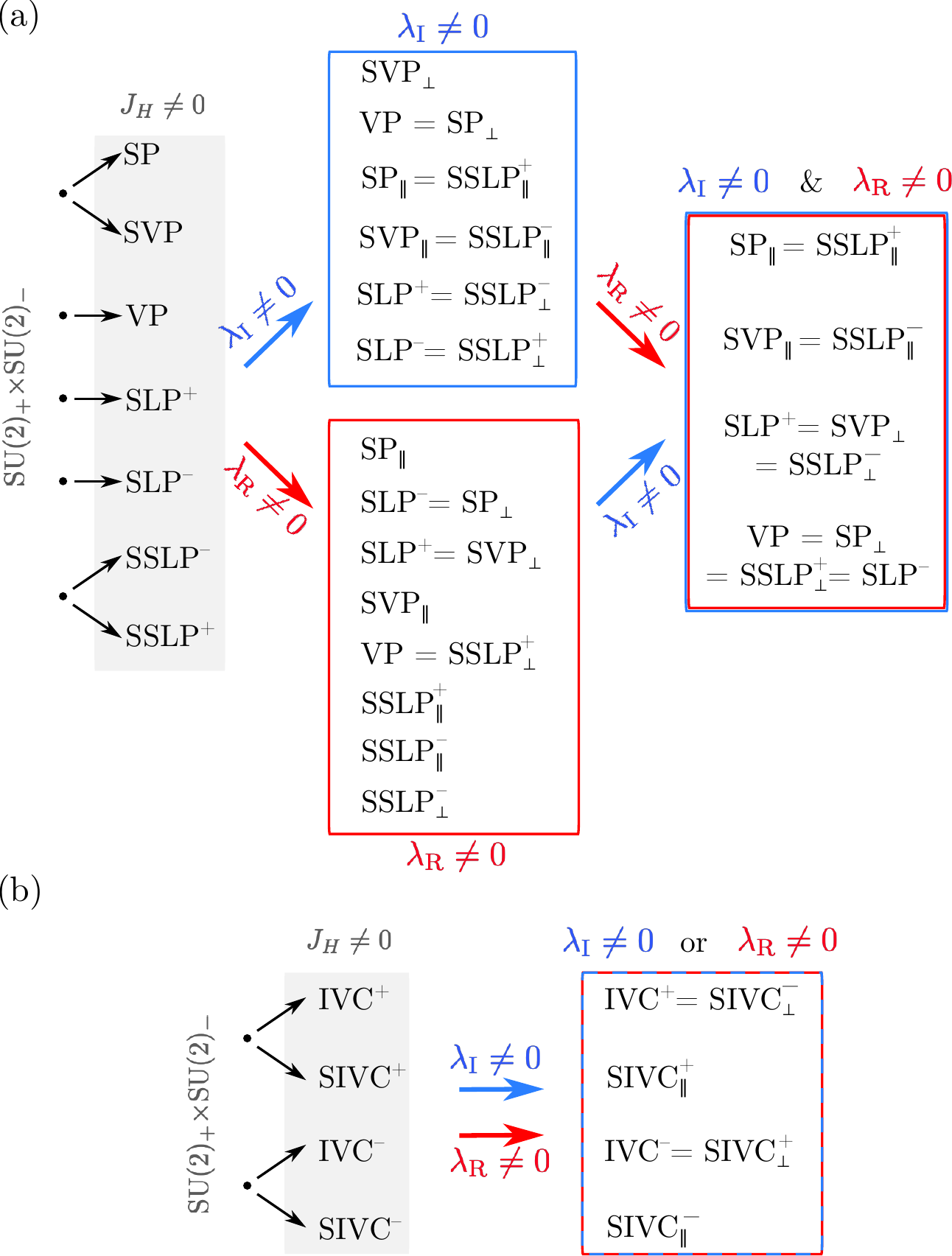}
    \caption{Evolution of normal state orders in \tableref{ListOfOrderParameters} upon turning on the SOC terms $\lambda_{\text{R}}$ and $\lambda_{\text{I}}$. As neither of these terms break the valley U(1)$_v$ symmetry, we can group these states into (a) U(1)$_v$-preserving and (b) intervalley coherent (IVC) phases.}
    \label{fig:evolutionofnormalstateorders}
\end{figure}

\subsection{Candidate normal-state orders}\label{CandidateNormalStateOrders}
To perform a systematic analysis of all possible normal state orders causing the ZFDE, we will start in the limit without SOC, setting all terms in \equref{SOCTerms} to zero. Then the system is equivalent to tTLG and we can use the set of candidate orders derived in \cite{2021arXiv210602063C}, where it was shown that all order parameters listed in \tableref{ListOfOrderParameters} constitute exact ground states in the chiral-flat-decoupled limit; the latter is defined by $w_0=0$ in $h^{\text{t}}$, $D_0=0$, and setting the bandwidth of the flat-bands in the mirror-even sector to zero. Tuning away from this limit will induce a multitude of different energetic contributions, removing the degeneracy between these candidate states, as detailed in \refcite{2021arXiv210602063C}. 

More precisely, for each of these states, the mirror-even sector of the theory will be a correlated, but symmetry unbroken, semi-metal, while the mirror-odd sector with twisted-bilayer-graphene-like bandstructure exhibits the respective order parameters $\Phi_j$, coupling to the electrons as 
\begin{equation}
    \Delta H_0 = \sum_{\vec{k}\in\text{MBZ}} b_{\vec{k}}^\dagger m_j b_{\vec{k}}^\pdagger \,\Phi_j. \label{SimpleDefinitionOfStates}
\end{equation}
In \equref{SimpleDefinitionOfStates}, $m_j$ are matrices in valley, spin, and band-space; the latter is spanned by the conduction (just above the charge-neutrality point) and valence (just below it) flat-bands of twist-bilayer graphene. The associated field operators are $b^\dagger_{\vec{k};p,\eta,s}$ which create Bloch electrons in the mirror-even conduction ($p=+$) or valence ($p=-$) flat band, in valley $\eta$ and spin $s$. The $m_j$ for each of these candidate orders can be found in the second column of \tableref{ListOfOrderParameters}, where we denote Pauli matrices in band space by $\sigma_j$ (index $p=\pm$ above). Note that we use this form of $m_j$ simply to characterize the different phases, in particular their symmetries, which can also be found in \tableref{ListOfOrderParameters}, and that the energetically most favorable version of each phase will have a $\vec{k}$-dependent order parameter that mixes different bands \cite{2021arXiv210602063C}.

Our main focus here will be on the impact of the SOC terms $\lambda_{\text{I}}$ and $\lambda_{\text{R}}$. Both of these terms will reduce the symmetries of the system, see \secref{DiscussionOfSymmetries} and \tableref{ActionOfSymmetries}, with two crucial consequence: first, all spin-polarized orders, which at $\lambda_{\text{I}}=\lambda_{\text{R}}=0$ belong to the three-dimensional IR of SO(3)$_s$ split into two different orders, associated with in-plane ($\parallel$) and out-of-plane spin polarizations ($\perp$). This increases the number of physically distinct candidate phases compared to tTLG \cite{2021arXiv210602063C}. At the same time, the reduction of symmetries reduces the number of IRs and previously distinct orders transform identically under all symmetries of the system. This means that they can mix and should be formally viewed as the same phase, reducing the number of distinct candidate orders. For instance, out-of-plane spin polarization (SP$_\perp$ in \tableref{ListOfOrderParameters}) and valley polarization (VP), while physically distinct for $\lambda_{\text{I}}=0$ due to their behavior under $C_{2z}^s$, become identical once Ising SOC is non-zero, $\lambda_{\text{I}}\neq 0$. Which states become equivalent once $\lambda_{\text{R}}$ or $\lambda_{\text{R}}$ is turned on is summarized in the last two columns in \tableref{ListOfOrderParameters}.

Since the ``evolution'' of the candidate orders is quite complex, we have also illustrated it graphically in \figref{fig:evolutionofnormalstateorders}. Note that U(1)$_v$ symmetry is always preserved in our description which is why intervalley coherent (IVC) states [breaking U(1)$_v$] cannot mix with states that preserve it. As can be seen in \figref{fig:evolutionofnormalstateorders}(a), we end up with only four distinct U(1)$_v$-preserving phases once $\lambda_{\text{I}}$ and $\lambda_{\text{R}}$ are non-zero. However, the diagram contains more relevant information if one of $\lambda_{\text{I}}$, $\lambda_{\text{R}}$ is small. For instance, if $\lambda_{\text{R}}$ is small and only provides a tiny perturbation to the energetics, while $\lambda_{\text{I}}$ is large (compared to the energetic differences between the candidate orders in \tableref{ListOfOrderParameters} at $\lambda_{\text{I}}=\lambda_{\text{R}}=0$, i.e., of order of a couple of meV \cite{2021arXiv210602063C}), \figref{fig:evolutionofnormalstateorders}(a) implies that one should not distinguish between VP and SP$_\perp$, as they will generically mix strongly. However, there is still an important distinction to be made between $\text{VP}=\text{SP}_\perp$ with a little bit of $\text{SSLP}^+_\perp = \text{SLP}^-$ admixed and vice versa [primarily $\text{SSLP}^+_\perp = \text{SLP}^-$ with a little bit of $\text{VP}=\text{SP}_\perp$]. If also $\lambda_{\text{R}}$ is large, this distinction will become irrelevant. As can be seen in \figref{fig:evolutionofnormalstateorders}(b), only four distinct IVC order are possible if $\lambda_{\text{R}}$ or $\lambda_{\text{I}}$ or both are large.

\subsection{Zero-field diode effect}\label{SymmetriesDiodeEffect}
We will next discuss whether a superconducting diode effect is possible by symmetry in the presence of any of the different normal state orders in \tableref{ListOfOrderParameters} and \figref{fig:evolutionofnormalstateorders}. We will here make the natural assumption that the superconductor that emerges out of this symmetry-broken normal state does not spontaneously break additional symmetries. The diode effect for the symmetry-breaking superconductors in the IR $E$, see \secref{PossiblePairingStates}, will be postponed to \secref{ExoticPairingCurrent} below.

To formalize the discussion, let us denote the magnitude of the critical current density for a current along the in-plane direction $\hat{n}$ by $J_c(\hat{n})$. The system exhibits a diode effect if there is some direction $\hat{n}$ for which the current asymmetry,
\begin{equation}
    \delta J_c(\hat{n}) := J_c(\hat{n}) - J_c(-\hat{n}), \label{CurrentAsymmetryDefinition}
\end{equation}
is non-zero. Inspection of the symmetries in \tableref{ActionOfSymmetries}, shows that the presence of at least one of the symmetries $\Theta$, $\Theta^s$, $C_{2z}$, $C_{2z}^s$, $C_{2z}^{s'}$, $I$, without or combined with a U(1)$_v$ transformation, implies that $J_c(\hat{n}) = J_c(-\hat{n})$ and no ZFDE is present.

It is straightforward to analyze for each of the 17 candidate orders defined in \tableref{ListOfOrderParameters} whether any of these symmetries is present as a function of whether any combination of $\lambda_{\text{I}}$, $\lambda_{\text{R}}$, and $\beta$ is non-zero. This allows us to deduce whether a ZFDE is possible at all and, if yes, which of $\lambda_{\text{I}}$, $\lambda_{\text{R}}$, $\beta$ the current asymmetry $\delta J_c$ has to be proportional to. For instance, SP$_\perp$ preserves $C_{2z}^s$ as long as $\lambda_{\text{I}}$ vanishes. As such, the diode effect can only be present if $\lambda_{\text{I}}\neq 0$ and $\delta J_c(\hat{n}) \propto \lambda_{\text{I}}$. For VP this is different, as it breaks all of the above-mentioned symmetries and hence exhibits a diode effect even if $\lambda_{\text{I}} = \lambda_{\text{R}} = \beta =0$. For any of the IVC states in \tableref{ListOfOrderParameters}, no diode effect is possible for any value of $\lambda_{\text{I}}$, $\lambda_{\text{R}}$, and $\beta$ since, in all cases, a combination of U(1)$_v$ and $\Theta^s$ remains a symmetry. In fact, out of 17 candidate orders only the six states listed in \tableref{ListOfPossibleDiodeeffects} are consistent with a ZFDE, with $\delta J_c(\hat{n})$ as indicated in the third column. We reiterate that some of these six states further become equivalent once $\lambda_{\text{I}}$ or $\lambda_{\text{R}}$ become sizeable as shown in \figref{fig:evolutionofnormalstateorders}. In the limit where both $\lambda_{\text{I}}$ or $\lambda_{\text{R}}$ are large (of order of a few meV \cite{2021arXiv210602063C}, which might very well be the case \cite{2021arXiv210806126N,DWPaper}), these six states decay into only two distinct phases (above and below the horizontal line in \tableref{ListOfOrderParameters}).

\begin{table*}[tb]
\begin{center}
\caption{Out of the in total 17 different states in \tableref{ListOfOrderParameters}, only the following six can lead to a diode effect. However, note that the first four of these state (the last two), above (below) the vertical line, have the same symmetries as long as $\lambda_{\text{I}},\lambda_{\text{R}}\neq 0$ and, hence, mix and formally constitute the same phase, see also \figref{fig:evolutionofnormalstateorders}. We list not only which of $\lambda_{\text{I}}$, $\lambda_{\text{R}}$ have to be non-zero for a diode effect, $J_c(\hat{n}) \neq J_c(-\hat{n})$, but also whether the critical current is $C_{3z}$ symmetric. Finally, the last four columns indicate whether the underlying normal-state order and, hence, the diode effect can be trained by an in-plane (out-of-plane) Zeeman $B^Z_{\parallel}$ ($B^Z_{\perp}$) and orbital magnetic field $B^O_{\parallel}$ ($B^O_{\perp}$). As with the critical current, we list which of $\lambda_{\text{I}}$, $\lambda_{\text{R}}$, $\beta$ the respective coupling has to be proportional to. Here $(D_0,\lambda_{\text{I}},\lambda_{\text{R}})$ indicates that one of the three is sufficient.}
\label{ListOfPossibleDiodeeffects}
\begin{ruledtabular}
 \begin{tabular} {cccccccc} 
\multicolumn{2}{c}{normal state} & \multicolumn{2}{c}{critical current} & \multicolumn{4}{c}{trainability} \\
\cline{1-2} \cline{3-4} \cline{5-8} 
type & $m_j$ & $J_c(\hat{n}) - J_c(-\hat{n})$ & $J_c(\hat{n}) - J_c(C_{3z}\hat{n})$ & $B^Z_{\perp}$ & $B^Z_{\parallel}$ & $B^O_{\perp}$ & $B^O_{\parallel}$  \\ \hline
SP$_\perp$ & $\sigma_0\eta_0s_z$ & $\propto\lambda_{\text{I}}$ & $\propto\beta$ & $\neq 0$ & $\propto\beta \lambda_{\text{I}} \lambda_{\text{R}}$ & $\propto\lambda_{\text{R}}$ & $\propto\beta\lambda_{\text{I}}$  \\
VP & $\sigma_0\eta_zs_0$ & $\neq 0$ & $\propto\beta$ & $\propto\lambda_{\text{I}}$ & $\propto\beta \lambda_{\text{R}}$ & $\propto\lambda_{\text{I}}\lambda_{\text{R}}$ & $ \propto(D_0,\lambda_{\text{I}},\lambda_{\text{R}})\beta$  \\
SSLP$^+_{\perp}$ & $\sigma_y\eta_zs_z$ & $\propto \lambda_{\text{R}}$ & $\propto\beta$ & $\propto\lambda_{\text{I}}$ & $\propto\beta \lambda_{\text{R}}$ & $\propto\lambda_{\text{I}}$ & $\propto\beta \lambda_{\text{R}}$  \\
SLP$^-$ & $\sigma_y\eta_0s_0$ & $\propto \lambda_{\text{R}}\lambda_{\text{I}}$ & $\propto\beta$ & $\propto\lambda_{\text{R}}$  & $\propto\beta\lambda_{\text{I}}$ & $\propto(D_0,\lambda_{\text{I}},\lambda_{\text{R}})$ & $\propto\beta \lambda_{\text{I}} \lambda_{\text{R}}$  \\ \hline
SP$_{\parallel}$ & $\sigma_0\eta_0(s_x,s_z)$ & $\propto \lambda_{\text{R}}$ & $\propto \lambda_{\text{R}}$ & $\propto\beta \lambda_{\text{I}} \lambda_{\text{R}}$ & $\neq 0$ & $\propto\beta \lambda_{\text{I}}\lambda_{\text{R}}$ & $\propto\lambda_{\text{R}}$ \\
SSLP$^+_{\parallel}$ & $\sigma_y\eta_z(s_x,s_y)$ & $\propto \lambda_{\text{R}}\lambda_{\text{I}}$ & $\propto \lambda_{\text{R}}$ &  $\propto\beta\lambda_{\text{R}}$ & $\propto\lambda_{\text{I}}$ & $\propto\beta\lambda_{\text{R}}$ & $\propto\lambda_{\text{I}}\lambda_{\text{R}}$
 \end{tabular}
 \end{ruledtabular}
\end{center}
\end{table*}

To be able to distinguish further between the remaining six microscopic candidate orders in \tableref{ListOfOrderParameters} driving the diode effect, let us analyze whether and under which conditions they can be trained by a magnetic field. An order parameter $m_j$ can be trained linearly by an external field $\mathcal{B}$, if and only if a linear coupling, $c(\lambda_{\text{I}},\lambda_{\text{R}},\beta) \mathcal{B} \Phi_j$, between the associated $\Phi_j$ in \equref{SimpleDefinitionOfStates} and $\mathcal{B}$ is allowed in the free energy. Whether $c$ can be non-zero and its behavior for small $\lambda_{\text{I}}$, $\lambda_{\text{R}}$, and $\beta$ can be deduced by symmetry. For $\mathcal{B}$ being either in-plane or out-of-plane Zeeman or orbital magnetic field, we list the respective $c(\lambda_{\text{I}},\lambda_{\text{R}},\beta)$ in the last four columns in \tableref{ListOfPossibleDiodeeffects} for all six states with a diode effect. 
Most strikingly, if $\beta$ vanishes, the four states above the horizontal line (the two below it) cannot be trained by an in-plane (out-of-plane) magnetic field.
As we expect $\beta$ to be rather weak in the samples of \refcite{SCPaper}, where the ZFDE can be trained much more effectively by an out-of-plane magnetic field, the four states above the horizontal line are much more likely behind the observed diode effect in tTLG on WSe$_2$. 

To learn more about the underlying mechanism of the diode effect, we also investigate under which conditions the $\hat{n}$-dependence of the critical current $J_c(\hat{n})$ is three-fold symmetric, $J_c(\hat{n}) = J_c(C_{3z}\hat{n})$, for all of the above six scenarios with a diode effect. The dependence of the associated asymmetry $J_c(\hat{n}) - J_c(C_{3z}\hat{n})$ for small $\lambda_{\text{I}}$, $\lambda_{\text{R}}$, $\beta$ is given in the fourth column of \tableref{ListOfPossibleDiodeeffects}. It reveals another crucial distinction between the first four and the last two candidate order parameters, that might be used in future experiment to probe the underlying physics: while $J_c(\hat{n})$ will be three-fold symmetric (unless $\beta\neq 0$, where $C_{3z}$ is trivially broken by the lattice or if it is broken spontaneously) for the first four states, it will not be three-fold invariant for the last two as long as $\lambda_{\text{R}}\neq 0$. We emphasize that this distinction has crucial consequences for the diode effect: in the three-fold symmetric case, the current asymmetry $\delta J_c(\hat{n})$ in \equref{CurrentAsymmetryDefinition} is required to have six zeros as $\hat{n}$ rotates by $2\pi$---a property that will not change if that rotational symmetry is only slightly broken by finite $\beta$; without that constraint, it is only required to have two. We will see examples of both in our explicit calculations in \secref{MBZmodels} below.

\subsection{Field-induced diode effect}\label{FieldInducedDiodeEffect}
Although the main focus of this work is on the ZFDE, we briefly comment on the Zeeman-field induced diode effect, which has already been studied previously in two-dimensional spin-orbit coupled systems \cite{DaidoSCDiode,Yuan2021diodes,HeSCDiode}. 
Upon noting that the in-plane Zeeman-field and the SP$_\parallel$ order parameter in \tableref{ListOfOrderParameters} transform identically under all symmetries of the system, we can immediately read off from \tableref{ListOfPossibleDiodeeffects} that Rashba SOC coupling is required to induce a diode effect with an in-plane Zeeman field for pairing in the IR $A$ and without additional normal-state order [$\Phi_j=0$ in \equref{SimpleDefinitionOfStates}].

Experimentally, it is found that the sample displaying the ZFDE does not exhibit a sizeable field-induced diode effect \cite{SCPaper} for in-plane fields. As we will demonstrate in \secref{s:tTLGdiode} below, this might be understood as a consequence of the additional Ising SOC. Note that the in-plane orbital coupling---associated with Peierls phases in the interlayer hopping---is not expected to yield a large contribution to the diode effect either since its impact on the current asymmetry $\delta J_c$ in \equref{CurrentAsymmetryDefinition} vanishes (approximately) when the mirror symmetry $\sigma_h$ is (approximately) conserved. This follows from the observation that the in-plane orbital coupling (in-plane current) is odd (even) under $\sigma_h$.

While the sample with ZFDE can be trained efficiently with out-of-plane fields, which makes the notion of a field-induced diode effect for this sample ill-defined for fields perpendicular to the plane, \refcite{SCPaper} also presents data for a sample without ZFDE. This sample shows a weak diode effect in the presence of a small out-of-plane field. Since an out-of-plane Zeeman field transforms as SP$_\perp$, we see in \tableref{ListOfPossibleDiodeeffects} that it will induce a diode effect as long as $\lambda_{\text{I}}$ is non-zero (while the orbital-coupling-induced diode effect will likely be sub-leading as its impact on the diode effect has to be proportional to $\lambda_{\text{I}}\lambda_{\text{R}}$). The behavior of the critical current of this sample is therefore consistent with a sizeable $\lambda_{\text{I}}$ and indicates that superconductivity does not coexist with any of the phases in \tableref{ListOfPossibleDiodeeffects} (in particular, those where $\delta J_c \neq 0$ for $\lambda_{\text{R}}= 0$).

\subsection{Diode effect without normal-state order}\label{ExoticPairingCurrent}
So far, we have assumed that the necessary symmetry-requirements, including broken time-reversal symmetry, stem from the normal state while the superconductor preserves all normal-state symmetries (pairing in IR $A$ in \figref{fig:states}). However, we have seen in \secref{PossiblePairingStates} that the other pairing channel, associated with the IR $E$ in \figref{fig:states}, does allow for a superconducting state, denoted by $E_{(1,i)}$ above, that spontaneously breaks time-reversal symmetry. In this case, even without any normal-state order, $\Phi_j=0$ in \equref{SimpleDefinitionOfStates}, the resulting superconducting phase breaks all symmetries in \tableref{ActionOfSymmetries} ($\Theta$, $\Theta^s$, $C_{2z}$, $C_{2z}^s$, $C_{2z}^{s'}$, $I$) which have to be broken for a ZFDE.

We first note that this state will still not lead to a ZFDE, if the superconductor always reaches the global energetic minimum in the current-carrying state; this is discussed and demonstrated in \appref{AppendixOnEPairingDiode} and is a direct consequence of the time-reversal symmetry of the normal state. To understand why this does not always have to be the case, let us consider the gauge-invariant quantity
    \begin{equation}
        \mathcal{C} := i \frac{1}{V}\int\diff\vec{x}\, (c_1^*c^{\phantom{*}}_2-c_2^*c^{\phantom{*}}_1) \label{VestigialEOP}
    \end{equation}
with $c_j$ introduced in \equref{EExpansionInBasisFunc} which we here also allow to be spatially varying; the integral in \equref{VestigialEOP} is over the entire system (with volume $V$). Importantly, $\mathcal{C}$ is odd under $\Theta_s$ but invariant under $C_{3z}$ and can thus be thought of as a composite order parameter measuring the broken time-reversal symmetry (absence thereof) in the $E_{(1,i)}$ ($E_{(1,0)}$) state. Note that $\braket{C}\neq 0$ does not break any continuous symmetry and, in particular, does not require superconducting phase coherence; instead, it is an Ising-like order parameter that can develop long-range order at a finite temperature $T^*$. In particular in the limit where $T^*$ is significantly larger than $T_c$, we can think of $\mathcal{C}$ as a magnetic order parameter, similar to $m_j$ in \equref{SimpleDefinitionOfStates}, defining a ``vestigial'' \cite{VestigialOrderReview} magnetic phase (for $T_c < T < T^*$) associated with the $E_{(1,i)}$ superconductor at lower $T<T_c$. If $\mathcal{C}$ exhibits a fixed sign when measuring the critical current, we can indeed obtain a diode effect (see \appref{AppendixOnEPairingDiode}). We point out that this mechanism of the diode effect is related to the one recently discussed in \cite{zinkl2021symmetry} for chiral $p$-wave pairing, to understand the asymmetric $I$-$V$ characteristics of the 3-K phase in eutectic samples of Sr$_2$RuO$_4$ \cite{PhysRevB.70.014510}. A crucial difference is that, in our case, no additional symmetry-breaking boundary conditions are required due to the reduced symmetry of tTLG on WSe$_2$.

In agreement with experiment \cite{SCPaper}, the intrinsic ZFDE of the $E_{(1,i)}$ state of tTLG/WSe$_2$ can (cannot) be trained linearly with a perpendicular (parallel) Zeeman or orbital magnetic field if $\beta=0$. This follows by noting that $\mathcal{C}$ in \equref{VestigialEOP} can couple linearly to out-of-plane but not to in-plane magnetic fields (due to $C_{3z}^s$). Furthermore, in this scenario, the vestigial phase associated with non-zero $\mathcal{C}$ is a possible origin of, or at least provides an additional contribution to, the enhanced transverse resistance above $T_c$ seen in experiment \cite{SCPaper}.
Nonetheless, the currently available experimental data is more naturally consistent with pairing in the IR $A$ together with a time-reversal-symmetry-breaking normal-state order $m_j$: so far, clear experimental signatures of magnetism in various graphene moir\'e systems have been reported, see, e.g., \cite{LiuReview2021,doi:10.1126/science.aaw3780,2021arXiv210206566L}, while a superconducting order parameter in a non-trivial IR of the spatial point group has not been clearly identified to date. Furthermore, this unconventional pairing state is expected to be more fragile against disorder on the moir\'e scale than the $A$ state \cite{OurMicroscopicTDBG}. For these reasons, we will focus on superconducting order parameters in the IR $A$ in our explicit calculations in the following sections.


\section{Model calculations}\label{ModelCalculations}
In this section we present the general Ginzburg-Landau formalism which allows for the diode effect, finite $\bm q$ pairing and nematicity, to be directly computed. To understand the salient features, we begin with a patch theory. We move onto 2D toy models which provide a description accounting for the entire MBZ. Finally, we perform direct computations for the full tTLG theory \eqref{ContinuumModelDefinition}. 

\subsection{General formalism}\label{GeneralExpression}
As we have argued above, valley polarization (even without SOC), which is symmetry-equivalent to out-of-plane spin polarization in the presence of Ising SOC, is the most natural cause of the reduced symmetry in the normal state that ultimately leads to the diode effect. To capture both of these scenarios simultaneously and in a way that identifies the key ingredients for the diode effect, we will neglect SOC for now and assume that there is an imbalance in the occupations of the different valleys. Let us for concreteness also first assume that this imbalance is not too strong such that Cooper pairs of electrons still form between electrons of different valleys (``intervalley pairing''). We will see that, for the purposes of computing the current (diode effect),  accounting for ``intra-valley'' pairing and/or SOC, follows immediately from the expressions provided here. 

To be concrete, consider the Hamiltonian
\begin{align}
    \notag &H = \sum_{\vec{k}} f^\dagger_{\vec{k},\eta,s} \vpe_{\vec{k},\eta} f^\pdagger_{\vec{k},\eta,s} \\ 
   & - \frac{g}{2} \sum_{\vec{k},\vec{k}',\vec{q}} f^\dagger_{\vec{k}+\vec{q},\eta,s} f^\dagger_{\vec{k}'-\vec{q},\eta',s'} f^\pdagger_{\vec{k}',\eta',s'}  f^\pdagger_{\vec{k},\eta,s}, \label{StartingHamiltonian}
\end{align}
where $f_{\vec{k},\eta,s}$ and $f^\dagger_{\vec{k},\eta,s}$ are annihilation and creation operators of electrons of spin $s$, valley $\eta$, in the low-energy bands crossing the Fermi level in the vicinity of $\vec{k}$. The band energies in valley $\eta$ are denoted by $\vpe_{\vec{k},\eta}$; without valley polarization, we have  $\vpe_{\vec{k},+} = \vpe_{-\vec{k},-}$ as follows from time-reversal symmetry. The simplest form to describe valley polarization is to introduce two different chemical potentials, $\vpe_{\vec{k},\eta} = \epsilon_{\eta\cdot\vec{k}}- \mu_\eta$, but we will keep it more general here. As there is no intervalley Hund's coupling, the model in \equref{StartingHamiltonian} is invariant under independent spin-rotations in the two valleys forming the group SU(2)$_+ \times$SU(2)$_-$.

Performing a mean-field decoupling in the intervalley channel, we obtain
\begin{align}
    \notag &H = \sum_{\vec{k}} f^\dagger_{\vec{k},\eta,s} \vpe_{\vec{k},\eta} f^\pdagger_{\vec{k},\eta,s}  + \frac{1}{g} \sum_{\vec{q}} \text{tr}\left[\Delta_{\vec{q}}^\dagger \Delta_{\vec{q}}^\pdagger\right]\\
    &+ \sum_{\vec{k},\vec{q}} \left[ f^\dagger_{\vec{k}+\vec{q},+,s}  (\Delta_{\vec{q}})_{s,s'} f^\dagger_{-\vec{k},-,s'} + \text{H.c.} \right]. \label{MeanFieldLowEnPairingModel}
\end{align}
Here, the $2\times 2$ matrix $\Delta_{\vec{q}}$ is the superconducting order parameter, which can be expanded in singlet and triplet as 
\begin{equation}
    \Delta_{\vec{q}} = \left( \Delta^s_{\vec{q}}s_0 + \vec{d}_{\vec{q}}\cdot\vec{s} \right)is_y, \label{SpinOnlyDefinitionOfSC}
\end{equation}
similar to \equref{FormOfSCOrderparameter}.
By integrating out the fermions, it is straightforward to derive the associated Ginzburg-Landau expansion which becomes (with number of sites $N$ in the system)
\begin{align}
\label{ExpressionForGamma}
    &\mathcal{F} \sim \sum_{\vec{q}} a_{\vec{q}} \text{tr}\left[\Delta_{\vec{q}}^\dagger \Delta_{\vec{q}}^\pdagger\right] +\mathcal{O}(\Delta^4), \   a_{\vec{q}} = \frac{1}{g} - \Gamma(\vec{q}),\\
  \notag  &\Gamma(\vec{q}) = \frac{1}{2N} \sum_{\vec{k}\in\text{MBZ}} \frac{\tanh\left(\frac{\vpe_{\vec{k}+\frac{\vec{q}}{2},+}}{2T}\right) + \tanh\left(\frac{\vpe_{-\vec{k}+\frac{\vec{q}}{2},-}}{2T}\right)}{\vpe_{\vec{k}+\frac{\vec{q}}{2},+}+\vpe_{-\vec{k}+\frac{\vec{q}}{2},-}}. 
\end{align}
Since $\text{tr}\left[\Delta_{\vec{q}}^\dagger \Delta_{\vec{q}}^\pdagger\right] = 2 (|\Delta^s_{\vec{q}}|^2 + \vec{d}_{\vec{q}}^\dagger \vec{d}_{\vec{q}}^\pdagger)$, we see that singlet and triplet are degenerate, which is a consequence of the aforementioned enhanced SU(2)$_+ \times$SU(2)$_-$ spin symmetry. 

To understand the relation to the diode effect, let us note that gauge invariance demands that a homogeneous vector potential $\vec{A}$ enters as
\begin{align}
    \notag &\mathcal{F} \sim \sum_{\vec{q}} a_{\vec{q}-2e\vec{A}} \,\text{tr}\left[\Delta_{\vec{q}}^\dagger \Delta_{\vec{q}}^\pdagger\right]\\
  \notag  & + \sum_{\vec{q}_1,\vec{q}_2,\vec{q}_3,\vec{q}_4}\delta_{\vec{q}_1+\vec{q}_3,\vec{q}_2+\vec{q}_4} \big\{b_1 \text{tr}\left[\Delta^\dagger_{\vec{q}_1}\Delta^\pdagger_{\vec{q}_2} \right]\, \text{tr}\left[\Delta^\dagger_{\vec{q}_3}\Delta^\pdagger_{\vec{q}_4} \right]\\
    &   \hspace{2cm} + b_2 \, \text{tr}\left[\Delta^\dagger_{\vec{q}_1}\Delta^\pdagger_{\vec{q}_2}\Delta^\dagger_{\vec{q}_3}\Delta^\pdagger_{\vec{q}_4}\right]  \big\},
\end{align}
where $e$ is the electron charge. Here we have also added terms quartic in the order parameter \cite{PhysRevResearch.2.033062}, neglecting the momentum dependence of $b_1$ and $b_2$.

Finding the minimum of $\mathcal{F}$ for $\vec{A}=0$ is straightforward: let $\vec{q}_0$ be the momentum of (one of) the minimum (minima) of $a_{\vec{q}}$ and $a_{\vec{q}_0}<0$; restricting the analysis to single-$\vec{q}$ superconducting order parameters, we then get $\Delta_{\vec{q}} = \delta_{\vec{q},\vec{q}_0} \Psi \hat{\Delta}$ with $\Psi\in\mathbb{C}$ and $\text{tr}[\hat{\Delta}^\dagger \hat{\Delta}]=1$. Depending on whether $b_2>0$ or $b_2<0$ we get $\hat{\Delta} = \sigma_0/\sqrt{2}$ or $\hat{\Delta} = (\sigma_0+\sigma_z)/2$, which corresponds to singlet/unitary triplet or singlet-triplet/non-unitary triplet, respectively; we refer to \cite{PhysRevResearch.2.033062} for a detailed discussion of these superconducting states in the vicinity of the SU(2)$_+ \times$SU(2)$_-$-symmetric point we focus on here. Furthermore, it holds $|\Psi|^2=-a_{\vec{q}_0}/(2b)$ with $b=b_1+b_2/2$ for $b_2>0$ and $b=b_1+b_2$ for $b_2<0$.

In real space, the order parameter is $\Delta(\vec{x}) = \Psi e^{i\vec{q}_0\cdot\vec{x} + i\phi} \hat{\Delta}$. 
For constant $\phi$, the equilibrium condensate forms at a wavevector $\bm q_0$ found from 
\begin{align}
\partial_{\bm q} a_{\bm q}\Big|_{\bm q_0}=0, \ \text{and} \ \det \partial_{q_i} \partial_{q_j} a_{\bm q}\Big|_{\bm q_0}>0. 
\end{align}
A supercurrent is imposed by taking a finite sample and setting a phase gradient, i.e., we generalize to spatially varying phases, $\phi \rightarrow \phi(\vec{x})$, in $\Delta(\vec{x})$ and impose twisted boundary conditions in $\phi$. 
The simplest scenario is to take $\phi(\bm x)=\delta \bm q\cdot \bm x$,
for some fixed $\delta  \bm q$. 
In this {\it nonequilibirum} case, the condensate forms at effective wavevector $\bm q = \bm q_0+\delta \bm q$, since $\Delta(\vec{x}) = \Psi e^{i(\vec{q}_0+\delta \vec{q})\cdot\vec{x}} \hat{\Delta}$. The definition of current follows from $\vec{J} = - \partial_{\vec{A}}\mathcal{F}\big|_{{\bm A}={\bm 0}} = 2e |\Psi|^2\partial_{\bm q} a_{\bm q}$, with charge $2e$ (see \cite{DaidoSCDiode} for a microscopic derivation of the current). Using the saddle point solution,  $|\Psi|^2= -a_{\bm q}/(2b)$, we arrive at 
\begin{equation}
    \vec{J}(\vec{q}) = -e \Theta(-a_{\vec{q}}) a_{\vec{q}} \partial_{\vec{q}}a_{\vec{q}}/b. \label{ExpressionForCurrent1}
\end{equation}
Here $\Theta(...)$ is a step-function ensuring that the saddle point condition $|\Psi|^2= -a_{\bm q}/(2b)\geq0$ is satisfied.

The critical current $J_c(\hat{n})$ along the direction $\hat{n}$ is now simply given by the maximum magnitude of $\vec{J}(\vec{q})$ for $\vec{q}\in\text{MBZ}$ which points along $\hat{n}$. From this, we conclude that $J_c(\hat{n}) = J_c(-\hat{n})$, and therefore no diode effect, if
\begin{equation}
   \exists \vec{q}_0:\,\, \Gamma(\vec{q}-\vec{q}_0) = \Gamma(-\vec{q}-\vec{q}_0). \label{ConditionForNoDiodeEffect}
\end{equation}
We emphasize that $\vec{q}_0$ does not have to be $\vec{0}$.

An instructive, albeit fine-tuned, example is $\vpe_{\vec{k},+} = \epsilon_{\vec{k}+\vec{Q}/2} - \mu$, $\vpe_{\vec{k},-} = \epsilon_{-\vec{k}-\vec{Q}/2} - \mu$ which also has broken time-reversal and $C_{2z}$ symmetry since $\vpe_{\vec{k},+} \neq \vpe_{-\vec{k},-}$, as long as $\vec{Q} \notin \text{RML}$. From \equref{ExpressionForGamma}, we get 
\begin{align}
  \notag \Gamma(\vec{q}) = \frac{1}{2N} \sum_{\vec{k}\in\text{MBZ}} \Bigg\{ \frac{\tanh\left(\frac{1}{2T}(\epsilon_{\vec{k}+\frac{\vec{q}+\vec{Q}}{2}}-\mu)\right)}{\epsilon_{\vec{k}+\frac{\vec{q}+\vec{Q}}{2}}+\epsilon_{\vec{k}-\frac{\vec{q}+\vec{Q}}{2}}-2\mu} \ \ 
    \\ + 
   \frac{\tanh\left(\frac{1}{2T}(\epsilon_{\vec{k}-\frac{\vec{q}+\vec{Q}}{2}}-\mu)\right)}{\epsilon_{\vec{k}+\frac{\vec{q}+\vec{Q}}{2}}+\epsilon_{\vec{k}-\frac{\vec{q}+\vec{Q}}{2}}-2\mu} \Bigg\}.   \label{TranslatedFunctionChi}
\end{align}
Clearly, $\Gamma(\vec{q})$ in \equref{TranslatedFunctionChi} obeys \equref{ConditionForNoDiodeEffect} with $\vec{q}_0=\vec{Q}$ and, hence, cannot exhibit a diode effect. Nonetheless, $\widetilde{\Gamma}(\vec{q}') := \Gamma(\vec{q}'-\vec{Q})$ is exactly equivalent to the $\Gamma$ of the time-reversal symmetric situation with $\vpe_{\vec{k},+} = \vpe_{-\vec{k},-} = \epsilon_{\vec{k}} - \mu$. As such, we know that $\widetilde{\Gamma}(\vec{q}')$ is maximal for $\vec{q}'=0$ and the resulting pairing occurs at finite wavevector $\vec{q}_0=\vec{Q}$.

Finally, to account for the cases with (i) strong SOC, or (ii) intravalley pairing, we allow for arbitrary quantum numbers $\alpha$ in \equref{ExpressionForGamma}; in case (i), the Bloch states at the Fermi surface can be labelled uniquely by their valley quantum number, $\alpha=\eta$, and exhibit momentum-dependent spin orientations, while $\alpha=\{\eta,s\}$ for case (ii). 
The dependence on index $\alpha$ enters via the particle-particle susceptibility, and in principle, the pairing interaction strength, $g$. As we derive in \appref{A:DerivationOfGamma}, the resulting $a_{\vec{q}}$ in \equref{ExpressionForGamma} has the same form as above,
\begin{align}
a_{\vec{q}} &= \frac{1}{g} - \Gamma(\vec{q}), \label{ForOfaGammamod}\\
     \notag     \Gamma(\vec{q}) &= \frac{1}{2N} \sum_{\vec{k}\in\text{MBZ}} \frac{\tanh\left(\frac{\xi_{\vec{k}+\frac{\vec{q}}{2},\alpha}}{2T}\right) + \tanh\left(\frac{\xi_{-\vec{k}+\frac{\vec{q}}{2},\alpha'}}{2T}\right)}{\xi_{\vec{k}+\frac{\vec{q}}{2},\alpha}+\xi_{-\vec{k}+\frac{\vec{q}}{2},\alpha'}}, 
\end{align}
with the only difference that $\vpe_{\vec{k},\eta}$ are replaced by the band energies $\xi_{\vec{k},\alpha}$ of the appropriate quantum numbers [cf.~\equsref{ExpressionForGammaSOC}{ExpressionForGammaIV}].

\subsection{Patch theory}\label{PatchSection}
Having established the general formalism, we begin our analysis with a simple patch-theory description as it allows for a particularly transparent analysis. For now, we neglect SOC, and treat singlet and triplet pairing on equal footing. Below in \secref{s:tTLGdiode}, SOC and the Fermi surfaces of the full continuum model (\ref{ContinuumModelDefinition})  will be taken into account.

\begin{figure}[t]
   \centering
    \includegraphics[width=\linewidth]{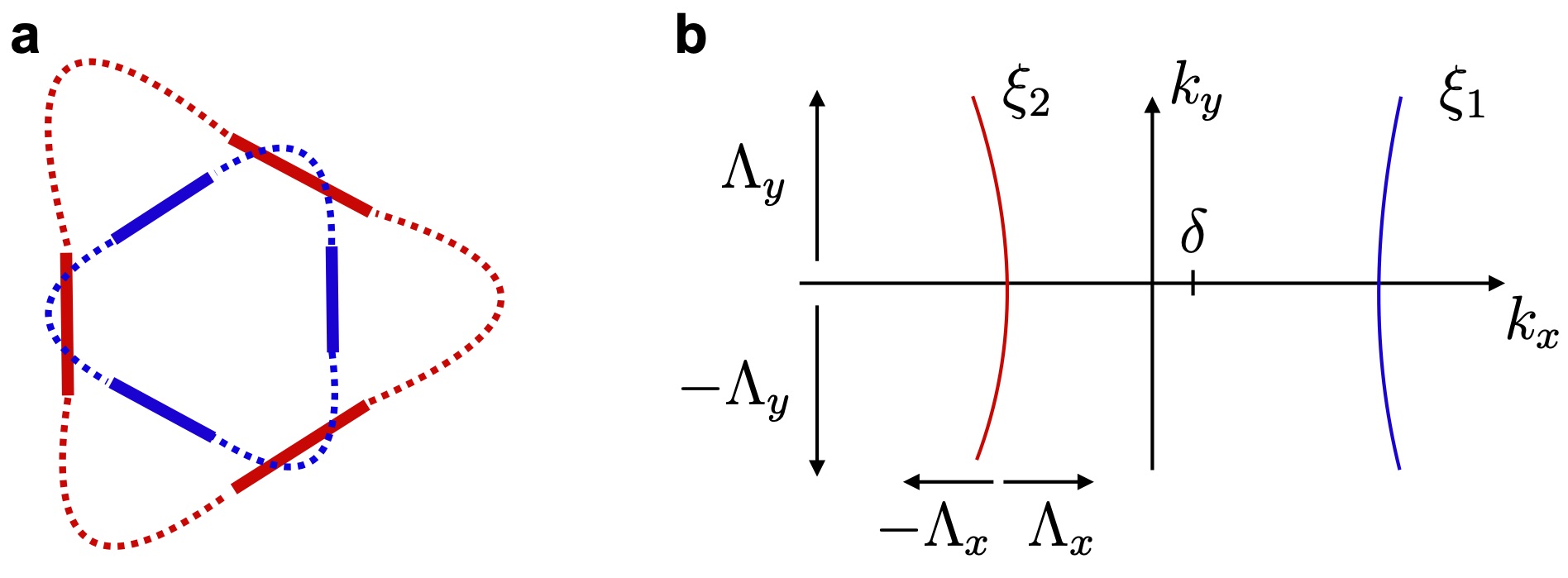}
    \caption{Definition of patch theory. (a) Schematic Fermi surfaces from each valley (red and blue), under a valley polarizing perturbation. Bold, straight segments correspond to the patches. (b) Definition of the effective band energies about a single patch. }
    \label{fig:patchtheory}
\end{figure}

Having in mind tTLG, we consider pairing of states at $\{\bm k, \eta\}$ and $\{-\bm k, -\eta\}$, whereby the Fermi surfaces at each valley are $C_{3z}$ symmetric, and we account for valley polarization. For the purpose of modelling, we consider the schematic Fermi surfaces shown in  \figref{fig:patchtheory}(a). Due to (approximate) nesting, the particle-particle susceptibility receives the largest contribution for momenta in the vicinity of the three (nearly) parallel segments. Based on this observation, we may reduce the full MBZ down to these three patches. For the single patch shown in \figref{fig:patchtheory}(b), we have (with $\delta q_x = q_x - \delta$)
\begin{align}
    \notag \Gamma^p(\vec{q}) &= \int_{-\Lambda_y}^{\Lambda_y} \frac{\diff  k_y}{2 \pi} \int_{-\Lambda_x}^{\Lambda_x} \frac{\diff \delta k_x}{4 \pi} \frac{\tanh\left[\frac{ \xi_1}{2 T}\right] + \tanh\left[\frac{\xi_2}{2T}\right]}{\xi_1+\xi_2}, \label{GammapExpression} \\
  \notag  \xi_1&= v(\delta k_x + \delta q_x/2) - \alpha_+(k_y+q_y/2)^2,\\
   \xi_2&= v(\delta k_x - \delta q_x/2)- \alpha_-(k_y-q_y/2)^2.
\end{align}
In this description, valley polarization enters both as the finite average momentum $\delta$ of the red and blue segments as well as in form of the imbalance $\alpha_+-\alpha_-$ of the curvature of the Fermi surfaces, see \figref{fig:patchtheory}.
Taking all three patches, we arrive at
\begin{equation}
    \Gamma(\vec{q}) = \sum_{j=-1,0,1}  \Gamma^p\left( R(2\pi j/3) \vec{q}\right), \label{FullGamma}
\end{equation}
where $R(\varphi)$ rotates two-dimensional vectors by angle $\varphi$. To obtain a simplified intuitive understanding for the conditions for (i) diode effect and (ii) finite-momentum pairing and nematic superconductivity, let us expand $\Gamma^p$ in \equref{GammapExpression} as
\begin{equation}
\label{GammaExpand}
    \Gamma^p(\vec{q}) \sim \Gamma_0 - \sum_{n=1}^4 a_n (q_x-\delta)^n - c  q_y^2. 
\end{equation}
It is deduced from \equref{GammapExpression}, that the coefficients $a_1, a_3\propto \alpha_+-\alpha_-$. Explicit expressions for $a_n$ and $c$ are presented in \appref{A:patch}. Consequently, to leading order, $a_1$, $a_3$, and $\delta$ will vary linearly with valley polarization. Using $\nu\in[-1,1]$ as a dimensionless measure of the valley polarization, we write $a_1 \sim \tilde{a}_1 \nu$, $a_3 \sim \tilde{a}_3 \nu$ and $\delta \sim \tilde{\delta} \nu$, as $\nu\rightarrow 0$, for later reference. 

First, to understand the emergence of a diode effect, we use $\Gamma(\vec{q}) -  \Gamma(-\vec{q})\neq0$ as a necessary condition for it. Considering the patch theory expansion in \equsref{FullGamma}{GammaExpand}, we find
\begin{equation}
     \Gamma(\vec{q}) -  \Gamma(-\vec{q}) = \frac{3}{2}  q_x \left(q_x^2-3 q_y^2\right) (4 a_4 \delta -a_3). \label{AsymmetryPartOfGamma}
\end{equation}
This asymmetry vanishes if $\nu=0$ and no diode effect is possible, in accordance with our symmetry analysis \secref{SymmetriesDiodeEffect}, as time-reversal (or $C_{2z}$) symmetry is preserved at $\nu=0$. Generically, it holds $4\tilde{a}_4\tilde{\delta} \neq \tilde{a}_3$ and we see that the asymmetic part in \equref{AsymmetryPartOfGamma} becomes non-zero immediately when $\nu$ is turned on. As such, the diode effect is expected to set in immediately when $\nu$ becomes non-zero; this will be confirmed by our explicit model calculations below. 

Second, to derive a sufficient condition for finite-$\bm q$ pairing, we expand \equref{FullGamma} to quadratic order in $\vec{q}$, yielding
\begin{subequations}\begin{align}
    \Gamma(\vec{q}) &\sim \Gamma(\vec{q}=0) + \gamma \vec{q}^2, \quad \vec{q}\rightarrow 0, \label{GammaExpansion}\\
    \gamma &= -\frac{3}{2}\left[ a_2 + c + (6 a_4 \delta^2 - 3 \delta a_3) \right].
\end{align}\end{subequations}
Recalling that $a_3 \sim \tilde{a}_3 \nu$ and $\delta \sim \tilde{\delta} \nu$, we have $\gamma|_{\nu=0} = - 3(a_2 + c)/2 <0$ (since $\vec{q}=0$ pairing must be favored at $\nu=0$) and there is a critical valley polarization,
\begin{equation}
    \nu_c \approx \sqrt{\frac{a_2+c}{3(\tilde{a}_3 - 2 \tilde{\delta}a_4) \tilde\delta}}, \label{nuCritical}
\end{equation}
that $|\nu|$ needs to exceed to turn the maximum at $\vec{q}=0$ into a (local) minimum. Note that the value of $\nu_c$ in \equref{nuCritical} is technically only an upper bound on the critical valley polarization, as the global minimum can occur at $\vec{q}\neq 0$ before the maximum at $\vec{q}=0$ turns into a local minimum. Nonetheless, the true critical $|\nu|$ must be finite since we expect $\Gamma$ to depend smoothly on $\nu$. We will revisit this conclusion in our treatment of full MBZ toy models, \secref{MBZmodels}. 

Once $|\nu|$ is larger than this critical value, pairing at finite momentum occurs. Intuitively, this behavior can be understood as follows: the role of valley polarization is to remove the degeneracy between a point at $\vec{k}$ on the blue Fermi surface and at $-\vec{k}$ on the red one in \figref{fig:patchtheory}(a); this reduces the condensation energy of the superconductor. Choosing a finite $\vec{q}$ for pairing appropriately can improve the energetics for the superconductor in only two of the three solid segments in \figref{fig:patchtheory}(a) [or, equivalently, terms in \equref{FullGamma}]. If valley polarization is sufficiently large, the energetic gain by two of the three patches or terms can overcompensate the disfavored one, leading to finite-momentum pairing.

\subsection{Full MBZ toy models}\label{MBZmodels}
We now extend the discussion from above to include the full MBZ. Our primary focus is to demonstrate the ZFDE, finite $\bm q$ paring and nematicity. To this end, we introduce a toy model to illuminate the role of valley polarization, and for completeness, strain. 

We construct a minimal model that captures the symmetries and basic form of the Fermi surfaces of tTLG: for each valley $\eta=\pm$, we consider a nearest-neighbor hopping ($t$) triangular-lattice model with staggered flux, $\phi_\eta$, that preserves translational and $C_{3z}$ rotational symmetry but breaks time-reversal and $C_{2z}$ in each valley. In order to describe finite strain, $\beta\neq 0$, we replace the hopping along two of the three nearest-neighbor bonds by $t(1-\beta)$. Explicitly, the dispersions take the form
\begin{align}
\label{ExplicitFormOfDispersion}
   \notag  &\vpe_{\vec{k},+} = \epsilon_{\vec{k};\phi_+} - \mu, \quad  \vpe_{\vec{k},-} = \epsilon_{-\vec{k};\phi_-} - (\mu+\delta\mu),\\
 \notag   &\epsilon_{\vec{k};\phi} \ = -t\cos \left(k_x-\frac{\phi }{3}\right)\\
     & -2t (1-\beta) \cos \left(\frac{\sqrt{3} k_y}{2}\right) \cos \left(\frac{1}{6} (3 k_x+2 \phi )\right). 
\end{align}

\begin{figure}[t]
   \centering
    {\includegraphics[width=0.45\textwidth,clip]{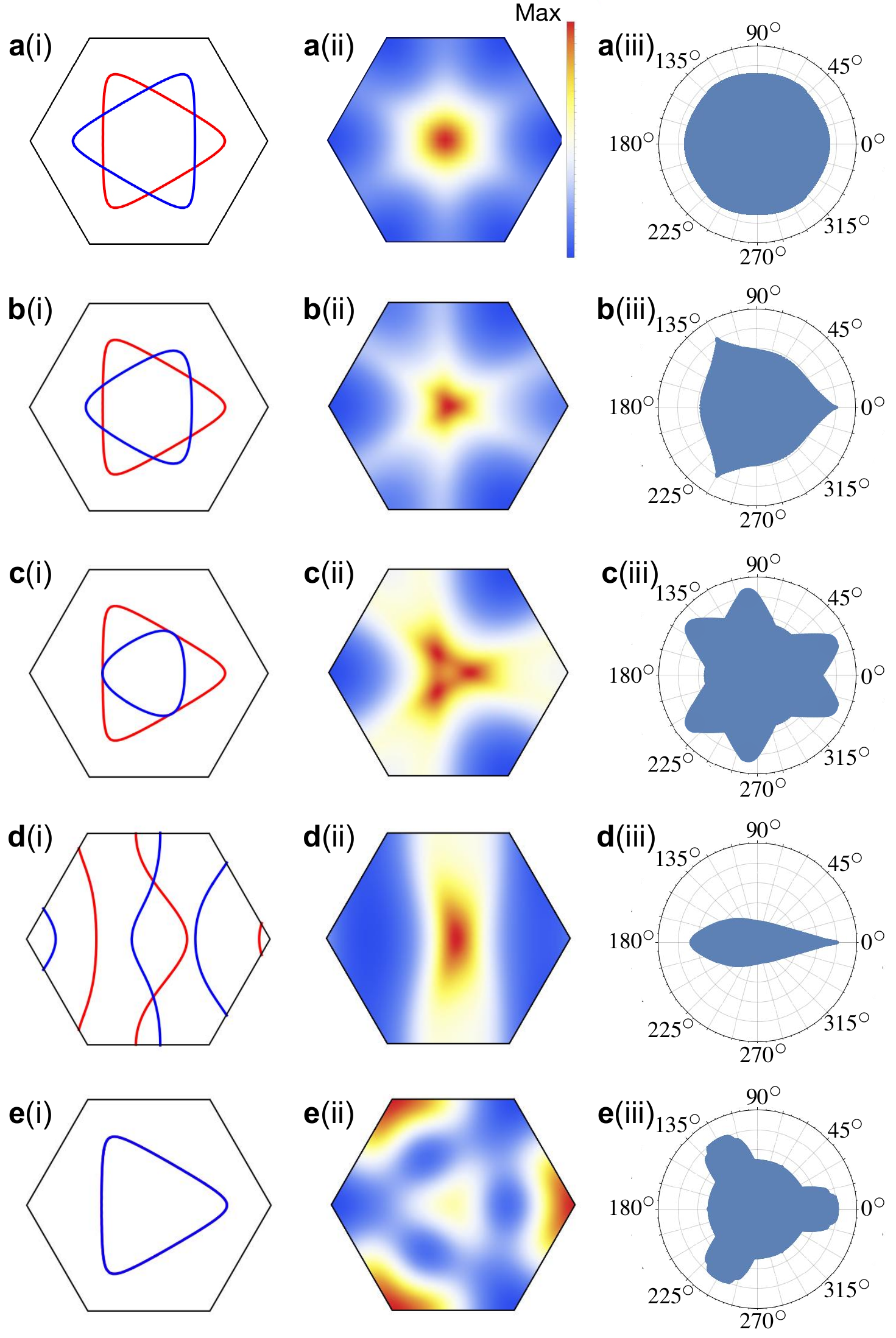}}
    \caption{Diode effect in the minimal model \equref{ExplicitFormOfDispersion}. Columns are marked (i), (ii), (iii), which show: (i) the Fermi surfaces of the states undergoing pairing, with valley $\eta=+$ ($\eta=-$) shown in red (blue), (ii) the susceptibility $\Gamma(\bm q)$,  (iii) the current $J(\hat{n})$ as a function of the direction of $\hat{n} = (\cos\varphi,\sin\varphi)$; the critical current for a given direction $\hat{n}$ is set by the boundary. For $\Gamma$ and the current, we have set $T=t/10$ and  $1/g=\text{max}_{\vec{q}}\Gamma(\vec{q})/2$ to evaluate $J(\hat{n})$. Each row represents a qualitatively different scenario arising from the model \equref{ExplicitFormOfDispersion}: (a) no valley polarization, no strain, (b) weak valley polarization, no strain, (c) strong valley polarization, no strain (d) (weak) valley polarization, with strain, (e) intravalley pairing, no strain. Explicit parameters are given by \cite{Caption}.}
    \label{fig:toymodels}
\end{figure}

Let us start first with the time-reversal, $C_{2z}$, and $C_{3z}$ symmetric limit by setting $\delta\mu=0$, $\phi_+=\phi_-$, and $\beta=0$. In that case, a state at the Fermi surface at $\vec{k}$ and one valley $\eta$ will be degenerate with a state at $-\vec{k}$ in the other, see \figref{fig:toymodels}(ai). As such, we expect $\Gamma(\vec{q})$ is peaked at $\vec{q}=0$, consistent with \figref{fig:toymodels}(aii), and exhibits the usual logarithmic divergence with temperature. By the same token ($C_{2z}$ or time-reversal symmetry, leading to $\vpe_{\vec{k},+}=\vpe_{-\vec{k},-}$), it follows from \equref{ExpressionForGamma} that $\Gamma(\vec{q})=\Gamma(-\vec{q})$ and there is no diode effect; this can be seen in \figref{fig:toymodels}(aiii).

To obtain a diode effect, let us assume $C_{2z}$ and time-reversal symmetry are broken due to finite valley polarization. In our model in \equref{ExplicitFormOfDispersion}, we capture this by setting $\delta\mu\neq0$ or $\phi_+\neq\phi_-$. It now holds $\vpe_{\vec{k},+}\neq\vpe_{-\vec{k},-}$, as is reflected in the Fermi surfaces of, e.g., \figref{fig:toymodels}(bi), and thus $\Gamma(\vec{q})\neq \Gamma(-\vec{q})$ supporting a diode effect, see \figref{fig:toymodels}(bii) and (biii), respectively. Moreover, since $C_{3z}$ is unbroken, the diode effect is seen to be three-fold rotational symmetric, \figref{fig:toymodels}(biii). Note, however, that the maximum of $\Gamma(\vec{q})$ still occurs at $\vec{q}=0$, such that the Cooper pairs still have vanishing center of mass momentum. In fact, this was generically expected since $C_{3z}$ symmetry implies $\Gamma(\vec{q})=\Gamma(C_{3z}\vec{q})$ and thus the form (\ref{GammaExpansion}) of the small-$\vec{q}$ expansion will still hold. 
We know that $\gamma < 0$ for $\delta\mu=\phi_+-\phi_-=0$ and, hence, $\gamma$ has to remain negative when these two quantities are turned on smoothly. This also agrees with our patch-theory analysis of \secref{PatchSection}, where the measure of valley polarization, $\nu$, had to surpass a critical value \equref{nuCritical} to induce finite-momentum pairing. Most importantly, this shows that finite-momentum pairing is not only not sufficient for a diode effect (as established above) but also not necessary.

To demonstrate explicitly that the MBZ model can support finite-momentum pairing, we repeat the analysis for larger valley polarization above the critical value (which, as we note in passing, depends on temperature). 
We indeed find, see \figref{fig:toymodels}(c), that for sufficiently large valley polarization, $\max\Gamma(\bm q)=\Gamma(\vec{q}_0\neq \bm 0)$ which supports finite momentum (i.e.~$\bm q_0$) pairing, in agreement with the patch model of \secref{PatchSection}. 

Once $C_{3z}$ is explicitly broken in the normal state above the superconducting transition, due to finite strain $\beta\neq0$ or electronic nematic order, we generically expect the critical value of valley polarization for finite-momentum-pairing to vanish: since the constraint $\Gamma(\vec{q})=\Gamma(C_{3z}\vec{q})$ is absent, linear-in-$\vec{q}$ terms are allowed in the expansion of $\Gamma$ for small $\vec{q}$, once $C_{2z}$ and time-reversal are broken by valley polarization. The maximum of $\Gamma(\vec{q})$ can then immediately occur at a non-zero momentum when valley polarization is introduced. 
In \figref{fig:toymodels}(d) we present results for a computation with strain $\beta$, where we apply quite large $\beta$ to make the effect and the resulting lack of $C_{3z}$ symmetry in the diode effect clearly visible.

In the limit of sufficiently strong valley polarization, or via other mechanisms, pairing may take place between states within a single valley, thereby breaking TRS and $C_{2z}$. For completeness, we also consider this {\it intravalley pairing} scenario within the minimal model \equref{ExplicitFormOfDispersion}. We indeed find, in \figref{fig:toymodels}(e), that it generates a ZFDE. It is also seen, from \figref{fig:toymodels}(eii), that $\max\Gamma(\bm q)=\Gamma(\vec{q}_0 \neq \bm 0)$; the intravalley state supports pairing at momentum $\bm K_i + \bm q_0$, where $\bm q_0\in$MBZ and $\bm K_i$ are the BZ (not MBZ) corners. As such, this exotic order parameter exhibits spatial (phase) modulations on the scale of the microscopic graphene layers, producing Kekul\'e-like patterns, see e.g. \cite{RoyHubert2010}.

Note that once $\Gamma(\vec{q})$ is maximal at a non-zero $\vec{q}=\vec{q}_0$, as in \figref{fig:toymodels}(cii), the superconductor will spontaneously break the rotational symmetry $C_{3z}^s$, 
in gauge-invariant observables. This can be most easily seen by defining the following composite order parameter
\begin{equation}
    \mathcal{N}_{j} = \frac{1}{V}\int\diff\vec{x}\, \text{tr}\left[\Delta^\dagger(\vec{x}) (-i\partial_j-2e A_j) \Delta(\vec{x})\right], \label{NematicVestOrderParameter}
\end{equation}
$j=x,y$, where the integral is over the volume of the system, $\Delta(\vec{x})$ is the Fourier transform of the superconducting order parameter $\Delta_{\vec{q}}$ in \equref{SpinOnlyDefinitionOfSC}, and $A_j$ the vector potential. Note that $ \mathcal{N}_{j}$ in \equref{NematicVestOrderParameter} is invariant under spin-rotations SO(3)$_s$ [in fact, even invariant under the full SU(2)$_+\times$SU(2)$_-$ symmetry], under U(1) gauge transformations, and under U(1)$_v$, but transforms as the vector $(x,y)$ under $C_{3z}$. 
As such, it can couple to physical observables such as the local density of states or the excitation spectrum of the Bogoliubov quasi-particles of the superconductor that will, in turn, exhibit broken $C_{3z}$ symmetry. In addition, by virtue of not breaking any continuous symmetry, $\mathcal{N}_j$ in \equref{NematicVestOrderParameter} can have long-range order at finite temperature in two dimensions (via a three-state Potts transition). It is possible that $\mathcal{N}_j$ condenses before the system exhibits significant quasi-long-range order in the superconducting phase. This ``vestigial'' nematic phase \cite{VestigialOrderReview} can provide a possible explanation of the observed nematic transport properties above but in the vicinity of the superconducting critical temperature \cite{SCPaper}.  
We note that the critical current $J_c(\hat{n})$ in \figref{fig:toymodels}(ciii) of this nematic state is still $C_{3z}$ symmetric. This is a consequence of the assumption in our calculation that the superconductor will always be able to minimize the free energy of the system (cf.~discussion of the $E_{(1,i)}$ in \secref{ExoticPairingCurrent}).

\subsection{Continuum model results}\label{s:tTLGdiode}
We now work directly with the continuum model for tTLG in \equref{ContinuumModelDefinition}, both with and without SOC coupling, \equref{SOCTerms}, which arises due to proximity coupling to the WSe$_2$ layer. As above, our primary focus is spontaneous valley polarization as the source of TRS and $C_{2z}$ breaking. To account for valley polarization within the tTLG model, we add to $h$ in \equref{DifferentPartsofContHam} the perturbation $h^V = V_0 \sigma_0 \eta_z s_0$, which acts simply as a valley-dependent shift of the chemical potential. 

\begin{figure}[t]
\centering
{\includegraphics[width=0.475\textwidth,clip]{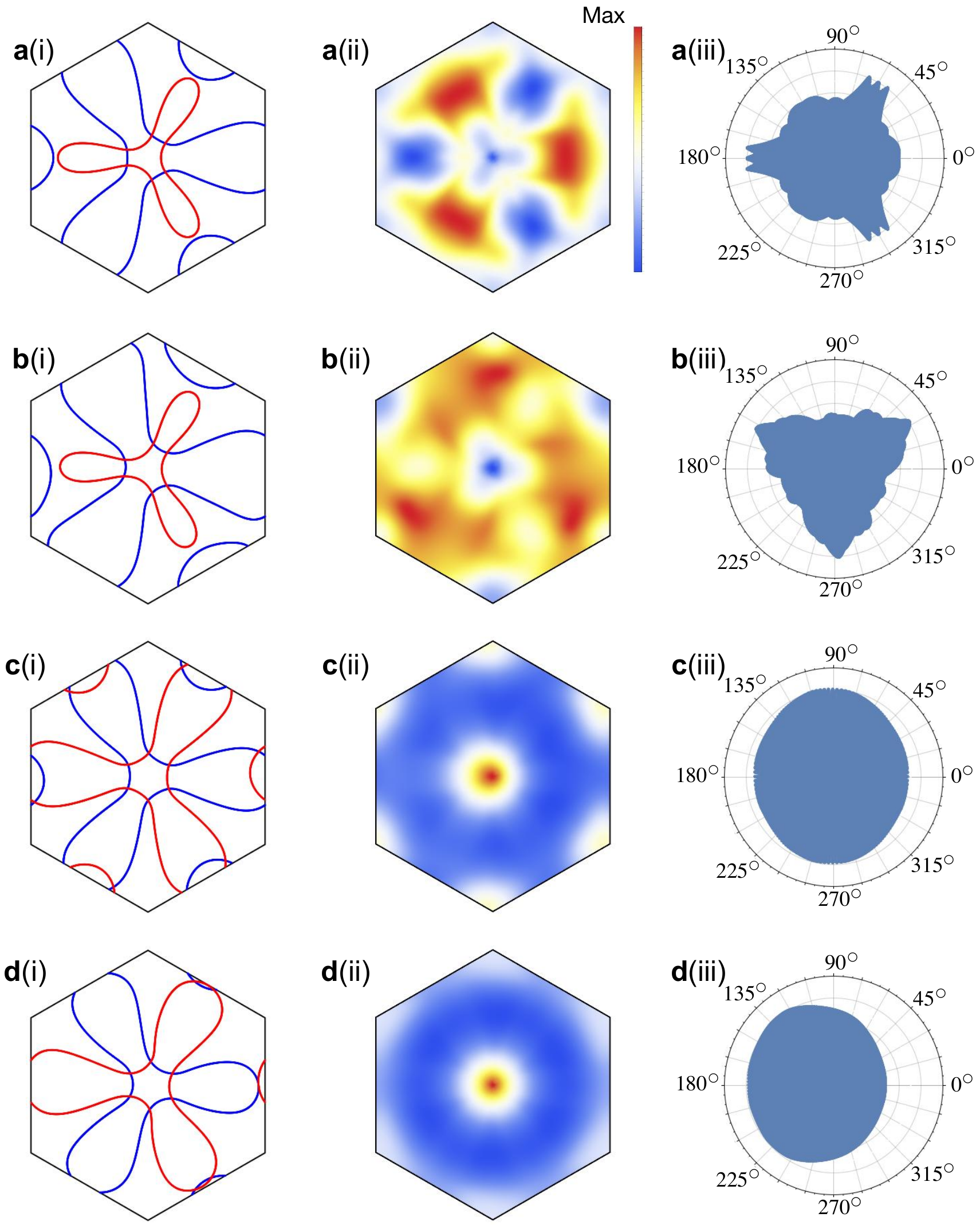}}
\caption{Diode effect in the tTLG model \equref{ContinuumModelDefinition}. Columns are marked (i), (ii), (iii), which show: (i) the Fermi surfaces of the states undergoing pairing, with valley $\eta=+$ ($\eta=-$) shown in red (blue), (ii) the susceptibility $\Gamma(\bm q)$,  (iii) the current $J(\hat{n})$ as a function of the direction of $\hat{n} = (\cos\varphi,\sin\varphi)$; the critical current for a given direction $\hat{n}$ is set by the boundary. Row (a) tTLG, no SOC, $\mu=8.5$ meV, and valley polarization  $V_0=1$ meV, (b) tTLG + WSe$_2$ with SOC $\lambda_{\text{R}}=\lambda_{\text{I}}=10$ meV, $\mu=12.3$ meV, and valley polarization  $V_0=1$ meV.  (c)  tTLG + WSe$_2$, $\lambda_{\text{R}}=1$, $\lambda_{\text{I}}=10$ meV, $\mu=12.7$ meV, and an in-plane field $B_x=10$ T, (d)  tTLG + WSe$_2$, $\lambda_{\text{R}}=\lambda_{\text{I}}=10$ meV, $\mu=12.7$ meV, and an in-plane field $B_x=10$ T. Everywhere, $T=0.1$ meV and $\theta=1.5$ degrees. }
\label{f:TTG_results}
\end{figure}

To understand the salient features, \figref{f:TTG_results}(a) and (b) present the Fermi surfaces, $\Gamma(\bm q)$, and critical current, showing the ZFDE, (a) without SOC and (b) with strong SOC, $\lambda_I=\lambda_R=10$ meV. In both cases, valley polarization is set to $V_0=1$ meV. Inclusion of SOC indeed has an effect on the critical current, compare \figref{f:TTG_results}(a)(iii) and (b)(iii); however, it is seen not to be a necessary ingredient for the ZFDE, in agreement with our toy model calculations of \secref{MBZmodels} and \tableref{ListOfPossibleDiodeeffects}, showing that VP order will induce a ZFDE without SOC. 
The results presented in \figref{f:TTG_results} are obtained using the formalism of \secref{GeneralExpression}. Moreover, the following assumptions are made: First, as discussed in \secref{PossiblePairingStates}, intervalley pairing is likely dominant, and is therefore assumed. Second, the chemical potential is selected to maintain an approximately equal area of the larger Fermi surface for all different parameter sets presented in \figref{f:TTG_results}. Third, noting that the free energy expansion, which leads to expression for $\Gamma(\bm q, T)$ \eqref{ExpressionForGamma}, is valid for $T\lesssim T_c$, we compute at $T=0.1$ meV $\lesssim T_C$; the scale for $T_c$ follows from \refcite{SCPaper}. Finally, for demonstration, we compute the current assuming an interaction coupling strength $g^{-1}=0.8\max_{\vec{q}}\Gamma(\vec{q})$.

As discussed in \cite{DaidoSCDiode,Yuan2021diodes,HeSCDiode}, having a non-zero Rashba SOC and in-plane field breaks $C_{2z}$, generates a finite momentum, {\it helical} pairing state, and is therefore expected to generate a diode effect. Experimentally, it was found, however, that the field-induced diode effect is very weak in the tTLG on WSe$_2$ \cite{SCPaper}. To demonstrate that this can be understood as a consequence of the different form of the SOC terms in the tTLG/WSe$_2$ heterostructure, we here consider the case where TRS breaking comes from an applied in-plane magnetic field, instead of valley polarization.
To account for the in-plane field, we add to \eqref{DifferentPartsofContHam} the Zeeman coupling $h^B=\sigma_0\eta_0\vec{B}_\parallel\cdot\vec{s}$. It has a more subtle impact on the bandstructure than valley polarization, which also crucially depends on $\lambda_{\text{R}}$, $\lambda_{\text{I}}$. 

\figref{f:TTG_results}(c) and (d) present results for tTLG with SOC and a large in-plane Zeeman field, $\vec{B}_\parallel=B\hat{x}$, with $B=10$ T. 
To demonstrate that, in accordance with our symmetry-analysis in \secref{FieldInducedDiodeEffect}, Rashba SOC is a necessary perturbation for the in-plane Zeeman-field diode mechanism, we consider two limits: In \figref{f:TTG_results}(c) a moderate Rashba coupling $\lambda_{\text{R}}=1$ meV,  while in \figref{f:TTG_results}(d), a large Rashba coupling $\lambda_{\text{R}}=10$ meV are assumed.  In both cases we fix $\lambda_{\text{I}}=10$ meV, which is motivated by the analysis in \refcite{DWPaper}. Comparison of \figref{f:TTG_results}(d)(iii) and (e)(iii) demonstrates the role played by Rashba SOC, and further that Ising SOC alone is insufficient to generate a diode effect. In particular, the diode effect in \figref{f:TTG_results}(ciii) is rather weak, despite the large magnetic field. This might explain why the untrained sample with ZFDE of \refcite{SCPaper} does not show any significant $\delta J_c(\hat{n})$ when in-plane fields are applied.

\section{Doping dependence of the diode effect}\label{DopingDependence}
One particularly striking observation of \refcite{SCPaper} is that the sign of the diode effect, i.e., the sign of $\delta J_c(\hat{n})$ in \equref{CurrentAsymmetryDefinition} for fixed direction $\hat{n}$, changes once the filling fraction $\nu_{\text{tTLG}}$ is tuned from electron, $\nu_{\text{tTLG}}>0$, to hole, $\nu_{\text{tTLG}}<0$, doping or vice versa. In this section, we will present a theoretical explanation for why this behavior might be expected and discuss implications for the doping dependence of the order parameter of the normal-state instability inducing the diode effect.  

As argued in \secref{SymmetriesDiodeEffect}, the diode effect can be induced by one of the six normal-state orders in \tableref{ListOfPossibleDiodeeffects}. The magnetic-field training behavior in experiment \cite{SCPaper} points towards one of the first four states, SP$_\perp$, VP, SSLP$^+_\perp$ or SLP$^-$, which start to mix once both $\lambda_{\text{I}}$ and $\lambda_{\text{R}}$ are non-zero. Therefore, to keep the discussion simple, we will focus on VP here and first neglect SOC altogether.

An effective mean-field model for electron doping, $\nu_{\text{tTLG}}>0$, in the presence of VP then reads as
\begin{equation}
    H^{\text{LE}}_{\text{eff}} = \sum_{\vec{k}} f^\dagger_{\vec{k},\eta,s} E_{\vec{k},\eta} f^\pdagger_{\vec{k},\eta,s}, \, E_{\vec{k},\pm} = \xi_{\pm\vec{k}} + V_z g_{\vec{k},\pm}, \label{EffectiveModel}
\end{equation}
where $f^\dagger_{\vec{k},\eta,s}$ ($f^\pdagger_{\vec{k},\eta,s}$) are the creation (annihilation) operators, already introduced in \equref{LowEnergyBandstructure}, for electrons with spin quantum number $s$, in valley $\eta$, in the band that is closest to the Fermi level at momentum $\vec{k}$. As in \equref{DispersionForm}, $\xi_{\vec{k}}$ parametrizes the associated bandstructure in the absence of VP, $V_z=0$, and $g_{\vec{k},\eta}$ is the form factor of the valley order; we will not further have to specify $g_{\vec{k},\eta}$ and only use that it has to be odd under time-reversal and, thus, obey 
\begin{equation}
    g_{\vec{k},\eta} = -g_{-\vec{k},-\eta}, \quad \eta=\pm, \,\forall \,\vec{k}. \label{TRCOnstraintOng}
\end{equation}

To establish a relation between the bandstructure for electron and hole doping, let us denote the twisted-bilayer-graphene-like flat bands \cite{2021arXiv210602063C} above ($p=+$) and below ($p=-$) the charge neutrality point in valley $\eta$ by $\epsilon_{\vec{k},p,\eta}$. As shown in \cite{2021arXiv210602063C}, a finite displacement field strongly violates the ``particle-hole-like'' symmetry $\epsilon_{\vec{k},p,\eta} = -\epsilon_{-\vec{k},-p,\eta}$ in tTLG (in contrast to the continuum model of twisted-bilayer graphene where it becomes exact in the limit of small twist angles). Instead, the ``chiral symmetry'', 
\begin{equation}
    \epsilon_{\vec{k},p,\eta} = -\epsilon_{\vec{k},-p,\eta}, \label{chiralsymmetry}
\end{equation}
turns out to be approximately obeyed for realistic parameters [becomes exact in the limit where the inter-layer tunneling between same sublattices, $w_0$ in $h^{\text{t}}$ in \equref{DifferentPartsofContHam}, is set to zero].

If we \textit{assume} that both the strength and sign of the VP $V_z$ and the functional form of $g_{\vec{k},\pm}$ are invariant under $\nu_{\text{tTLG}}\rightarrow -\nu_{\text{tTLG}}$, we conclude from \equref{chiralsymmetry} that the effective model for the hole-doped region is given by \equref{EffectiveModel} with
\begin{equation}
    E_{\vec{k},\pm} \, \longrightarrow \,\, \bar{E}_{\vec{k},\pm} = -\xi_{\pm\vec{k}} + V_z g_{\vec{k},\pm} = -E_{-\vec{k},\mp},
\end{equation}
where we have used \equref{TRCOnstraintOng} in the last equality. Inspection of \equref{ExpressionForGamma} yields that $\Gamma(\vec{q})$ for electrons ($\Gamma^e$) and for holes ($\Gamma^h$) are, thus, related by
\begin{equation}
    \Gamma^e(\vec{q}) = \Gamma^h(-\vec{q}).
\end{equation}
From \equref{ExpressionForCurrent1}, we immediately get the relation $\vec{J}^e(\vec{q}) = - \vec{J}^h(-\vec{q})$ and thus $J^e_c(\hat{n}) = J^h_c(-\hat{n})$; this, in turn, directly implies 
\begin{equation}
    \delta J^e_c(\hat{n}) = -\delta J^h_c(\hat{n}).
\end{equation}
This shows that the sign reversal of the diode effect between electron and hole doping can be readily understood from the approximate chiral symmetry, \equref{chiralsymmetry}, of the bandstructure. Within this picture, it also follows that the sign of the order parameter, $V_z$, of valley-polarization does not change in experiment \cite{SCPaper} when sweeping $\nu_{\text{tTLG}}$ between electron and hole doping.
The mechanism fixing the effective sign of $V_z$ when changing the electron density at zero external field might be related to the interpretation of recent observations on twisted monolayer-bilayer graphene \cite{Polshyn2020, ZhuSu2020}.

Of course, in the realistic system neither \equref{chiralsymmetry} is obeyed exactly nor will $V_z$ and $g_{\vec{k},\pm}$ be exactly the same for particle and hole doping (and SOC represents another perturbation), which would explain the reason why the measured magnitude of $\delta J^e_c(\hat{n})$ and $\delta J^h_c(\hat{n})$ are not exactly the same \cite{SCPaper}.

\section{Conclusion and Outlook}\label{Outlook}

We presented a microscopic theory, and detailed analysis of the necessary conditions, for the ZFDE observed \cite{SCPaper} in the tTLG-WSe$_2$ heterostructure in \figref{fig:overview}. We use a combination of general symmetry arguments and explicit model computations, determine the possible superconducting (summarized in \figref{fig:states}) and normal-state instabilities (see summary in \figref{fig:evolutionofnormalstateorders} and \tableref{ListOfOrderParameters}) of the system, study the emergence of vestigial orders [cf.~\equsref{VestigialEOP}{NematicVestOrderParameter}], and the influence of SOC and external magnetic fields on the ZFDE. Taken together, our results offer an explanation of several key findings reported in \cite{SCPaper}---in particular, the field trainability and doping dependence of the ZFDE, as well as the enhanced transverse resistance above the superconducting transition.

We discussed two different microscopic origins of the ZFDE: either (a) time-reversal symmetry is preserved in the normal state but broken spontaneously by the superconducting phase (see \secref{ExoticPairingCurrent}) or (b) it is already broken in the normal state as a result of one of the candidate particle-hole instabilities summarized in \tableref{ListOfOrderParameters}. In case of the latter, we showed that only the states listed in \tableref{ListOfPossibleDiodeeffects} can yield a ZFDE, where the first four (last two) states become symmetry-equivalent in the presence of strong SOC. We also derived the field trainability of these candidate states, showing that the first set of four states in \tableref{ListOfPossibleDiodeeffects} is more consistent with experiment \cite{SCPaper}. Motivated by the fact that all of these states exhibit valley polarization in the presence of SOC, it would be interesting to explicitly control valley polarization in future experiments through the use of combined strain-induced artificial magnetic fields and real magnetic fields, as demonstrated in single-layer graphene \cite{LiLin2020}. 

Invoking the approximate chiral symmetry of the system, we have provided in \secref{DopingDependence} an explanation of the observed sign change of the current asymmetry $\delta J_c$ in \equref{CurrentAsymmetryDefinition} with doping---from electron to hole filling. Since moir\'e systems host ultralow carrier density and narrow bandwidths, electrostatic gating is able to {\it in situ} control the doping and therefore the diode effect. Hence, the ZFDE in moir\'e systems is both generated and manipulated without recourse to external magnetic fields, and thereby offers an interesting platform for future technological applications.

We considered composite order parameters, defined in \equsref{VestigialEOP}{NematicVestOrderParameter}, that capture, respectively, the broken time-reversal and rotational symmetry of the superconducting phases in scenario (a) and (b) above.
Moreover, their condensation above the resistive superconducting transition defines vestigial phases that provide an appealing interpretation for the enhanced transverse resistance measurements in the vicinity of the critical temperature $T\gtrsim T_c$ \cite{SCPaper}. Relatedly, considering the region $T\gtrsim T_c$ there have been earlier works reporting non-reciprocal paraconductivity for Rashba superconductors in a magnetic field \cite{WakatsukiSciAdv2017, WakatsukiPRL2018, HoshinoPRB2018}. It would be interesting to extend the theory presented here to include superconducting fluctuations to examine the possibility of zero-field, paraconducting, non-reciprocal charge transport.

We point out an {\it extreme} diode effect was observed for certain electron fillings in \cite{SCPaper}, whereby a current is needed to stabilize superconductivity. Within our theory, this might be most naturally understood by noting that the underlying magnetic order inducing the ZFDE also weakens superconductivity at the same time; if an applied current acts to weaken the magnetic order, this will, in turn, promote superconductivity that was previously destabilized by the magnetic order parameter.
For the case of valley polarization, this is certainly plausible since current switching of valley polarization in twisted bilayer graphene was demonstrated recently \cite{doi:10.1126/science.aaw3780,CurrentSwitching,YingBalents2021}.

The interplay of topology and the phenomena considered here is worthy of further investigation, as both inter- and intra-valley pairing in related systems have been shown to host first and higher-order topology \cite{chew2021higherorder, LiHOTS}, including in the presence of spin-orbit coupling \cite{scammell2021intrinsic}. 
Furthermore, depending on the precise form of the Fermi surfaces in the magnetically ordered phase of the system, it would also be interesting to generalize the analysis to multiple-$\vec{q}$ superconducting order parameters.

Our results straightforwardly apply to other twisted graphene systems, yet in light of the recent observation of spin-polarized superconductivity in rhombohedral trilayer graphene \cite{Zhou2021RTG}, it would be interesting to extend our analysis to establish the conditions for zero-field diode effect in that system.

\begin{acknowledgments}
M.S.S. thanks Peter P.~Orth for helpful discussions. We also acknowledge discussions with Jiang-Xiazi Lin and Phum Siriviboon in the context of the companion experimental works \cite{SCPaper,DWPaper}.
J.I.A.L. acknowledges support from Brown University. H.D.S. acknowledges funding from ARC Centre of Excellence FLEET.

\end{acknowledgments}

\bibliography{draft_Refs}

\begin{thebibliography}{73}%
\makeatletter
\providecommand \@ifxundefined [1]{%
 \@ifx{#1\undefined}
}%
\providecommand \@ifnum [1]{%
 \ifnum #1\expandafter \@firstoftwo
 \else \expandafter \@secondoftwo
 \fi
}%
\providecommand \@ifx [1]{%
 \ifx #1\expandafter \@firstoftwo
 \else \expandafter \@secondoftwo
 \fi
}%
\providecommand \natexlab [1]{#1}%
\providecommand \enquote  [1]{``#1''}%
\providecommand \bibnamefont  [1]{#1}%
\providecommand \bibfnamefont [1]{#1}%
\providecommand \citenamefont [1]{#1}%
\providecommand \href@noop [0]{\@secondoftwo}%
\providecommand \href [0]{\begingroup \@sanitize@url \@href}%
\providecommand \@href[1]{\@@startlink{#1}\@@href}%
\providecommand \@@href[1]{\endgroup#1\@@endlink}%
\providecommand \@sanitize@url [0]{\catcode `\\12\catcode `\$12\catcode
  `\&12\catcode `\#12\catcode `\^12\catcode `\_12\catcode `\%12\relax}%
\providecommand \@@startlink[1]{}%
\providecommand \@@endlink[0]{}%
\providecommand \url  [0]{\begingroup\@sanitize@url \@url }%
\providecommand \@url [1]{\endgroup\@href {#1}{\urlprefix }}%
\providecommand \urlprefix  [0]{URL }%
\providecommand \Eprint [0]{\href }%
\providecommand \doibase [0]{http://dx.doi.org/}%
\providecommand \selectlanguage [0]{\@gobble}%
\providecommand \bibinfo  [0]{\@secondoftwo}%
\providecommand \bibfield  [0]{\@secondoftwo}%
\providecommand \translation [1]{[#1]}%
\providecommand \BibitemOpen [0]{}%
\providecommand \bibitemStop [0]{}%
\providecommand \bibitemNoStop [0]{.\EOS\space}%
\providecommand \EOS [0]{\spacefactor3000\relax}%
\providecommand \BibitemShut  [1]{\csname bibitem#1\endcsname}%
\let\auto@bib@innerbib\@empty
\bibitem [{\citenamefont {Kitai}(2011)}]{kitai2011principles}%
  \BibitemOpen
  \bibfield  {author} {\bibinfo {author} {\bibfnamefont {A.}~\bibnamefont
  {Kitai}},\ }\href@noop {} {\emph {\bibinfo {title} {Principles of Solar
  Cells, LEDs and Diodes: The role of the PN junction}}}\ (\bibinfo
  {publisher} {John Wiley \& Sons},\ \bibinfo {year} {2011})\BibitemShut
  {NoStop}%
\bibitem [{\citenamefont {Ando}\ \emph {et~al.}(2020)\citenamefont {Ando},
  \citenamefont {Miyasaka}, \citenamefont {Li}, \citenamefont {Ishizuka},
  \citenamefont {Arakawa}, \citenamefont {Shiota}, \citenamefont {Moriyama},
  \citenamefont {Yanase},\ and\ \citenamefont {Ono}}]{Ando2020diodes}%
  \BibitemOpen
  \bibfield  {author} {\bibinfo {author} {\bibfnamefont {F.}~\bibnamefont
  {Ando}}, \bibinfo {author} {\bibfnamefont {Y.}~\bibnamefont {Miyasaka}},
  \bibinfo {author} {\bibfnamefont {T.}~\bibnamefont {Li}}, \bibinfo {author}
  {\bibfnamefont {J.}~\bibnamefont {Ishizuka}}, \bibinfo {author}
  {\bibfnamefont {T.}~\bibnamefont {Arakawa}}, \bibinfo {author} {\bibfnamefont
  {Y.}~\bibnamefont {Shiota}}, \bibinfo {author} {\bibfnamefont
  {T.}~\bibnamefont {Moriyama}}, \bibinfo {author} {\bibfnamefont
  {Y.}~\bibnamefont {Yanase}}, \ and\ \bibinfo {author} {\bibfnamefont
  {T.}~\bibnamefont {Ono}},\ }\bibfield  {title} {\enquote {\bibinfo {title}
  {Observation of superconducting diode effect},}\ }\href@noop {} {\bibfield
  {journal} {\bibinfo  {journal} {Nature}\ }\textbf {\bibinfo {volume} {584}},\
  \bibinfo {pages} {373} (\bibinfo {year} {2020})}\BibitemShut {NoStop}%
\bibitem [{\citenamefont {{Daido}}\ \emph {et~al.}(2021)\citenamefont
  {{Daido}}, \citenamefont {{Ikeda}},\ and\ \citenamefont
  {{Yanase}}}]{DaidoSCDiode}%
  \BibitemOpen
  \bibfield  {author} {\bibinfo {author} {\bibfnamefont {A.}~\bibnamefont
  {{Daido}}}, \bibinfo {author} {\bibfnamefont {Y.}~\bibnamefont {{Ikeda}}}, \
  and\ \bibinfo {author} {\bibfnamefont {Y.}~\bibnamefont {{Yanase}}},\
  }\bibfield  {title} {\enquote {\bibinfo {title} {{Intrinsic Superconducting
  Diode Effect}},}\ }\href@noop {} {\bibfield  {journal} {\bibinfo  {journal}
  {arXiv e-prints}\ } (\bibinfo {year} {2021})},\ \Eprint
  {http://arxiv.org/abs/2106.03326} {arXiv:2106.03326 [cond-mat.supr-con]}
  \BibitemShut {NoStop}%
\bibitem [{\citenamefont {{Yuan}}\ and\ \citenamefont
  {{Fu}}(2021)}]{Yuan2021diodes}%
  \BibitemOpen
  \bibfield  {author} {\bibinfo {author} {\bibfnamefont {N.~F.~Q.}\
  \bibnamefont {{Yuan}}}\ and\ \bibinfo {author} {\bibfnamefont
  {L.}~\bibnamefont {{Fu}}},\ }\bibfield  {title} {\enquote {\bibinfo {title}
  {{Supercurrent diode effect and finite momentum superconductivity}},}\
  }\href@noop {} {\bibfield  {journal} {\bibinfo  {journal} {arXiv e-prints}\ }
  (\bibinfo {year} {2021})},\ \Eprint {http://arxiv.org/abs/2106.01909}
  {arXiv:2106.01909 [cond-mat.supr-con]} \BibitemShut {NoStop}%
\bibitem [{\citenamefont {{He}}\ \emph {et~al.}(2021)\citenamefont {{He}},
  \citenamefont {{Tanaka}},\ and\ \citenamefont {{Nagaosa}}}]{HeSCDiode}%
  \BibitemOpen
  \bibfield  {author} {\bibinfo {author} {\bibfnamefont {J.~J.}\ \bibnamefont
  {{He}}}, \bibinfo {author} {\bibfnamefont {Y.}~\bibnamefont {{Tanaka}}}, \
  and\ \bibinfo {author} {\bibfnamefont {N.}~\bibnamefont {{Nagaosa}}},\
  }\bibfield  {title} {\enquote {\bibinfo {title} {{A Phenomenological Theory
  of Superconductor Diodes in Presence of Magnetochiral Anisotropy}},}\
  }\href@noop {} {\bibfield  {journal} {\bibinfo  {journal} {arXiv e-prints}\ }
  (\bibinfo {year} {2021})},\ \Eprint {http://arxiv.org/abs/2106.03575}
  {arXiv:2106.03575 [cond-mat.supr-con]} \BibitemShut {NoStop}%
\bibitem [{\citenamefont {Lyu}\ \emph {et~al.}(2021)\citenamefont {Lyu},
  \citenamefont {Jiang}, \citenamefont {Wang}, \citenamefont {Xiao},
  \citenamefont {Dong}, \citenamefont {Chen}, \citenamefont
  {Milo{\v{s}}evi{\'{c}}}, \citenamefont {Wang}, \citenamefont {Divan},
  \citenamefont {Pearson}, \citenamefont {Wu}, \citenamefont {Peeters},\ and\
  \citenamefont {Kwok}}]{Lyu2021}%
  \BibitemOpen
  \bibfield  {author} {\bibinfo {author} {\bibfnamefont {Y.-Y.}\ \bibnamefont
  {Lyu}}, \bibinfo {author} {\bibfnamefont {J.}~\bibnamefont {Jiang}}, \bibinfo
  {author} {\bibfnamefont {Y.-L.}\ \bibnamefont {Wang}}, \bibinfo {author}
  {\bibfnamefont {Z.-L.}\ \bibnamefont {Xiao}}, \bibinfo {author}
  {\bibfnamefont {S.}~\bibnamefont {Dong}}, \bibinfo {author} {\bibfnamefont
  {Q.-H.}\ \bibnamefont {Chen}}, \bibinfo {author} {\bibfnamefont {M.~V.}\
  \bibnamefont {Milo{\v{s}}evi{\'{c}}}}, \bibinfo {author} {\bibfnamefont
  {H.}~\bibnamefont {Wang}}, \bibinfo {author} {\bibfnamefont {R.}~\bibnamefont
  {Divan}}, \bibinfo {author} {\bibfnamefont {J.~E.}\ \bibnamefont {Pearson}},
  \bibinfo {author} {\bibfnamefont {P.}~\bibnamefont {Wu}}, \bibinfo {author}
  {\bibfnamefont {F.~M.}\ \bibnamefont {Peeters}}, \ and\ \bibinfo {author}
  {\bibfnamefont {W.-K.}\ \bibnamefont {Kwok}},\ }\bibfield  {title} {\enquote
  {\bibinfo {title} {Superconducting diode effect via conformal-mapped
  nanoholes},}\ }\href {\doibase 10.1038/s41467-021-23077-0} {\bibfield
  {journal} {\bibinfo  {journal} {Nature Communications}\ }\textbf {\bibinfo
  {volume} {12}},\ \bibinfo {pages} {2703} (\bibinfo {year}
  {2021})}\BibitemShut {NoStop}%
\bibitem [{\citenamefont {Bauriedl}\ \emph {et~al.}(2021)\citenamefont
  {Bauriedl}, \citenamefont {Bäuml}, \citenamefont {Fuchs}, \citenamefont
  {Baumgartner}, \citenamefont {Paulik}, \citenamefont {Bauer}, \citenamefont
  {Lin}, \citenamefont {Lupton}, \citenamefont {Taniguchi}, \citenamefont
  {Watanabe}, \citenamefont {Strunk},\ and\ \citenamefont
  {Paradiso}}]{bauriedl2021supercurrent}%
  \BibitemOpen
  \bibfield  {author} {\bibinfo {author} {\bibfnamefont {L.}~\bibnamefont
  {Bauriedl}}, \bibinfo {author} {\bibfnamefont {C.}~\bibnamefont {Bäuml}},
  \bibinfo {author} {\bibfnamefont {L.}~\bibnamefont {Fuchs}}, \bibinfo
  {author} {\bibfnamefont {C.}~\bibnamefont {Baumgartner}}, \bibinfo {author}
  {\bibfnamefont {N.}~\bibnamefont {Paulik}}, \bibinfo {author} {\bibfnamefont
  {J.~M.}\ \bibnamefont {Bauer}}, \bibinfo {author} {\bibfnamefont {K.-Q.}\
  \bibnamefont {Lin}}, \bibinfo {author} {\bibfnamefont {J.~M.}\ \bibnamefont
  {Lupton}}, \bibinfo {author} {\bibfnamefont {T.}~\bibnamefont {Taniguchi}},
  \bibinfo {author} {\bibfnamefont {K.}~\bibnamefont {Watanabe}}, \bibinfo
  {author} {\bibfnamefont {C.}~\bibnamefont {Strunk}}, \ and\ \bibinfo {author}
  {\bibfnamefont {N.}~\bibnamefont {Paradiso}},\ }\href@noop {} {\enquote
  {\bibinfo {title} {Supercurrent diode effect and magnetochiral anisotropy in
  few-layer {NbSe$_2$} nanowires},}\ } (\bibinfo {year} {2021}),\ \Eprint
  {http://arxiv.org/abs/2110.15752} {arXiv:2110.15752 [cond-mat.supr-con]}
  \BibitemShut {NoStop}%
\bibitem [{\citenamefont {{Ili{\'c}}}\ and\ \citenamefont
  {{Bergeret}}(2021)}]{2021arXiv210800209I}%
  \BibitemOpen
  \bibfield  {author} {\bibinfo {author} {\bibfnamefont {S.}~\bibnamefont
  {{Ili{\'c}}}}\ and\ \bibinfo {author} {\bibfnamefont {F.~S.}\ \bibnamefont
  {{Bergeret}}},\ }\bibfield  {title} {\enquote {\bibinfo {title} {{Effect of
  disorder on superconducting diodes}},}\ }\href@noop {} {\bibfield  {journal}
  {\bibinfo  {journal} {arXiv e-prints}\ } (\bibinfo {year} {2021})},\ \Eprint
  {http://arxiv.org/abs/2108.00209} {arXiv:2108.00209 [cond-mat.supr-con]}
  \BibitemShut {NoStop}%
\bibitem [{\citenamefont {Shin}\ \emph {et~al.}(2021)\citenamefont {Shin},
  \citenamefont {Son}, \citenamefont {Yun}, \citenamefont {Park}, \citenamefont
  {Zhang}, \citenamefont {Shin}, \citenamefont {Park},\ and\ \citenamefont
  {Kim}}]{shin2021magnetic}%
  \BibitemOpen
  \bibfield  {author} {\bibinfo {author} {\bibfnamefont {J.}~\bibnamefont
  {Shin}}, \bibinfo {author} {\bibfnamefont {S.}~\bibnamefont {Son}}, \bibinfo
  {author} {\bibfnamefont {J.}~\bibnamefont {Yun}}, \bibinfo {author}
  {\bibfnamefont {G.}~\bibnamefont {Park}}, \bibinfo {author} {\bibfnamefont
  {K.}~\bibnamefont {Zhang}}, \bibinfo {author} {\bibfnamefont {Y.~J.}\
  \bibnamefont {Shin}}, \bibinfo {author} {\bibfnamefont {J.-G.}\ \bibnamefont
  {Park}}, \ and\ \bibinfo {author} {\bibfnamefont {D.}~\bibnamefont {Kim}},\
  }\href@noop {} {\enquote {\bibinfo {title} {Magnetic proximity-induced
  superconducting diode effect and infinite magnetoresistance in van der waals
  heterostructure},}\ } (\bibinfo {year} {2021}),\ \Eprint
  {http://arxiv.org/abs/2111.05627} {arXiv:2111.05627 [cond-mat.supr-con]}
  \BibitemShut {NoStop}%
\bibitem [{\citenamefont {Hu}\ \emph {et~al.}(2007)\citenamefont {Hu},
  \citenamefont {Wu},\ and\ \citenamefont {Dai}}]{PhysRevLett.99.067004}%
  \BibitemOpen
  \bibfield  {author} {\bibinfo {author} {\bibfnamefont {J.}~\bibnamefont
  {Hu}}, \bibinfo {author} {\bibfnamefont {C.}~\bibnamefont {Wu}}, \ and\
  \bibinfo {author} {\bibfnamefont {X.}~\bibnamefont {Dai}},\ }\bibfield
  {title} {\enquote {\bibinfo {title} {Proposed design of a josephson diode},}\
  }\href {\doibase 10.1103/PhysRevLett.99.067004} {\bibfield  {journal}
  {\bibinfo  {journal} {Phys. Rev. Lett.}\ }\textbf {\bibinfo {volume} {99}},\
  \bibinfo {pages} {067004} (\bibinfo {year} {2007})}\BibitemShut {NoStop}%
\bibitem [{\citenamefont {Buzdin}(2008)}]{Buzdin2008}%
  \BibitemOpen
  \bibfield  {author} {\bibinfo {author} {\bibfnamefont {A.}~\bibnamefont
  {Buzdin}},\ }\bibfield  {title} {\enquote {\bibinfo {title} {Direct coupling
  between magnetism and superconducting current in the josephson
  ${\ensuremath{\varphi}}_{0}$ junction},}\ }\href {\doibase
  10.1103/PhysRevLett.101.107005} {\bibfield  {journal} {\bibinfo  {journal}
  {Phys. Rev. Lett.}\ }\textbf {\bibinfo {volume} {101}},\ \bibinfo {pages}
  {107005} (\bibinfo {year} {2008})}\BibitemShut {NoStop}%
\bibitem [{\citenamefont {Szombati}\ \emph {et~al.}(2016)\citenamefont
  {Szombati}, \citenamefont {Nadj-Perge}, \citenamefont {Car}, \citenamefont
  {Plissard}, \citenamefont {Bakkers},\ and\ \citenamefont
  {Kouwenhoven}}]{Szombati2016}%
  \BibitemOpen
  \bibfield  {author} {\bibinfo {author} {\bibfnamefont {D.~B.}\ \bibnamefont
  {Szombati}}, \bibinfo {author} {\bibfnamefont {S.}~\bibnamefont
  {Nadj-Perge}}, \bibinfo {author} {\bibfnamefont {D.}~\bibnamefont {Car}},
  \bibinfo {author} {\bibfnamefont {S.~R.}\ \bibnamefont {Plissard}}, \bibinfo
  {author} {\bibfnamefont {E.~P. A.~M.}\ \bibnamefont {Bakkers}}, \ and\
  \bibinfo {author} {\bibfnamefont {L.~P.}\ \bibnamefont {Kouwenhoven}},\
  }\bibfield  {title} {\enquote {\bibinfo {title} {Josephson $\phi_0$-junction
  in nanowire quantum dots},}\ }\href {\doibase 10.1038/nphys3742} {\bibfield
  {journal} {\bibinfo  {journal} {Nature Physics}\ }\textbf {\bibinfo {volume}
  {12}},\ \bibinfo {pages} {568} (\bibinfo {year} {2016})}\BibitemShut
  {NoStop}%
\bibitem [{\citenamefont {Kopasov}\ \emph {et~al.}(2021)\citenamefont
  {Kopasov}, \citenamefont {Kutlin},\ and\ \citenamefont
  {Mel'nikov}}]{Kopasov2021}%
  \BibitemOpen
  \bibfield  {author} {\bibinfo {author} {\bibfnamefont {A.~A.}\ \bibnamefont
  {Kopasov}}, \bibinfo {author} {\bibfnamefont {A.~G.}\ \bibnamefont {Kutlin}},
  \ and\ \bibinfo {author} {\bibfnamefont {A.~S.}\ \bibnamefont {Mel'nikov}},\
  }\bibfield  {title} {\enquote {\bibinfo {title} {Geometry controlled
  superconducting diode and anomalous josephson effect triggered by the
  topological phase transition in curved proximitized nanowires},}\ }\href
  {\doibase 10.1103/PhysRevB.103.144520} {\bibfield  {journal} {\bibinfo
  {journal} {Phys. Rev. B}\ }\textbf {\bibinfo {volume} {103}},\ \bibinfo
  {pages} {144520} (\bibinfo {year} {2021})}\BibitemShut {NoStop}%
\bibitem [{\citenamefont {Baumgartner}\ \emph
  {et~al.}(2021{\natexlab{a}})\citenamefont {Baumgartner}, \citenamefont
  {Fuchs}, \citenamefont {Costa}, \citenamefont {Reinhardt}, \citenamefont
  {Gronin}, \citenamefont {Gardner}, \citenamefont {Lindemann}, \citenamefont
  {Manfra}, \citenamefont {Faria~Junior}, \citenamefont {Kochan}, \citenamefont
  {Fabian}, \citenamefont {Paradiso},\ and\ \citenamefont
  {Strunk}}]{Baumgartner2021}%
  \BibitemOpen
  \bibfield  {author} {\bibinfo {author} {\bibfnamefont {C.}~\bibnamefont
  {Baumgartner}}, \bibinfo {author} {\bibfnamefont {L.}~\bibnamefont {Fuchs}},
  \bibinfo {author} {\bibfnamefont {A.}~\bibnamefont {Costa}}, \bibinfo
  {author} {\bibfnamefont {S.}~\bibnamefont {Reinhardt}}, \bibinfo {author}
  {\bibfnamefont {S.}~\bibnamefont {Gronin}}, \bibinfo {author} {\bibfnamefont
  {G.~C.}\ \bibnamefont {Gardner}}, \bibinfo {author} {\bibfnamefont
  {T.}~\bibnamefont {Lindemann}}, \bibinfo {author} {\bibfnamefont {M.~J.}\
  \bibnamefont {Manfra}}, \bibinfo {author} {\bibfnamefont {P.~E.}\
  \bibnamefont {Faria~Junior}}, \bibinfo {author} {\bibfnamefont
  {D.}~\bibnamefont {Kochan}}, \bibinfo {author} {\bibfnamefont
  {J.}~\bibnamefont {Fabian}}, \bibinfo {author} {\bibfnamefont
  {N.}~\bibnamefont {Paradiso}}, \ and\ \bibinfo {author} {\bibfnamefont
  {C.}~\bibnamefont {Strunk}},\ }\bibfield  {title} {\enquote {\bibinfo {title}
  {Supercurrent rectification and magnetochiral effects in symmetric josephson
  junctions},}\ }\href {\doibase 10.1038/s41565-021-01009-9} {\bibfield
  {journal} {\bibinfo  {journal} {Nature Nanotechnology}\ } (\bibinfo {year}
  {2021}{\natexlab{a}}),\ 10.1038/s41565-021-01009-9}\BibitemShut {NoStop}%
\bibitem [{\citenamefont {Diez-Merida}\ \emph {et~al.}(2021)\citenamefont
  {Diez-Merida}, \citenamefont {Diez-Carlon}, \citenamefont {Yang},
  \citenamefont {Xie}, \citenamefont {Gao}, \citenamefont {Watanabe},
  \citenamefont {Taniguchi}, \citenamefont {Lu}, \citenamefont {Law},\ and\
  \citenamefont {Efetov}}]{Diez2021magnetic}%
  \BibitemOpen
  \bibfield  {author} {\bibinfo {author} {\bibfnamefont {J.}~\bibnamefont
  {Diez-Merida}}, \bibinfo {author} {\bibfnamefont {A.}~\bibnamefont
  {Diez-Carlon}}, \bibinfo {author} {\bibfnamefont {S.}~\bibnamefont {Yang}},
  \bibinfo {author} {\bibfnamefont {Y.-M.}\ \bibnamefont {Xie}}, \bibinfo
  {author} {\bibfnamefont {X.-J.}\ \bibnamefont {Gao}}, \bibinfo {author}
  {\bibfnamefont {K.}~\bibnamefont {Watanabe}}, \bibinfo {author}
  {\bibfnamefont {T.}~\bibnamefont {Taniguchi}}, \bibinfo {author}
  {\bibfnamefont {X.}~\bibnamefont {Lu}}, \bibinfo {author} {\bibfnamefont
  {K.}~\bibnamefont {Law}}, \ and\ \bibinfo {author} {\bibfnamefont {D.~K.}\
  \bibnamefont {Efetov}},\ }\bibfield  {title} {\enquote {\bibinfo {title}
  {Magnetic josephson junctions and superconducting diodes in magic angle
  twisted bilayer graphene},}\ }\href@noop {} {\bibfield  {journal} {\bibinfo
  {journal} {arXiv preprint arXiv:2110.01067}\ } (\bibinfo {year}
  {2021})}\BibitemShut {NoStop}%
\bibitem [{\citenamefont {Baumgartner}\ \emph
  {et~al.}(2021{\natexlab{b}})\citenamefont {Baumgartner}, \citenamefont
  {Fuchs}, \citenamefont {Costa}, \citenamefont {Cortes}, \citenamefont
  {Reinhardt}, \citenamefont {Gronin}, \citenamefont {Gardner}, \citenamefont
  {Lindemann}, \citenamefont {Manfra}, \citenamefont {Junior}, \citenamefont
  {Kochan}, \citenamefont {Fabian}, \citenamefont {Paradiso},\ and\
  \citenamefont {Strunk}}]{baumgartner2021effect}%
  \BibitemOpen
  \bibfield  {author} {\bibinfo {author} {\bibfnamefont {C.}~\bibnamefont
  {Baumgartner}}, \bibinfo {author} {\bibfnamefont {L.}~\bibnamefont {Fuchs}},
  \bibinfo {author} {\bibfnamefont {A.}~\bibnamefont {Costa}}, \bibinfo
  {author} {\bibfnamefont {J.~P.}\ \bibnamefont {Cortes}}, \bibinfo {author}
  {\bibfnamefont {S.}~\bibnamefont {Reinhardt}}, \bibinfo {author}
  {\bibfnamefont {S.}~\bibnamefont {Gronin}}, \bibinfo {author} {\bibfnamefont
  {G.~C.}\ \bibnamefont {Gardner}}, \bibinfo {author} {\bibfnamefont
  {T.}~\bibnamefont {Lindemann}}, \bibinfo {author} {\bibfnamefont {M.~J.}\
  \bibnamefont {Manfra}}, \bibinfo {author} {\bibfnamefont {P.~E.~F.}\
  \bibnamefont {Junior}}, \bibinfo {author} {\bibfnamefont {D.}~\bibnamefont
  {Kochan}}, \bibinfo {author} {\bibfnamefont {J.}~\bibnamefont {Fabian}},
  \bibinfo {author} {\bibfnamefont {N.}~\bibnamefont {Paradiso}}, \ and\
  \bibinfo {author} {\bibfnamefont {C.}~\bibnamefont {Strunk}},\ }\href@noop {}
  {\enquote {\bibinfo {title} {Effect of rashba and dresselhaus spin-orbit
  coupling on supercurrent rectification and magnetochiral anisotropy of
  ballistic josephson junctions},}\ } (\bibinfo {year} {2021}{\natexlab{b}}),\
  \Eprint {http://arxiv.org/abs/2111.13983} {arXiv:2111.13983
  [cond-mat.supr-con]} \BibitemShut {NoStop}%
\bibitem [{\citenamefont {Wu}\ \emph {et~al.}(2021)\citenamefont {Wu},
  \citenamefont {Wang}, \citenamefont {Sivakumar}, \citenamefont {Pasco},
  \citenamefont {Parkin}, \citenamefont {Zeng}, \citenamefont {McQueen},\ and\
  \citenamefont {Ali}}]{wu2021realization}%
  \BibitemOpen
  \bibfield  {author} {\bibinfo {author} {\bibfnamefont {H.}~\bibnamefont
  {Wu}}, \bibinfo {author} {\bibfnamefont {Y.}~\bibnamefont {Wang}}, \bibinfo
  {author} {\bibfnamefont {P.~K.}\ \bibnamefont {Sivakumar}}, \bibinfo {author}
  {\bibfnamefont {C.}~\bibnamefont {Pasco}}, \bibinfo {author} {\bibfnamefont
  {S.~S.~P.}\ \bibnamefont {Parkin}}, \bibinfo {author} {\bibfnamefont {Y.-J.}\
  \bibnamefont {Zeng}}, \bibinfo {author} {\bibfnamefont {T.}~\bibnamefont
  {McQueen}}, \ and\ \bibinfo {author} {\bibfnamefont {M.~N.}\ \bibnamefont
  {Ali}},\ }\href@noop {} {\enquote {\bibinfo {title} {Realization of the
  field-free josephson diode},}\ } (\bibinfo {year} {2021}),\ \Eprint
  {http://arxiv.org/abs/2103.15809} {arXiv:2103.15809 [cond-mat.supr-con]}
  \BibitemShut {NoStop}%
\bibitem [{\citenamefont {Strambini}\ \emph {et~al.}(2021)\citenamefont
  {Strambini}, \citenamefont {Spies}, \citenamefont {Ligato}, \citenamefont
  {Ilic}, \citenamefont {Rouco}, \citenamefont {Orellana}, \citenamefont
  {Ilyn}, \citenamefont {Rogero}, \citenamefont {Bergeret}, \citenamefont
  {Moodera}, \citenamefont {Virtanen}, \citenamefont {Heikkilä},\ and\
  \citenamefont {Giazotto}}]{strambini2021rectification}%
  \BibitemOpen
  \bibfield  {author} {\bibinfo {author} {\bibfnamefont {E.}~\bibnamefont
  {Strambini}}, \bibinfo {author} {\bibfnamefont {M.}~\bibnamefont {Spies}},
  \bibinfo {author} {\bibfnamefont {N.}~\bibnamefont {Ligato}}, \bibinfo
  {author} {\bibfnamefont {S.}~\bibnamefont {Ilic}}, \bibinfo {author}
  {\bibfnamefont {M.}~\bibnamefont {Rouco}}, \bibinfo {author} {\bibfnamefont
  {C.~G.}\ \bibnamefont {Orellana}}, \bibinfo {author} {\bibfnamefont
  {M.}~\bibnamefont {Ilyn}}, \bibinfo {author} {\bibfnamefont {C.}~\bibnamefont
  {Rogero}}, \bibinfo {author} {\bibfnamefont {F.~S.}\ \bibnamefont
  {Bergeret}}, \bibinfo {author} {\bibfnamefont {J.~S.}\ \bibnamefont
  {Moodera}}, \bibinfo {author} {\bibfnamefont {P.}~\bibnamefont {Virtanen}},
  \bibinfo {author} {\bibfnamefont {T.~T.}\ \bibnamefont {Heikkilä}}, \ and\
  \bibinfo {author} {\bibfnamefont {F.}~\bibnamefont {Giazotto}},\ }\href@noop
  {} {\enquote {\bibinfo {title} {Rectification in a eu-chalcogenide-based
  superconducting diode},}\ } (\bibinfo {year} {2021}),\ \Eprint
  {http://arxiv.org/abs/2109.01061} {arXiv:2109.01061 [cond-mat.supr-con]}
  \BibitemShut {NoStop}%
\bibitem [{\citenamefont {{Halterman}}\ \emph {et~al.}(2021)\citenamefont
  {{Halterman}}, \citenamefont {{Alidoust}}, \citenamefont {{Smith}},\ and\
  \citenamefont {{Starr}}}]{2021arXiv211101242H}%
  \BibitemOpen
  \bibfield  {author} {\bibinfo {author} {\bibfnamefont {K.}~\bibnamefont
  {{Halterman}}}, \bibinfo {author} {\bibfnamefont {M.}~\bibnamefont
  {{Alidoust}}}, \bibinfo {author} {\bibfnamefont {R.}~\bibnamefont {{Smith}}},
  \ and\ \bibinfo {author} {\bibfnamefont {S.}~\bibnamefont {{Starr}}},\
  }\bibfield  {title} {\enquote {\bibinfo {title} {{Supercurrent Diode Effect,
  Spin Torques, and Robust Zero-Energy Peak in Planar Half-Metallic
  Trilayers}},}\ }\href@noop {} {\bibfield  {journal} {\bibinfo  {journal}
  {arXiv e-prints}\ } (\bibinfo {year} {2021})},\ \Eprint
  {http://arxiv.org/abs/2111.01242} {arXiv:2111.01242 [cond-mat.supr-con]}
  \BibitemShut {NoStop}%
\bibitem [{\citenamefont {Lin}\ \emph {et~al.}(2021)\citenamefont {Lin},
  \citenamefont {Siriviboon}, \citenamefont {Scammell}, \citenamefont {Liu},
  \citenamefont {Rhodes}, \citenamefont {Watanabe}, \citenamefont {Taniguchi},
  \citenamefont {Hone}, \citenamefont {Scheurer},\ and\ \citenamefont
  {Li}}]{SCPaper}%
  \BibitemOpen
  \bibfield  {author} {\bibinfo {author} {\bibfnamefont {J.-X.}\ \bibnamefont
  {Lin}}, \bibinfo {author} {\bibfnamefont {P.}~\bibnamefont {Siriviboon}},
  \bibinfo {author} {\bibfnamefont {H.~D.}\ \bibnamefont {Scammell}}, \bibinfo
  {author} {\bibfnamefont {S.}~\bibnamefont {Liu}}, \bibinfo {author}
  {\bibfnamefont {D.}~\bibnamefont {Rhodes}}, \bibinfo {author} {\bibfnamefont
  {K.}~\bibnamefont {Watanabe}}, \bibinfo {author} {\bibfnamefont
  {T.}~\bibnamefont {Taniguchi}}, \bibinfo {author} {\bibfnamefont
  {J.}~\bibnamefont {Hone}}, \bibinfo {author} {\bibfnamefont {M.~S.}\
  \bibnamefont {Scheurer}}, \ and\ \bibinfo {author} {\bibfnamefont
  {J.}~\bibnamefont {Li}},\ }\bibfield  {title} {\enquote {\bibinfo {title}
  {{Zero-field superconducting diode effect in twisted trilayer graphene}},}\
  }\href@noop {} {\bibfield  {journal} {\bibinfo  {journal} {arXiv e-prints}\ }
  (\bibinfo {year} {2021})},\ \Eprint {http://arxiv.org/abs/2112.07841}
  {arXiv:2112.07841 [cond-mat.str-el]} \BibitemShut {NoStop}%
\bibitem [{\citenamefont {Park}\ \emph {et~al.}(2021)\citenamefont {Park},
  \citenamefont {Cao}, \citenamefont {Watanabe}, \citenamefont {Taniguchi},\
  and\ \citenamefont {Jarillo-Herrero}}]{Park_2021}%
  \BibitemOpen
  \bibfield  {author} {\bibinfo {author} {\bibfnamefont {J.~M.}\ \bibnamefont
  {Park}}, \bibinfo {author} {\bibfnamefont {Y.}~\bibnamefont {Cao}}, \bibinfo
  {author} {\bibfnamefont {K.}~\bibnamefont {Watanabe}}, \bibinfo {author}
  {\bibfnamefont {T.}~\bibnamefont {Taniguchi}}, \ and\ \bibinfo {author}
  {\bibfnamefont {P.}~\bibnamefont {Jarillo-Herrero}},\ }\bibfield  {title}
  {\enquote {\bibinfo {title} {Tunable strongly coupled superconductivity in
  magic-angle twisted trilayer graphene},}\ }\href {\doibase
  10.1038/s41586-021-03192-0} {\bibfield  {journal} {\bibinfo  {journal}
  {Nature}\ }\textbf {\bibinfo {volume} {590}},\ \bibinfo {pages} {249–255}
  (\bibinfo {year} {2021})}\BibitemShut {NoStop}%
\bibitem [{\citenamefont {Hao}\ \emph {et~al.}(2021)\citenamefont {Hao},
  \citenamefont {Zimmerman}, \citenamefont {Ledwith}, \citenamefont {Khalaf},
  \citenamefont {Najafabadi}, \citenamefont {Watanabe}, \citenamefont
  {Taniguchi}, \citenamefont {Vishwanath},\ and\ \citenamefont
  {Kim}}]{Hao_2021}%
  \BibitemOpen
  \bibfield  {author} {\bibinfo {author} {\bibfnamefont {Z.}~\bibnamefont
  {Hao}}, \bibinfo {author} {\bibfnamefont {A.~M.}\ \bibnamefont {Zimmerman}},
  \bibinfo {author} {\bibfnamefont {P.}~\bibnamefont {Ledwith}}, \bibinfo
  {author} {\bibfnamefont {E.}~\bibnamefont {Khalaf}}, \bibinfo {author}
  {\bibfnamefont {D.~H.}\ \bibnamefont {Najafabadi}}, \bibinfo {author}
  {\bibfnamefont {K.}~\bibnamefont {Watanabe}}, \bibinfo {author}
  {\bibfnamefont {T.}~\bibnamefont {Taniguchi}}, \bibinfo {author}
  {\bibfnamefont {A.}~\bibnamefont {Vishwanath}}, \ and\ \bibinfo {author}
  {\bibfnamefont {P.}~\bibnamefont {Kim}},\ }\bibfield  {title} {\enquote
  {\bibinfo {title} {Electric field–tunable superconductivity in
  alternating-twist magic-angle trilayer graphene},}\ }\href {\doibase
  10.1126/science.abg0399} {\bibfield  {journal} {\bibinfo  {journal}
  {Science}\ }\textbf {\bibinfo {volume} {371}},\ \bibinfo {pages}
  {1133–1138} (\bibinfo {year} {2021})}\BibitemShut {NoStop}%
\bibitem [{\citenamefont {{Cao}}\ \emph {et~al.}(2021)\citenamefont {{Cao}},
  \citenamefont {{Park}}, \citenamefont {{Watanabe}}, \citenamefont
  {{Taniguchi}},\ and\ \citenamefont
  {{Jarillo-Herrero}}}]{2021arXiv210312083C}%
  \BibitemOpen
  \bibfield  {author} {\bibinfo {author} {\bibfnamefont {Y.}~\bibnamefont
  {{Cao}}}, \bibinfo {author} {\bibfnamefont {J.~M.}\ \bibnamefont {{Park}}},
  \bibinfo {author} {\bibfnamefont {K.}~\bibnamefont {{Watanabe}}}, \bibinfo
  {author} {\bibfnamefont {T.}~\bibnamefont {{Taniguchi}}}, \ and\ \bibinfo
  {author} {\bibfnamefont {P.}~\bibnamefont {{Jarillo-Herrero}}},\ }\bibfield
  {title} {\enquote {\bibinfo {title} {{Large Pauli Limit Violation and
  Reentrant Superconductivity in Magic-Angle Twisted Trilayer Graphene}},}\
  }\href@noop {} {\bibfield  {journal} {\bibinfo  {journal} {arXiv e-prints}\
  ,\ \bibinfo {eid} {arXiv:2103.12083}} (\bibinfo {year} {2021})},\ \Eprint
  {http://arxiv.org/abs/2103.12083} {arXiv:2103.12083 [cond-mat.mes-hall]}
  \BibitemShut {NoStop}%
\bibitem [{\citenamefont {{Kim}}\ \emph {et~al.}(2021)\citenamefont {{Kim}},
  \citenamefont {{Choi}}, \citenamefont {{Lewandowski}}, \citenamefont
  {{Thomson}}, \citenamefont {{Zhang}}, \citenamefont {{Polski}}, \citenamefont
  {{Watanabe}}, \citenamefont {{Taniguchi}}, \citenamefont {{Alicea}},\ and\
  \citenamefont {{Nadj-Perge}}}]{2021arXiv210912127K}%
  \BibitemOpen
  \bibfield  {author} {\bibinfo {author} {\bibfnamefont {H.}~\bibnamefont
  {{Kim}}}, \bibinfo {author} {\bibfnamefont {Y.}~\bibnamefont {{Choi}}},
  \bibinfo {author} {\bibfnamefont {C.}~\bibnamefont {{Lewandowski}}}, \bibinfo
  {author} {\bibfnamefont {A.}~\bibnamefont {{Thomson}}}, \bibinfo {author}
  {\bibfnamefont {Y.}~\bibnamefont {{Zhang}}}, \bibinfo {author} {\bibfnamefont
  {R.}~\bibnamefont {{Polski}}}, \bibinfo {author} {\bibfnamefont
  {K.}~\bibnamefont {{Watanabe}}}, \bibinfo {author} {\bibfnamefont
  {T.}~\bibnamefont {{Taniguchi}}}, \bibinfo {author} {\bibfnamefont
  {J.}~\bibnamefont {{Alicea}}}, \ and\ \bibinfo {author} {\bibfnamefont
  {S.}~\bibnamefont {{Nadj-Perge}}},\ }\bibfield  {title} {\enquote {\bibinfo
  {title} {{Spectroscopic Signatures of Strong Correlations and Unconventional
  Superconductivity in Twisted Trilayer Graphene}},}\ }\href@noop {} {\bibfield
   {journal} {\bibinfo  {journal} {arXiv e-prints}\ } (\bibinfo {year}
  {2021})},\ \Eprint {http://arxiv.org/abs/2109.12127} {arXiv:2109.12127
  [cond-mat.mes-hall]} \BibitemShut {NoStop}%
\bibitem [{\citenamefont {Turkel}\ \emph {et~al.}(2022)\citenamefont {Turkel},
  \citenamefont {Swann}, \citenamefont {Zhu}, \citenamefont {Christos},
  \citenamefont {Watanabe}, \citenamefont {Taniguchi}, \citenamefont {Sachdev},
  \citenamefont {Scheurer}, \citenamefont {Kaxiras}, \citenamefont {Dean},\
  and\ \citenamefont {Pasupathy}}]{2021arXiv210912631T}%
  \BibitemOpen
  \bibfield  {author} {\bibinfo {author} {\bibfnamefont {S.}~\bibnamefont
  {Turkel}}, \bibinfo {author} {\bibfnamefont {J.}~\bibnamefont {Swann}},
  \bibinfo {author} {\bibfnamefont {Z.}~\bibnamefont {Zhu}}, \bibinfo {author}
  {\bibfnamefont {M.}~\bibnamefont {Christos}}, \bibinfo {author}
  {\bibfnamefont {K.}~\bibnamefont {Watanabe}}, \bibinfo {author}
  {\bibfnamefont {T.}~\bibnamefont {Taniguchi}}, \bibinfo {author}
  {\bibfnamefont {S.}~\bibnamefont {Sachdev}}, \bibinfo {author} {\bibfnamefont
  {M.~S.}\ \bibnamefont {Scheurer}}, \bibinfo {author} {\bibfnamefont
  {E.}~\bibnamefont {Kaxiras}}, \bibinfo {author} {\bibfnamefont {C.~R.}\
  \bibnamefont {Dean}}, \ and\ \bibinfo {author} {\bibfnamefont {A.~N.}\
  \bibnamefont {Pasupathy}},\ }\bibfield  {title} {\enquote {\bibinfo {title}
  {Orderly disorder in magic-angle twisted trilayer graphene},}\ }\href
  {\doibase 10.1126/science.abk1895} {\bibfield  {journal} {\bibinfo  {journal}
  {Science}\ }\textbf {\bibinfo {volume} {376}},\ \bibinfo {pages} {193}
  (\bibinfo {year} {2022})}\BibitemShut {NoStop}%
\bibitem [{\citenamefont {{Liu}}\ \emph {et~al.}(2021)\citenamefont {{Liu}},
  \citenamefont {{Zhang}}, \citenamefont {{Watanabe}}, \citenamefont
  {{Taniguchi}},\ and\ \citenamefont {{Li}}}]{2021arXiv210803338L}%
  \BibitemOpen
  \bibfield  {author} {\bibinfo {author} {\bibfnamefont {X.}~\bibnamefont
  {{Liu}}}, \bibinfo {author} {\bibfnamefont {N.~J.}\ \bibnamefont {{Zhang}}},
  \bibinfo {author} {\bibfnamefont {K.}~\bibnamefont {{Watanabe}}}, \bibinfo
  {author} {\bibfnamefont {T.}~\bibnamefont {{Taniguchi}}}, \ and\ \bibinfo
  {author} {\bibfnamefont {J.~I.~A.}\ \bibnamefont {{Li}}},\ }\bibfield
  {title} {\enquote {\bibinfo {title} {{Coulomb screening and thermodynamic
  measurements in magic-angle twisted trilayer graphene}},}\ }\href@noop {}
  {\bibfield  {journal} {\bibinfo  {journal} {arXiv e-prints}\ } (\bibinfo
  {year} {2021})},\ \Eprint {http://arxiv.org/abs/2108.03338} {arXiv:2108.03338
  [cond-mat.mes-hall]} \BibitemShut {NoStop}%
\bibitem [{\citenamefont {Dos~Santos}\ \emph {et~al.}(2007)\citenamefont
  {Dos~Santos}, \citenamefont {Peres},\ and\ \citenamefont
  {Neto}}]{dos2007graphene}%
  \BibitemOpen
  \bibfield  {author} {\bibinfo {author} {\bibfnamefont {J.~M. B.~L.}\
  \bibnamefont {Dos~Santos}}, \bibinfo {author} {\bibfnamefont {N.~M.~R.}\
  \bibnamefont {Peres}}, \ and\ \bibinfo {author} {\bibfnamefont {A.~H.~C.}\
  \bibnamefont {Neto}},\ }\bibfield  {title} {\enquote {\bibinfo {title}
  {Graphene bilayer with a twist: electronic structure},}\ }\href {\doibase
  10.1103/PhysRevLett.99.256802} {\bibfield  {journal} {\bibinfo  {journal}
  {Phys. Rev. Lett.}\ }\textbf {\bibinfo {volume} {99}},\ \bibinfo {pages}
  {256802} (\bibinfo {year} {2007})}\BibitemShut {NoStop}%
\bibitem [{\citenamefont {Bistritzer}\ and\ \citenamefont
  {MacDonald}(2011)}]{bistritzer2011moire}%
  \BibitemOpen
  \bibfield  {author} {\bibinfo {author} {\bibfnamefont {R.}~\bibnamefont
  {Bistritzer}}\ and\ \bibinfo {author} {\bibfnamefont {A.~H.}\ \bibnamefont
  {MacDonald}},\ }\bibfield  {title} {\enquote {\bibinfo {title} {Moir{\'e}
  bands in twisted double-layer graphene},}\ }\href {\doibase
  10.1073/pnas.1108174108} {\bibfield  {journal} {\bibinfo  {journal} {Proc.
  Natl. Acad. Sci. U.S.A.}\ }\textbf {\bibinfo {volume} {108}},\ \bibinfo
  {pages} {12233} (\bibinfo {year} {2011})}\BibitemShut {NoStop}%
\bibitem [{\citenamefont {Dos~Santos}\ \emph {et~al.}(2012)\citenamefont
  {Dos~Santos}, \citenamefont {Peres},\ and\ \citenamefont
  {Neto}}]{dos2012continuum}%
  \BibitemOpen
  \bibfield  {author} {\bibinfo {author} {\bibfnamefont {J.~M. B.~L.}\
  \bibnamefont {Dos~Santos}}, \bibinfo {author} {\bibfnamefont {N.~M.~R.}\
  \bibnamefont {Peres}}, \ and\ \bibinfo {author} {\bibfnamefont {A.~H.~C.}\
  \bibnamefont {Neto}},\ }\bibfield  {title} {\enquote {\bibinfo {title}
  {Continuum model of the twisted graphene bilayer},}\ }\href {\doibase
  10.1103/PhysRevB.86.155449} {\bibfield  {journal} {\bibinfo  {journal} {Phys.
  Rev. B}\ }\textbf {\bibinfo {volume} {86}},\ \bibinfo {pages} {155449}
  (\bibinfo {year} {2012})}\BibitemShut {NoStop}%
\bibitem [{\citenamefont {Gmitra}\ and\ \citenamefont
  {Fabian}(2015)}]{Gmitra2015}%
  \BibitemOpen
  \bibfield  {author} {\bibinfo {author} {\bibfnamefont {M.}~\bibnamefont
  {Gmitra}}\ and\ \bibinfo {author} {\bibfnamefont {J.}~\bibnamefont
  {Fabian}},\ }\bibfield  {title} {\enquote {\bibinfo {title} {Graphene on
  transition-metal dichalcogenides: A platform for proximity spin-orbit physics
  and optospintronics},}\ }\href {\doibase 10.1103/PhysRevB.92.155403}
  {\bibfield  {journal} {\bibinfo  {journal} {Phys. Rev. B}\ }\textbf {\bibinfo
  {volume} {92}},\ \bibinfo {pages} {155403} (\bibinfo {year}
  {2015})}\BibitemShut {NoStop}%
\bibitem [{\citenamefont {{Naimer}}\ \emph {et~al.}(2021)\citenamefont
  {{Naimer}}, \citenamefont {{Zollner}}, \citenamefont {{Gmitra}},\ and\
  \citenamefont {{Fabian}}}]{2021arXiv210806126N}%
  \BibitemOpen
  \bibfield  {author} {\bibinfo {author} {\bibfnamefont {T.}~\bibnamefont
  {{Naimer}}}, \bibinfo {author} {\bibfnamefont {K.}~\bibnamefont {{Zollner}}},
  \bibinfo {author} {\bibfnamefont {M.}~\bibnamefont {{Gmitra}}}, \ and\
  \bibinfo {author} {\bibfnamefont {J.}~\bibnamefont {{Fabian}}},\ }\bibfield
  {title} {\enquote {\bibinfo {title} {{Twist-angle dependent proximity induced
  spin-orbit coupling in graphene/transition-metal dichalcogenide
  heterostructures}},}\ }\href@noop {} {\bibfield  {journal} {\bibinfo
  {journal} {arXiv e-prints}\ } (\bibinfo {year} {2021})},\ \Eprint
  {http://arxiv.org/abs/2108.06126} {arXiv:2108.06126 [cond-mat.mes-hall]}
  \BibitemShut {NoStop}%
\bibitem [{\citenamefont {Khalaf}\ \emph
  {et~al.}(2019{\natexlab{a}})\citenamefont {Khalaf}, \citenamefont {Kruchkov},
  \citenamefont {Tarnopolsky},\ and\ \citenamefont
  {Vishwanath}}]{KhalafKruchkov2019}%
  \BibitemOpen
  \bibfield  {author} {\bibinfo {author} {\bibfnamefont {E.}~\bibnamefont
  {Khalaf}}, \bibinfo {author} {\bibfnamefont {A.~J.}\ \bibnamefont
  {Kruchkov}}, \bibinfo {author} {\bibfnamefont {G.}~\bibnamefont
  {Tarnopolsky}}, \ and\ \bibinfo {author} {\bibfnamefont {A.}~\bibnamefont
  {Vishwanath}},\ }\bibfield  {title} {\enquote {\bibinfo {title} {Magic angle
  hierarchy in twisted graphene multilayers},}\ }\href {\doibase
  10.1103/PhysRevB.100.085109} {\bibfield  {journal} {\bibinfo  {journal}
  {Phys. Rev. B}\ }\textbf {\bibinfo {volume} {100}},\ \bibinfo {pages}
  {085109} (\bibinfo {year} {2019}{\natexlab{a}})}\BibitemShut {NoStop}%
\bibitem [{\citenamefont {Carr}\ \emph {et~al.}(2020)\citenamefont {Carr},
  \citenamefont {Li}, \citenamefont {Zhu}, \citenamefont {Kaxiras},
  \citenamefont {Sachdev},\ and\ \citenamefont {Kruchkov}}]{CarrKruchkov2020}%
  \BibitemOpen
  \bibfield  {author} {\bibinfo {author} {\bibfnamefont {S.}~\bibnamefont
  {Carr}}, \bibinfo {author} {\bibfnamefont {C.}~\bibnamefont {Li}}, \bibinfo
  {author} {\bibfnamefont {Z.}~\bibnamefont {Zhu}}, \bibinfo {author}
  {\bibfnamefont {E.}~\bibnamefont {Kaxiras}}, \bibinfo {author} {\bibfnamefont
  {S.}~\bibnamefont {Sachdev}}, \ and\ \bibinfo {author} {\bibfnamefont
  {A.}~\bibnamefont {Kruchkov}},\ }\bibfield  {title} {\enquote {\bibinfo
  {title} {Ultraheavy and ultrarelativistic dirac quasiparticles in sandwiched
  graphenes},}\ }\href {\doibase 10.1021/acs.nanolett.9b04979} {\bibfield
  {journal} {\bibinfo  {journal} {Nano Letters}\ }\textbf {\bibinfo {volume}
  {20}},\ \bibinfo {pages} {3030} (\bibinfo {year} {2020})}\BibitemShut
  {NoStop}%
\bibitem [{\citenamefont {Mora}\ \emph {et~al.}(2019)\citenamefont {Mora},
  \citenamefont {Regnault},\ and\ \citenamefont {Bernevig}}]{MoraRegnault2019}%
  \BibitemOpen
  \bibfield  {author} {\bibinfo {author} {\bibfnamefont {C.}~\bibnamefont
  {Mora}}, \bibinfo {author} {\bibfnamefont {N.}~\bibnamefont {Regnault}}, \
  and\ \bibinfo {author} {\bibfnamefont {B.~A.}\ \bibnamefont {Bernevig}},\
  }\bibfield  {title} {\enquote {\bibinfo {title} {Flatbands and perfect metal
  in trilayer moir\'e graphene},}\ }\href {\doibase
  10.1103/PhysRevLett.123.026402} {\bibfield  {journal} {\bibinfo  {journal}
  {Phys. Rev. Lett.}\ }\textbf {\bibinfo {volume} {123}},\ \bibinfo {pages}
  {026402} (\bibinfo {year} {2019})}\BibitemShut {NoStop}%
\bibitem [{\citenamefont {{Siriviboon}}\ \emph {et~al.}(2021)\citenamefont
  {{Siriviboon}}, \citenamefont {{Lin}}, \citenamefont {{Scammell}},
  \citenamefont {{Liu}}, \citenamefont {{Rhodes}}, \citenamefont {{Watanabe}},
  \citenamefont {{Taniguchi}}, \citenamefont {{Hone}}, \citenamefont
  {{Scheurer}},\ and\ \citenamefont {{Li}}}]{DWPaper}%
  \BibitemOpen
  \bibfield  {author} {\bibinfo {author} {\bibfnamefont {P.}~\bibnamefont
  {{Siriviboon}}}, \bibinfo {author} {\bibfnamefont {J.-X.}\ \bibnamefont
  {{Lin}}}, \bibinfo {author} {\bibfnamefont {H.~D.}\ \bibnamefont
  {{Scammell}}}, \bibinfo {author} {\bibfnamefont {S.}~\bibnamefont {{Liu}}},
  \bibinfo {author} {\bibfnamefont {D.}~\bibnamefont {{Rhodes}}}, \bibinfo
  {author} {\bibfnamefont {K.}~\bibnamefont {{Watanabe}}}, \bibinfo {author}
  {\bibfnamefont {T.}~\bibnamefont {{Taniguchi}}}, \bibinfo {author}
  {\bibfnamefont {J.}~\bibnamefont {{Hone}}}, \bibinfo {author} {\bibfnamefont
  {M.~S.}\ \bibnamefont {{Scheurer}}}, \ and\ \bibinfo {author} {\bibfnamefont
  {J.~I.~A.}\ \bibnamefont {{Li}}},\ }\bibfield  {title} {\enquote {\bibinfo
  {title} {{Abundance of density wave phases in twisted trilayer graphene on
  WSe$_2$}},}\ }\href@noop {} {\bibfield  {journal} {\bibinfo  {journal} {arXiv
  e-prints}\ } (\bibinfo {year} {2021})},\ \Eprint
  {http://arxiv.org/abs/2112.07127} {arXiv:2112.07127 [cond-mat.mes-hall]}
  \BibitemShut {NoStop}%
\bibitem [{\citenamefont {Christos}\ \emph {et~al.}(2020)\citenamefont
  {Christos}, \citenamefont {Sachdev},\ and\ \citenamefont
  {Scheurer}}]{Christos29543}%
  \BibitemOpen
  \bibfield  {author} {\bibinfo {author} {\bibfnamefont {M.}~\bibnamefont
  {Christos}}, \bibinfo {author} {\bibfnamefont {S.}~\bibnamefont {Sachdev}}, \
  and\ \bibinfo {author} {\bibfnamefont {M.~S.}\ \bibnamefont {Scheurer}},\
  }\bibfield  {title} {\enquote {\bibinfo {title} {{Superconductivity,
  correlated insulators, and Wess--Zumino--Witten terms in twisted bilayer
  graphene}},}\ }\href {\doibase 10.1073/pnas.2014691117} {\bibfield  {journal}
  {\bibinfo  {journal} {Proc. Natl. Acad. Sci. U.S.A.}\ }\textbf {\bibinfo
  {volume} {117}},\ \bibinfo {pages} {29543} (\bibinfo {year}
  {2020})}\BibitemShut {NoStop}%
\bibitem [{\citenamefont {{C{\v{a}}lug{\v{a}}ru}}\ \emph
  {et~al.}(2021)\citenamefont {{C{\v{a}}lug{\v{a}}ru}}, \citenamefont {{Xie}},
  \citenamefont {{Song}}, \citenamefont {{Lian}}, \citenamefont {{Regnault}},\
  and\ \citenamefont {{Bernevig}}}]{2021PhRvB.103s5411C}%
  \BibitemOpen
  \bibfield  {author} {\bibinfo {author} {\bibfnamefont {D.}~\bibnamefont
  {{C{\v{a}}lug{\v{a}}ru}}}, \bibinfo {author} {\bibfnamefont {F.}~\bibnamefont
  {{Xie}}}, \bibinfo {author} {\bibfnamefont {Z.-D.}\ \bibnamefont {{Song}}},
  \bibinfo {author} {\bibfnamefont {B.}~\bibnamefont {{Lian}}}, \bibinfo
  {author} {\bibfnamefont {N.}~\bibnamefont {{Regnault}}}, \ and\ \bibinfo
  {author} {\bibfnamefont {B.~A.}\ \bibnamefont {{Bernevig}}},\ }\bibfield
  {title} {\enquote {\bibinfo {title} {{Twisted symmetric trilayer graphene:
  Single-particle and many-body Hamiltonians and hidden nonlocal symmetries of
  trilayer moir{\'e} systems with and without displacement field}},}\ }\href
  {\doibase 10.1103/PhysRevB.103.195411} {\bibfield  {journal} {\bibinfo
  {journal} {\prb}\ }\textbf {\bibinfo {volume} {103}},\ \bibinfo {eid}
  {195411} (\bibinfo {year} {2021})},\ \Eprint
  {http://arxiv.org/abs/2102.06201} {arXiv:2102.06201 [cond-mat.str-el]}
  \BibitemShut {NoStop}%
\bibitem [{\citenamefont {Khalaf}\ \emph
  {et~al.}(2019{\natexlab{b}})\citenamefont {Khalaf}, \citenamefont {Kruchkov},
  \citenamefont {Tarnopolsky},\ and\ \citenamefont {Vishwanath}}]{Khalaf_2019}%
  \BibitemOpen
  \bibfield  {author} {\bibinfo {author} {\bibfnamefont {E.}~\bibnamefont
  {Khalaf}}, \bibinfo {author} {\bibfnamefont {A.~J.}\ \bibnamefont
  {Kruchkov}}, \bibinfo {author} {\bibfnamefont {G.}~\bibnamefont
  {Tarnopolsky}}, \ and\ \bibinfo {author} {\bibfnamefont {A.}~\bibnamefont
  {Vishwanath}},\ }\bibfield  {title} {\enquote {\bibinfo {title} {Magic angle
  hierarchy in twisted graphene multilayers},}\ }\href {\doibase
  10.1103/physrevb.100.085109} {\bibfield  {journal} {\bibinfo  {journal}
  {Physical Review B}\ }\textbf {\bibinfo {volume} {100}} (\bibinfo {year}
  {2019}{\natexlab{b}}),\ 10.1103/physrevb.100.085109}\BibitemShut {NoStop}%
\bibitem [{\citenamefont {Huder}\ \emph {et~al.}(2018)\citenamefont {Huder},
  \citenamefont {Artaud}, \citenamefont {Le~Quang}, \citenamefont
  {de~Laissardi\`ere}, \citenamefont {Jansen}, \citenamefont {Lapertot},
  \citenamefont {Chapelier},\ and\ \citenamefont {Renard}}]{Huder2018}%
  \BibitemOpen
  \bibfield  {author} {\bibinfo {author} {\bibfnamefont {L.}~\bibnamefont
  {Huder}}, \bibinfo {author} {\bibfnamefont {A.}~\bibnamefont {Artaud}},
  \bibinfo {author} {\bibfnamefont {T.}~\bibnamefont {Le~Quang}}, \bibinfo
  {author} {\bibfnamefont {G.~T.}\ \bibnamefont {de~Laissardi\`ere}}, \bibinfo
  {author} {\bibfnamefont {A.~G.~M.}\ \bibnamefont {Jansen}}, \bibinfo {author}
  {\bibfnamefont {G.}~\bibnamefont {Lapertot}}, \bibinfo {author}
  {\bibfnamefont {C.}~\bibnamefont {Chapelier}}, \ and\ \bibinfo {author}
  {\bibfnamefont {V.~T.}\ \bibnamefont {Renard}},\ }\bibfield  {title}
  {\enquote {\bibinfo {title} {Electronic spectrum of twisted graphene layers
  under heterostrain},}\ }\href {\doibase 10.1103/PhysRevLett.120.156405}
  {\bibfield  {journal} {\bibinfo  {journal} {Phys. Rev. Lett.}\ }\textbf
  {\bibinfo {volume} {120}},\ \bibinfo {pages} {156405} (\bibinfo {year}
  {2018})}\BibitemShut {NoStop}%
\bibitem [{\citenamefont {Kazmierczak}\ \emph {et~al.}(2021)\citenamefont
  {Kazmierczak}, \citenamefont {Van~Winkle}, \citenamefont {Ophus},
  \citenamefont {Bustillo}, \citenamefont {Carr}, \citenamefont {Brown},
  \citenamefont {Ciston}, \citenamefont {Taniguchi}, \citenamefont {Watanabe},\
  and\ \citenamefont {Bediako}}]{Kazmierczak2021}%
  \BibitemOpen
  \bibfield  {author} {\bibinfo {author} {\bibfnamefont {N.~P.}\ \bibnamefont
  {Kazmierczak}}, \bibinfo {author} {\bibfnamefont {M.}~\bibnamefont
  {Van~Winkle}}, \bibinfo {author} {\bibfnamefont {C.}~\bibnamefont {Ophus}},
  \bibinfo {author} {\bibfnamefont {K.~C.}\ \bibnamefont {Bustillo}}, \bibinfo
  {author} {\bibfnamefont {S.}~\bibnamefont {Carr}}, \bibinfo {author}
  {\bibfnamefont {H.~G.}\ \bibnamefont {Brown}}, \bibinfo {author}
  {\bibfnamefont {J.}~\bibnamefont {Ciston}}, \bibinfo {author} {\bibfnamefont
  {T.}~\bibnamefont {Taniguchi}}, \bibinfo {author} {\bibfnamefont
  {K.}~\bibnamefont {Watanabe}}, \ and\ \bibinfo {author} {\bibfnamefont
  {D.~K.}\ \bibnamefont {Bediako}},\ }\bibfield  {title} {\enquote {\bibinfo
  {title} {Strain fields in twisted bilayer graphene},}\ }\href {\doibase
  10.1038/s41563-021-00973-w} {\bibfield  {journal} {\bibinfo  {journal}
  {Nature Materials}\ }\textbf {\bibinfo {volume} {20}},\ \bibinfo {pages}
  {956} (\bibinfo {year} {2021})}\BibitemShut {NoStop}%
\bibitem [{\citenamefont {Kerelsky}\ \emph {et~al.}(2019)\citenamefont
  {Kerelsky}, \citenamefont {McGilly}, \citenamefont {Kennes}, \citenamefont
  {Xian}, \citenamefont {Yankowitz}, \citenamefont {Chen}, \citenamefont
  {Watanabe}, \citenamefont {Taniguchi}, \citenamefont {Hone}, \citenamefont
  {Dean}, \citenamefont {Rubio},\ and\ \citenamefont
  {Pasupathy}}]{Kerelsky2019}%
  \BibitemOpen
  \bibfield  {author} {\bibinfo {author} {\bibfnamefont {A.}~\bibnamefont
  {Kerelsky}}, \bibinfo {author} {\bibfnamefont {L.~J.}\ \bibnamefont
  {McGilly}}, \bibinfo {author} {\bibfnamefont {D.~M.}\ \bibnamefont {Kennes}},
  \bibinfo {author} {\bibfnamefont {L.}~\bibnamefont {Xian}}, \bibinfo {author}
  {\bibfnamefont {M.}~\bibnamefont {Yankowitz}}, \bibinfo {author}
  {\bibfnamefont {S.}~\bibnamefont {Chen}}, \bibinfo {author} {\bibfnamefont
  {K.}~\bibnamefont {Watanabe}}, \bibinfo {author} {\bibfnamefont
  {T.}~\bibnamefont {Taniguchi}}, \bibinfo {author} {\bibfnamefont
  {J.}~\bibnamefont {Hone}}, \bibinfo {author} {\bibfnamefont {C.}~\bibnamefont
  {Dean}}, \bibinfo {author} {\bibfnamefont {A.}~\bibnamefont {Rubio}}, \ and\
  \bibinfo {author} {\bibfnamefont {A.~N.}\ \bibnamefont {Pasupathy}},\
  }\bibfield  {title} {\enquote {\bibinfo {title} {Maximized electron
  interactions at the magic angle in twisted bilayer graphene},}\ }\href
  {\doibase 10.1038/s41586-019-1431-9} {\bibfield  {journal} {\bibinfo
  {journal} {Nature}\ }\textbf {\bibinfo {volume} {572}},\ \bibinfo {pages}
  {95} (\bibinfo {year} {2019})}\BibitemShut {NoStop}%
\bibitem [{\citenamefont {Jiang}\ \emph {et~al.}(2019)\citenamefont {Jiang},
  \citenamefont {Lai}, \citenamefont {Watanabe}, \citenamefont {Taniguchi},
  \citenamefont {Haule}, \citenamefont {Mao},\ and\ \citenamefont
  {Andrei}}]{Jiang2019}%
  \BibitemOpen
  \bibfield  {author} {\bibinfo {author} {\bibfnamefont {Y.}~\bibnamefont
  {Jiang}}, \bibinfo {author} {\bibfnamefont {X.}~\bibnamefont {Lai}}, \bibinfo
  {author} {\bibfnamefont {K.}~\bibnamefont {Watanabe}}, \bibinfo {author}
  {\bibfnamefont {T.}~\bibnamefont {Taniguchi}}, \bibinfo {author}
  {\bibfnamefont {K.}~\bibnamefont {Haule}}, \bibinfo {author} {\bibfnamefont
  {J.}~\bibnamefont {Mao}}, \ and\ \bibinfo {author} {\bibfnamefont {E.~Y.}\
  \bibnamefont {Andrei}},\ }\bibfield  {title} {\enquote {\bibinfo {title}
  {Charge order and broken rotational symmetry in magic-angle twisted bilayer
  graphene},}\ }\href {\doibase 10.1038/s41586-019-1460-4} {\bibfield
  {journal} {\bibinfo  {journal} {Nature}\ }\textbf {\bibinfo {volume} {573}},\
  \bibinfo {pages} {91} (\bibinfo {year} {2019})}\BibitemShut {NoStop}%
\bibitem [{\citenamefont {Choi}\ \emph {et~al.}(2019)\citenamefont {Choi},
  \citenamefont {Kemmer}, \citenamefont {Peng}, \citenamefont {Thomson},
  \citenamefont {Arora}, \citenamefont {Polski}, \citenamefont {Zhang},
  \citenamefont {Ren}, \citenamefont {Alicea}, \citenamefont {Refael},
  \citenamefont {von Oppen}, \citenamefont {Watanabe}, \citenamefont
  {Taniguchi},\ and\ \citenamefont {Nadj-Perge}}]{Choi2019}%
  \BibitemOpen
  \bibfield  {author} {\bibinfo {author} {\bibfnamefont {Y.}~\bibnamefont
  {Choi}}, \bibinfo {author} {\bibfnamefont {J.}~\bibnamefont {Kemmer}},
  \bibinfo {author} {\bibfnamefont {Y.}~\bibnamefont {Peng}}, \bibinfo {author}
  {\bibfnamefont {A.}~\bibnamefont {Thomson}}, \bibinfo {author} {\bibfnamefont
  {H.}~\bibnamefont {Arora}}, \bibinfo {author} {\bibfnamefont
  {R.}~\bibnamefont {Polski}}, \bibinfo {author} {\bibfnamefont
  {Y.}~\bibnamefont {Zhang}}, \bibinfo {author} {\bibfnamefont
  {H.}~\bibnamefont {Ren}}, \bibinfo {author} {\bibfnamefont {J.}~\bibnamefont
  {Alicea}}, \bibinfo {author} {\bibfnamefont {G.}~\bibnamefont {Refael}},
  \bibinfo {author} {\bibfnamefont {F.}~\bibnamefont {von Oppen}}, \bibinfo
  {author} {\bibfnamefont {K.}~\bibnamefont {Watanabe}}, \bibinfo {author}
  {\bibfnamefont {T.}~\bibnamefont {Taniguchi}}, \ and\ \bibinfo {author}
  {\bibfnamefont {S.}~\bibnamefont {Nadj-Perge}},\ }\bibfield  {title}
  {\enquote {\bibinfo {title} {Electronic correlations in twisted bilayer
  graphene near the magic angle},}\ }\href {\doibase 10.1038/s41567-019-0606-5}
  {\bibfield  {journal} {\bibinfo  {journal} {Nature Physics}\ }\textbf
  {\bibinfo {volume} {15}},\ \bibinfo {pages} {1174} (\bibinfo {year}
  {2019})}\BibitemShut {NoStop}%
\bibitem [{\citenamefont {Cao}\ \emph {et~al.}(2021)\citenamefont {Cao},
  \citenamefont {Rodan-Legrain}, \citenamefont {Park}, \citenamefont {Yuan},
  \citenamefont {Watanabe}, \citenamefont {Taniguchi}, \citenamefont
  {Fernandes}, \citenamefont {Fu},\ and\ \citenamefont
  {Jarillo-Herrero}}]{CaoNematicity2021}%
  \BibitemOpen
  \bibfield  {author} {\bibinfo {author} {\bibfnamefont {Y.}~\bibnamefont
  {Cao}}, \bibinfo {author} {\bibfnamefont {D.}~\bibnamefont {Rodan-Legrain}},
  \bibinfo {author} {\bibfnamefont {J.~M.}\ \bibnamefont {Park}}, \bibinfo
  {author} {\bibfnamefont {N.~F.~Q.}\ \bibnamefont {Yuan}}, \bibinfo {author}
  {\bibfnamefont {K.}~\bibnamefont {Watanabe}}, \bibinfo {author}
  {\bibfnamefont {T.}~\bibnamefont {Taniguchi}}, \bibinfo {author}
  {\bibfnamefont {R.~M.}\ \bibnamefont {Fernandes}}, \bibinfo {author}
  {\bibfnamefont {L.}~\bibnamefont {Fu}}, \ and\ \bibinfo {author}
  {\bibfnamefont {P.}~\bibnamefont {Jarillo-Herrero}},\ }\bibfield  {title}
  {\enquote {\bibinfo {title} {Nematicity and competing orders in
  superconducting magic-angle graphene},}\ }\href {\doibase
  10.1126/science.abc2836} {\bibfield  {journal} {\bibinfo  {journal}
  {Science}\ }\textbf {\bibinfo {volume} {372}},\ \bibinfo {pages} {264}
  (\bibinfo {year} {2021})}\BibitemShut {NoStop}%
\bibitem [{\citenamefont {Rubio-Verd{\'u}}\ \emph {et~al.}(2022)\citenamefont
  {Rubio-Verd{\'u}}, \citenamefont {Turkel}, \citenamefont {Song},
  \citenamefont {Klebl}, \citenamefont {Samajdar}, \citenamefont {Scheurer},
  \citenamefont {Venderbos}, \citenamefont {Watanabe}, \citenamefont
  {Taniguchi}, \citenamefont {Ochoa}, \citenamefont {Xian}, \citenamefont
  {Kennes}, \citenamefont {Fernandes}, \citenamefont {Rubio},\ and\
  \citenamefont {Pasupathy}}]{rubioverdu2020universal}%
  \BibitemOpen
  \bibfield  {author} {\bibinfo {author} {\bibfnamefont {C.}~\bibnamefont
  {Rubio-Verd{\'u}}}, \bibinfo {author} {\bibfnamefont {S.}~\bibnamefont
  {Turkel}}, \bibinfo {author} {\bibfnamefont {Y.}~\bibnamefont {Song}},
  \bibinfo {author} {\bibfnamefont {L.}~\bibnamefont {Klebl}}, \bibinfo
  {author} {\bibfnamefont {R.}~\bibnamefont {Samajdar}}, \bibinfo {author}
  {\bibfnamefont {M.~S.}\ \bibnamefont {Scheurer}}, \bibinfo {author}
  {\bibfnamefont {J.~W.~F.}\ \bibnamefont {Venderbos}}, \bibinfo {author}
  {\bibfnamefont {K.}~\bibnamefont {Watanabe}}, \bibinfo {author}
  {\bibfnamefont {T.}~\bibnamefont {Taniguchi}}, \bibinfo {author}
  {\bibfnamefont {H.}~\bibnamefont {Ochoa}}, \bibinfo {author} {\bibfnamefont
  {L.}~\bibnamefont {Xian}}, \bibinfo {author} {\bibfnamefont {D.~M.}\
  \bibnamefont {Kennes}}, \bibinfo {author} {\bibfnamefont {R.~M.}\
  \bibnamefont {Fernandes}}, \bibinfo {author} {\bibfnamefont
  {{\'A}.}~\bibnamefont {Rubio}}, \ and\ \bibinfo {author} {\bibfnamefont
  {A.~N.}\ \bibnamefont {Pasupathy}},\ }\bibfield  {title} {\enquote {\bibinfo
  {title} {Moir{\'e}nematic phase in twisted double bilayer graphene},}\ }\href
  {\doibase 10.1038/s41567-021-01438-2} {\bibfield  {journal} {\bibinfo
  {journal} {Nature Physics}\ }\textbf {\bibinfo {volume} {18}},\ \bibinfo
  {pages} {196} (\bibinfo {year} {2022})}\BibitemShut {NoStop}%
\bibitem [{\citenamefont {Bi}\ \emph {et~al.}(2019)\citenamefont {Bi},
  \citenamefont {Yuan},\ and\ \citenamefont {Fu}}]{PhysRevB.100.035448}%
  \BibitemOpen
  \bibfield  {author} {\bibinfo {author} {\bibfnamefont {Z.}~\bibnamefont
  {Bi}}, \bibinfo {author} {\bibfnamefont {N.~F.~Q.}\ \bibnamefont {Yuan}}, \
  and\ \bibinfo {author} {\bibfnamefont {L.}~\bibnamefont {Fu}},\ }\bibfield
  {title} {\enquote {\bibinfo {title} {Designing flat bands by strain},}\
  }\href {\doibase 10.1103/PhysRevB.100.035448} {\bibfield  {journal} {\bibinfo
   {journal} {Phys. Rev. B}\ }\textbf {\bibinfo {volume} {100}},\ \bibinfo
  {pages} {035448} (\bibinfo {year} {2019})}\BibitemShut {NoStop}%
\bibitem [{\citenamefont {Samajdar}\ \emph {et~al.}(2021)\citenamefont
  {Samajdar}, \citenamefont {Scheurer}, \citenamefont {Turkel}, \citenamefont
  {Rubio-Verd{\'{u}}}, \citenamefont {Pasupathy}, \citenamefont {Venderbos},\
  and\ \citenamefont {Fernandes}}]{10.1088/2053-1583/abfcd6}%
  \BibitemOpen
  \bibfield  {author} {\bibinfo {author} {\bibfnamefont {R.}~\bibnamefont
  {Samajdar}}, \bibinfo {author} {\bibfnamefont {M.}~\bibnamefont {Scheurer}},
  \bibinfo {author} {\bibfnamefont {S.}~\bibnamefont {Turkel}}, \bibinfo
  {author} {\bibfnamefont {C.}~\bibnamefont {Rubio-Verd{\'{u}}}}, \bibinfo
  {author} {\bibfnamefont {A.}~\bibnamefont {Pasupathy}}, \bibinfo {author}
  {\bibfnamefont {J.}~\bibnamefont {Venderbos}}, \ and\ \bibinfo {author}
  {\bibfnamefont {R.~M.}\ \bibnamefont {Fernandes}},\ }\bibfield  {title}
  {\enquote {\bibinfo {title} {Electric-field-tunable electronic nematic order
  in twisted double-bilayer graphene},}\ }\href
  {https://doi.org/10.1088/2053-1583/abfcd6} {\bibfield  {journal} {\bibinfo
  {journal} {2D Materials}\ }\textbf {\bibinfo {volume} {8}} (\bibinfo {year}
  {2021})}\BibitemShut {NoStop}%
\bibitem [{\citenamefont {Christos}\ \emph {et~al.}(2022)\citenamefont
  {Christos}, \citenamefont {Sachdev},\ and\ \citenamefont
  {Scheurer}}]{2021arXiv210602063C}%
  \BibitemOpen
  \bibfield  {author} {\bibinfo {author} {\bibfnamefont {M.}~\bibnamefont
  {Christos}}, \bibinfo {author} {\bibfnamefont {S.}~\bibnamefont {Sachdev}}, \
  and\ \bibinfo {author} {\bibfnamefont {M.~S.}\ \bibnamefont {Scheurer}},\
  }\bibfield  {title} {\enquote {\bibinfo {title} {Correlated insulators,
  semimetals, and superconductivity in twisted trilayer graphene},}\ }\href
  {\doibase 10.1103/PhysRevX.12.021018} {\bibfield  {journal} {\bibinfo
  {journal} {Phys. Rev. X}\ }\textbf {\bibinfo {volume} {12}},\ \bibinfo
  {pages} {021018} (\bibinfo {year} {2022})}\BibitemShut {NoStop}%
\bibitem [{\citenamefont {{Gonzalez}}\ and\ \citenamefont
  {{Stauber}}(2021)}]{2021arXiv211011294G}%
  \BibitemOpen
  \bibfield  {author} {\bibinfo {author} {\bibfnamefont {J.}~\bibnamefont
  {{Gonzalez}}}\ and\ \bibinfo {author} {\bibfnamefont {T.}~\bibnamefont
  {{Stauber}}},\ }\bibfield  {title} {\enquote {\bibinfo {title} {{$p$-wave
  superconductivity induced from valley symmetry breaking in twisted trilayer
  graphene}},}\ }\href@noop {} {\bibfield  {journal} {\bibinfo  {journal}
  {arXiv e-prints}\ } (\bibinfo {year} {2021})},\ \Eprint
  {http://arxiv.org/abs/2110.11294} {arXiv:2110.11294 [cond-mat.supr-con]}
  \BibitemShut {NoStop}%
\bibitem [{\citenamefont {Scheurer}\ and\ \citenamefont
  {Samajdar}(2020)}]{PhysRevResearch.2.033062}%
  \BibitemOpen
  \bibfield  {author} {\bibinfo {author} {\bibfnamefont {M.~S.}\ \bibnamefont
  {Scheurer}}\ and\ \bibinfo {author} {\bibfnamefont {R.}~\bibnamefont
  {Samajdar}},\ }\bibfield  {title} {\enquote {\bibinfo {title} {Pairing in
  graphene-based moir\'e superlattices},}\ }\href {\doibase
  10.1103/PhysRevResearch.2.033062} {\bibfield  {journal} {\bibinfo  {journal}
  {Phys. Rev. Research}\ }\textbf {\bibinfo {volume} {2}},\ \bibinfo {pages}
  {033062} (\bibinfo {year} {2020})}\BibitemShut {NoStop}%
\bibitem [{\citenamefont {Scheurer}(2016)}]{PhysRevB.93.174509}%
  \BibitemOpen
  \bibfield  {author} {\bibinfo {author} {\bibfnamefont {M.~S.}\ \bibnamefont
  {Scheurer}},\ }\bibfield  {title} {\enquote {\bibinfo {title} {Mechanism,
  time-reversal symmetry, and topology of superconductivity in
  noncentrosymmetric systems},}\ }\href {\doibase 10.1103/PhysRevB.93.174509}
  {\bibfield  {journal} {\bibinfo  {journal} {Phys. Rev. B}\ }\textbf {\bibinfo
  {volume} {93}},\ \bibinfo {pages} {174509} (\bibinfo {year}
  {2016})}\BibitemShut {NoStop}%
\bibitem [{\citenamefont {Samajdar}\ and\ \citenamefont
  {Scheurer}(2020)}]{OurMicroscopicTDBG}%
  \BibitemOpen
  \bibfield  {author} {\bibinfo {author} {\bibfnamefont {R.}~\bibnamefont
  {Samajdar}}\ and\ \bibinfo {author} {\bibfnamefont {M.~S.}\ \bibnamefont
  {Scheurer}},\ }\bibfield  {title} {\enquote {\bibinfo {title} {{Microscopic
  pairing mechanism, order parameter, and disorder sensitivity in moir\'e
  superlattices: Applications to twisted double-bilayer graphene}},}\ }\href
  {\doibase 10.1103/PhysRevB.102.064501} {\bibfield  {journal} {\bibinfo
  {journal} {Phys. Rev. B}\ }\textbf {\bibinfo {volume} {102}},\ \bibinfo
  {pages} {064501} (\bibinfo {year} {2020})}\BibitemShut {NoStop}%
\bibitem [{\citenamefont {Scheurer}\ \emph {et~al.}(2017)\citenamefont
  {Scheurer}, \citenamefont {Agterberg},\ and\ \citenamefont
  {Schmalian}}]{SelectionRulePaper}%
  \BibitemOpen
  \bibfield  {author} {\bibinfo {author} {\bibfnamefont {M.~S.}\ \bibnamefont
  {Scheurer}}, \bibinfo {author} {\bibfnamefont {D.~F.}\ \bibnamefont
  {Agterberg}}, \ and\ \bibinfo {author} {\bibfnamefont {J.}~\bibnamefont
  {Schmalian}},\ }\bibfield  {title} {\enquote {\bibinfo {title} {Selection
  rules for cooper pairing in two-dimensional interfaces and sheets},}\ }\href
  {\doibase 10.1038/s41535-016-0008-1} {\bibfield  {journal} {\bibinfo
  {journal} {npj Quantum Materials}\ }\textbf {\bibinfo {volume} {2}},\
  \bibinfo {pages} {9} (\bibinfo {year} {2017})}\BibitemShut {NoStop}%
\bibitem [{\citenamefont {Fernandes}\ \emph {et~al.}(2019)\citenamefont
  {Fernandes}, \citenamefont {Orth},\ and\ \citenamefont
  {Schmalian}}]{VestigialOrderReview}%
  \BibitemOpen
  \bibfield  {author} {\bibinfo {author} {\bibfnamefont {R.~M.}\ \bibnamefont
  {Fernandes}}, \bibinfo {author} {\bibfnamefont {P.~P.}\ \bibnamefont {Orth}},
  \ and\ \bibinfo {author} {\bibfnamefont {J.}~\bibnamefont {Schmalian}},\
  }\bibfield  {title} {\enquote {\bibinfo {title} {Intertwined vestigial order
  in quantum materials: Nematicity and beyond},}\ }\href {\doibase
  10.1146/annurev-conmatphys-031218-013200} {\bibfield  {journal} {\bibinfo
  {journal} {Annual Review of Condensed Matter Physics}\ }\textbf {\bibinfo
  {volume} {10}},\ \bibinfo {pages} {133} (\bibinfo {year} {2019})}\BibitemShut
  {NoStop}%
\bibitem [{\citenamefont {Zinkl}\ \emph {et~al.}(2021)\citenamefont {Zinkl},
  \citenamefont {Hamamoto},\ and\ \citenamefont {Sigrist}}]{zinkl2021symmetry}%
  \BibitemOpen
  \bibfield  {author} {\bibinfo {author} {\bibfnamefont {B.}~\bibnamefont
  {Zinkl}}, \bibinfo {author} {\bibfnamefont {K.}~\bibnamefont {Hamamoto}}, \
  and\ \bibinfo {author} {\bibfnamefont {M.}~\bibnamefont {Sigrist}},\
  }\href@noop {} {\enquote {\bibinfo {title} {Symmetry conditions for the
  superconducting diode effect in chiral superconductors},}\ } (\bibinfo {year}
  {2021}),\ \Eprint {http://arxiv.org/abs/2111.05340} {arXiv:2111.05340
  [cond-mat.supr-con]} \BibitemShut {NoStop}%
\bibitem [{\citenamefont {Hooper}\ \emph {et~al.}(2004)\citenamefont {Hooper},
  \citenamefont {Mao}, \citenamefont {Nelson}, \citenamefont {Liu},
  \citenamefont {Wada},\ and\ \citenamefont {Maeno}}]{PhysRevB.70.014510}%
  \BibitemOpen
  \bibfield  {author} {\bibinfo {author} {\bibfnamefont {J.}~\bibnamefont
  {Hooper}}, \bibinfo {author} {\bibfnamefont {Z.~Q.}\ \bibnamefont {Mao}},
  \bibinfo {author} {\bibfnamefont {K.~D.}\ \bibnamefont {Nelson}}, \bibinfo
  {author} {\bibfnamefont {Y.}~\bibnamefont {Liu}}, \bibinfo {author}
  {\bibfnamefont {M.}~\bibnamefont {Wada}}, \ and\ \bibinfo {author}
  {\bibfnamefont {Y.}~\bibnamefont {Maeno}},\ }\bibfield  {title} {\enquote
  {\bibinfo {title} {Anomalous josephson network in the
  $\mathrm{Ru}\text{\penalty1000-\hskip0pt}{\mathrm{sr}}_{2}{\mathrm{ru}\mathrm{o}}_{4}$
  eutectic system},}\ }\href {\doibase 10.1103/PhysRevB.70.014510} {\bibfield
  {journal} {\bibinfo  {journal} {Phys. Rev. B}\ }\textbf {\bibinfo {volume}
  {70}},\ \bibinfo {pages} {014510} (\bibinfo {year} {2004})}\BibitemShut
  {NoStop}%
\bibitem [{\citenamefont {Liu}\ and\ \citenamefont
  {Dai}(2021)}]{LiuReview2021}%
  \BibitemOpen
  \bibfield  {author} {\bibinfo {author} {\bibfnamefont {J.}~\bibnamefont
  {Liu}}\ and\ \bibinfo {author} {\bibfnamefont {X.}~\bibnamefont {Dai}},\
  }\bibfield  {title} {\enquote {\bibinfo {title} {Orbital magnetic states in
  moir{\'e} graphene systems},}\ }\href {\doibase 10.1038/s42254-021-00297-3}
  {\bibfield  {journal} {\bibinfo  {journal} {Nature Reviews Physics}\ }\textbf
  {\bibinfo {volume} {3}},\ \bibinfo {pages} {367} (\bibinfo {year}
  {2021})}\BibitemShut {NoStop}%
\bibitem [{\citenamefont {Sharpe}\ \emph {et~al.}(2019)\citenamefont {Sharpe},
  \citenamefont {Fox}, \citenamefont {Barnard}, \citenamefont {Finney},
  \citenamefont {Watanabe}, \citenamefont {Taniguchi}, \citenamefont
  {Kastner},\ and\ \citenamefont
  {Goldhaber-Gordon}}]{doi:10.1126/science.aaw3780}%
  \BibitemOpen
  \bibfield  {author} {\bibinfo {author} {\bibfnamefont {A.~L.}\ \bibnamefont
  {Sharpe}}, \bibinfo {author} {\bibfnamefont {E.~J.}\ \bibnamefont {Fox}},
  \bibinfo {author} {\bibfnamefont {A.~W.}\ \bibnamefont {Barnard}}, \bibinfo
  {author} {\bibfnamefont {J.}~\bibnamefont {Finney}}, \bibinfo {author}
  {\bibfnamefont {K.}~\bibnamefont {Watanabe}}, \bibinfo {author}
  {\bibfnamefont {T.}~\bibnamefont {Taniguchi}}, \bibinfo {author}
  {\bibfnamefont {M.~A.}\ \bibnamefont {Kastner}}, \ and\ \bibinfo {author}
  {\bibfnamefont {D.}~\bibnamefont {Goldhaber-Gordon}},\ }\bibfield  {title}
  {\enquote {\bibinfo {title} {Emergent ferromagnetism near three-quarters
  filling in twisted bilayer graphene},}\ }\href {\doibase
  10.1126/science.aaw3780} {\bibfield  {journal} {\bibinfo  {journal}
  {Science}\ }\textbf {\bibinfo {volume} {365}},\ \bibinfo {pages} {605}
  (\bibinfo {year} {2019})}\BibitemShut {NoStop}%
\bibitem [{\citenamefont {{Lin}}\ \emph {et~al.}(2021)\citenamefont {{Lin}},
  \citenamefont {{Zhang}}, \citenamefont {{Morissette}}, \citenamefont
  {{Wang}}, \citenamefont {{Liu}}, \citenamefont {{Rhodes}}, \citenamefont
  {{Watanabe}}, \citenamefont {{Taniguchi}}, \citenamefont {{Hone}},\ and\
  \citenamefont {{Li}}}]{2021arXiv210206566L}%
  \BibitemOpen
  \bibfield  {author} {\bibinfo {author} {\bibfnamefont {J.-X.}\ \bibnamefont
  {{Lin}}}, \bibinfo {author} {\bibfnamefont {Y.-H.}\ \bibnamefont {{Zhang}}},
  \bibinfo {author} {\bibfnamefont {E.}~\bibnamefont {{Morissette}}}, \bibinfo
  {author} {\bibfnamefont {Z.}~\bibnamefont {{Wang}}}, \bibinfo {author}
  {\bibfnamefont {S.}~\bibnamefont {{Liu}}}, \bibinfo {author} {\bibfnamefont
  {D.}~\bibnamefont {{Rhodes}}}, \bibinfo {author} {\bibfnamefont
  {K.}~\bibnamefont {{Watanabe}}}, \bibinfo {author} {\bibfnamefont
  {T.}~\bibnamefont {{Taniguchi}}}, \bibinfo {author} {\bibfnamefont
  {J.}~\bibnamefont {{Hone}}}, \ and\ \bibinfo {author} {\bibfnamefont
  {J.~I.~A.}\ \bibnamefont {{Li}}},\ }\bibfield  {title} {\enquote {\bibinfo
  {title} {{Spin-orbit driven ferromagnetism at half moir{\'e} filling in
  magic-angle twisted bilayer graphene}},}\ }\href@noop {} {\bibfield
  {journal} {\bibinfo  {journal} {arXiv e-prints}\ } (\bibinfo {year}
  {2021})},\ \Eprint {http://arxiv.org/abs/2102.06566} {arXiv:2102.06566
  [cond-mat.mes-hall]} \BibitemShut {NoStop}%
\bibitem [{Cap()}]{Caption}%
  \BibitemOpen
  \href@noop {} {}\bibinfo {howpublished} {(a) without strain or valley
  polarization $\mu/t=-0.68; \phi_+=\phi_-=-0.7\pi$; $\delta\mu=\beta=0$, (b)
  with {\it weak} valley polarization but without strain ($\mu/t=-0.68;
  \phi_+=\phi_=-0.7\pi$; $\delta\mu/t=-0.2$; $\beta=0$), (c) with {\it strong}
  valley polarization but without strain ($\mu/t=-0.68; \phi_+=\phi_=-0.7\pi$;
  $\delta\mu/t=-0.6$; $\beta=0$), (d) {\it weak} valley polarization in the
  presence of strain ($\mu/t=-0.42; \phi_+=-0.8\pi, \phi_-=-1.1\pi$;
  $\delta\mu/t=-0.1$; $\beta=0.5$), (e) intravalley pairing, no strain
  ($\mu/t=-0.68; \phi_+=\phi_-=-0.7\pi$; $\beta=0$).}\BibitemShut {Stop}%
\bibitem [{\citenamefont {Roy}\ and\ \citenamefont
  {Herbut}(2010)}]{RoyHubert2010}%
  \BibitemOpen
  \bibfield  {author} {\bibinfo {author} {\bibfnamefont {B.}~\bibnamefont
  {Roy}}\ and\ \bibinfo {author} {\bibfnamefont {I.~F.}\ \bibnamefont
  {Herbut}},\ }\bibfield  {title} {\enquote {\bibinfo {title} {Unconventional
  superconductivity on honeycomb lattice: Theory of kekule order parameter},}\
  }\href {\doibase 10.1103/PhysRevB.82.035429} {\bibfield  {journal} {\bibinfo
  {journal} {Phys. Rev. B}\ }\textbf {\bibinfo {volume} {82}},\ \bibinfo
  {pages} {035429} (\bibinfo {year} {2010})}\BibitemShut {NoStop}%
\bibitem [{\citenamefont {Polshyn}\ \emph {et~al.}(2020)\citenamefont
  {Polshyn}, \citenamefont {Zhu}, \citenamefont {Kumar}, \citenamefont {Zhang},
  \citenamefont {Yang}, \citenamefont {Tschirhart}, \citenamefont {Serlin},
  \citenamefont {Watanabe}, \citenamefont {Taniguchi}, \citenamefont
  {MacDonald},\ and\ \citenamefont {Young}}]{Polshyn2020}%
  \BibitemOpen
  \bibfield  {author} {\bibinfo {author} {\bibfnamefont {H.}~\bibnamefont
  {Polshyn}}, \bibinfo {author} {\bibfnamefont {J.}~\bibnamefont {Zhu}},
  \bibinfo {author} {\bibfnamefont {M.~A.}\ \bibnamefont {Kumar}}, \bibinfo
  {author} {\bibfnamefont {Y.}~\bibnamefont {Zhang}}, \bibinfo {author}
  {\bibfnamefont {F.}~\bibnamefont {Yang}}, \bibinfo {author} {\bibfnamefont
  {C.~L.}\ \bibnamefont {Tschirhart}}, \bibinfo {author} {\bibfnamefont
  {M.}~\bibnamefont {Serlin}}, \bibinfo {author} {\bibfnamefont
  {K.}~\bibnamefont {Watanabe}}, \bibinfo {author} {\bibfnamefont
  {T.}~\bibnamefont {Taniguchi}}, \bibinfo {author} {\bibfnamefont {A.~H.}\
  \bibnamefont {MacDonald}}, \ and\ \bibinfo {author} {\bibfnamefont {A.~F.}\
  \bibnamefont {Young}},\ }\bibfield  {title} {\enquote {\bibinfo {title}
  {Electrical switching of magnetic order in an orbital chern insulator},}\
  }\href {\doibase 10.1038/s41586-020-2963-8} {\bibfield  {journal} {\bibinfo
  {journal} {Nature}\ }\textbf {\bibinfo {volume} {588}},\ \bibinfo {pages}
  {66} (\bibinfo {year} {2020})}\BibitemShut {NoStop}%
\bibitem [{\citenamefont {Zhu}\ \emph {et~al.}(2020)\citenamefont {Zhu},
  \citenamefont {Su},\ and\ \citenamefont {MacDonald}}]{ZhuSu2020}%
  \BibitemOpen
  \bibfield  {author} {\bibinfo {author} {\bibfnamefont {J.}~\bibnamefont
  {Zhu}}, \bibinfo {author} {\bibfnamefont {J.-J.}\ \bibnamefont {Su}}, \ and\
  \bibinfo {author} {\bibfnamefont {A.~H.}\ \bibnamefont {MacDonald}},\
  }\bibfield  {title} {\enquote {\bibinfo {title} {Voltage-controlled magnetic
  reversal in orbital chern insulators},}\ }\href {\doibase
  10.1103/PhysRevLett.125.227702} {\bibfield  {journal} {\bibinfo  {journal}
  {Phys. Rev. Lett.}\ }\textbf {\bibinfo {volume} {125}},\ \bibinfo {pages}
  {227702} (\bibinfo {year} {2020})}\BibitemShut {NoStop}%
\bibitem [{\citenamefont {Li}\ \emph {et~al.}(2020)\citenamefont {Li},
  \citenamefont {Su}, \citenamefont {Ren},\ and\ \citenamefont
  {He}}]{LiLin2020}%
  \BibitemOpen
  \bibfield  {author} {\bibinfo {author} {\bibfnamefont {S.-Y.}\ \bibnamefont
  {Li}}, \bibinfo {author} {\bibfnamefont {Y.}~\bibnamefont {Su}}, \bibinfo
  {author} {\bibfnamefont {Y.-N.}\ \bibnamefont {Ren}}, \ and\ \bibinfo
  {author} {\bibfnamefont {L.}~\bibnamefont {He}},\ }\bibfield  {title}
  {\enquote {\bibinfo {title} {Valley polarization and inversion in strained
  graphene via pseudo-landau levels, valley splitting of real landau levels,
  and confined states},}\ }\href {\doibase 10.1103/PhysRevLett.124.106802}
  {\bibfield  {journal} {\bibinfo  {journal} {Phys. Rev. Lett.}\ }\textbf
  {\bibinfo {volume} {124}},\ \bibinfo {pages} {106802} (\bibinfo {year}
  {2020})}\BibitemShut {NoStop}%
\bibitem [{\citenamefont {Wakatsuki}\ \emph {et~al.}(2017)\citenamefont
  {Wakatsuki}, \citenamefont {Saito}, \citenamefont {Hoshino}, \citenamefont
  {Itahashi}, \citenamefont {Ideue}, \citenamefont {Ezawa}, \citenamefont
  {Iwasa},\ and\ \citenamefont {Nagaosa}}]{WakatsukiSciAdv2017}%
  \BibitemOpen
  \bibfield  {author} {\bibinfo {author} {\bibfnamefont {R.}~\bibnamefont
  {Wakatsuki}}, \bibinfo {author} {\bibfnamefont {Y.}~\bibnamefont {Saito}},
  \bibinfo {author} {\bibfnamefont {S.}~\bibnamefont {Hoshino}}, \bibinfo
  {author} {\bibfnamefont {Y.~M.}\ \bibnamefont {Itahashi}}, \bibinfo {author}
  {\bibfnamefont {T.}~\bibnamefont {Ideue}}, \bibinfo {author} {\bibfnamefont
  {M.}~\bibnamefont {Ezawa}}, \bibinfo {author} {\bibfnamefont
  {Y.}~\bibnamefont {Iwasa}}, \ and\ \bibinfo {author} {\bibfnamefont
  {N.}~\bibnamefont {Nagaosa}},\ }\bibfield  {title} {\enquote {\bibinfo
  {title} {Nonreciprocal charge transport in noncentrosymmetric
  superconductors},}\ }\href {\doibase 10.1126/sciadv.1602390} {\bibfield
  {journal} {\bibinfo  {journal} {Science Advances}\ }\textbf {\bibinfo
  {volume} {3}},\ \bibinfo {pages} {e1602390} (\bibinfo {year}
  {2017})}\BibitemShut {NoStop}%
\bibitem [{\citenamefont {Wakatsuki}\ and\ \citenamefont
  {Nagaosa}(2018)}]{WakatsukiPRL2018}%
  \BibitemOpen
  \bibfield  {author} {\bibinfo {author} {\bibfnamefont {R.}~\bibnamefont
  {Wakatsuki}}\ and\ \bibinfo {author} {\bibfnamefont {N.}~\bibnamefont
  {Nagaosa}},\ }\bibfield  {title} {\enquote {\bibinfo {title} {Nonreciprocal
  current in noncentrosymmetric rashba superconductors},}\ }\href {\doibase
  10.1103/PhysRevLett.121.026601} {\bibfield  {journal} {\bibinfo  {journal}
  {Phys. Rev. Lett.}\ }\textbf {\bibinfo {volume} {121}},\ \bibinfo {pages}
  {026601} (\bibinfo {year} {2018})}\BibitemShut {NoStop}%
\bibitem [{\citenamefont {Hoshino}\ \emph {et~al.}(2018)\citenamefont
  {Hoshino}, \citenamefont {Wakatsuki}, \citenamefont {Hamamoto},\ and\
  \citenamefont {Nagaosa}}]{HoshinoPRB2018}%
  \BibitemOpen
  \bibfield  {author} {\bibinfo {author} {\bibfnamefont {S.}~\bibnamefont
  {Hoshino}}, \bibinfo {author} {\bibfnamefont {R.}~\bibnamefont {Wakatsuki}},
  \bibinfo {author} {\bibfnamefont {K.}~\bibnamefont {Hamamoto}}, \ and\
  \bibinfo {author} {\bibfnamefont {N.}~\bibnamefont {Nagaosa}},\ }\bibfield
  {title} {\enquote {\bibinfo {title} {Nonreciprocal charge transport in
  two-dimensional noncentrosymmetric superconductors},}\ }\href {\doibase
  10.1103/PhysRevB.98.054510} {\bibfield  {journal} {\bibinfo  {journal} {Phys.
  Rev. B}\ }\textbf {\bibinfo {volume} {98}},\ \bibinfo {pages} {054510}
  (\bibinfo {year} {2018})}\BibitemShut {NoStop}%
\bibitem [{\citenamefont {Serlin}\ \emph {et~al.}(2020)\citenamefont {Serlin},
  \citenamefont {Tschirhart}, \citenamefont {Polshyn}, \citenamefont {Zhang},
  \citenamefont {Zhu}, \citenamefont {Watanabe}, \citenamefont {Taniguchi},
  \citenamefont {Balents},\ and\ \citenamefont {Young}}]{CurrentSwitching}%
  \BibitemOpen
  \bibfield  {author} {\bibinfo {author} {\bibfnamefont {M.}~\bibnamefont
  {Serlin}}, \bibinfo {author} {\bibfnamefont {C.~L.}\ \bibnamefont
  {Tschirhart}}, \bibinfo {author} {\bibfnamefont {H.}~\bibnamefont {Polshyn}},
  \bibinfo {author} {\bibfnamefont {Y.}~\bibnamefont {Zhang}}, \bibinfo
  {author} {\bibfnamefont {J.}~\bibnamefont {Zhu}}, \bibinfo {author}
  {\bibfnamefont {K.}~\bibnamefont {Watanabe}}, \bibinfo {author}
  {\bibfnamefont {T.}~\bibnamefont {Taniguchi}}, \bibinfo {author}
  {\bibfnamefont {L.}~\bibnamefont {Balents}}, \ and\ \bibinfo {author}
  {\bibfnamefont {A.~F.}\ \bibnamefont {Young}},\ }\bibfield  {title} {\enquote
  {\bibinfo {title} {Intrinsic quantized anomalous hall effect in a moire
  heterostructure},}\ }\href {\doibase 10.1126/science.aay5533} {\bibfield
  {journal} {\bibinfo  {journal} {Science}\ }\textbf {\bibinfo {volume}
  {367}},\ \bibinfo {pages} {900} (\bibinfo {year} {2020})}\BibitemShut
  {NoStop}%
\bibitem [{\citenamefont {Ying}\ \emph {et~al.}(2021)\citenamefont {Ying},
  \citenamefont {Ye},\ and\ \citenamefont {Balents}}]{YingBalents2021}%
  \BibitemOpen
  \bibfield  {author} {\bibinfo {author} {\bibfnamefont {X.}~\bibnamefont
  {Ying}}, \bibinfo {author} {\bibfnamefont {M.}~\bibnamefont {Ye}}, \ and\
  \bibinfo {author} {\bibfnamefont {L.}~\bibnamefont {Balents}},\ }\bibfield
  {title} {\enquote {\bibinfo {title} {Current switching of valley polarization
  in twisted bilayer graphene},}\ }\href {\doibase 10.1103/PhysRevB.103.115436}
  {\bibfield  {journal} {\bibinfo  {journal} {Phys. Rev. B}\ }\textbf {\bibinfo
  {volume} {103}},\ \bibinfo {pages} {115436} (\bibinfo {year}
  {2021})}\BibitemShut {NoStop}%
\bibitem [{\citenamefont {Chew}\ \emph {et~al.}(2021)\citenamefont {Chew},
  \citenamefont {Wang}, \citenamefont {Bernevig},\ and\ \citenamefont
  {Song}}]{chew2021higherorder}%
  \BibitemOpen
  \bibfield  {author} {\bibinfo {author} {\bibfnamefont {A.}~\bibnamefont
  {Chew}}, \bibinfo {author} {\bibfnamefont {Y.}~\bibnamefont {Wang}}, \bibinfo
  {author} {\bibfnamefont {B.~A.}\ \bibnamefont {Bernevig}}, \ and\ \bibinfo
  {author} {\bibfnamefont {Z.-D.}\ \bibnamefont {Song}},\ }\href@noop {}
  {\enquote {\bibinfo {title} {Higher-order topological superconductivity in
  twisted bilayer graphene},}\ } (\bibinfo {year} {2021}),\ \Eprint
  {http://arxiv.org/abs/2108.05373} {arXiv:2108.05373 [cond-mat.supr-con]}
  \BibitemShut {NoStop}%
\bibitem [{\citenamefont {Li}\ \emph {et~al.}(2021)\citenamefont {Li},
  \citenamefont {Geier}, \citenamefont {Ingham},\ and\ \citenamefont
  {Scammell}}]{LiHOTS}%
  \BibitemOpen
  \bibfield  {author} {\bibinfo {author} {\bibfnamefont {T.}~\bibnamefont
  {Li}}, \bibinfo {author} {\bibfnamefont {M.}~\bibnamefont {Geier}}, \bibinfo
  {author} {\bibfnamefont {J.}~\bibnamefont {Ingham}}, \ and\ \bibinfo {author}
  {\bibfnamefont {H.}~\bibnamefont {Scammell}},\ }\bibfield  {title} {\enquote
  {\bibinfo {title} {Higher-order topological superconductivity from repulsive
  interactions in kagome and honeycomb systems},}\ }\href
  {http://iopscience.iop.org/article/10.1088/2053-1583/ac4060} {\bibfield
  {journal} {\bibinfo  {journal} {2D Materials}\ } (\bibinfo {year}
  {2021})}\BibitemShut {NoStop}%
\bibitem [{\citenamefont {Scammell}\ \emph {et~al.}(2021)\citenamefont
  {Scammell}, \citenamefont {Ingham}, \citenamefont {Geier},\ and\
  \citenamefont {Li}}]{scammell2021intrinsic}%
  \BibitemOpen
  \bibfield  {author} {\bibinfo {author} {\bibfnamefont {H.~D.}\ \bibnamefont
  {Scammell}}, \bibinfo {author} {\bibfnamefont {J.}~\bibnamefont {Ingham}},
  \bibinfo {author} {\bibfnamefont {M.}~\bibnamefont {Geier}}, \ and\ \bibinfo
  {author} {\bibfnamefont {T.}~\bibnamefont {Li}},\ }\href@noop {} {\enquote
  {\bibinfo {title} {Intrinsic first and higher-order topological
  superconductivity in a doped topological insulator},}\ } (\bibinfo {year}
  {2021}),\ \Eprint {http://arxiv.org/abs/2111.07252} {arXiv:2111.07252
  [cond-mat.supr-con]} \BibitemShut {NoStop}%
\bibitem [{\citenamefont {Zhou}\ \emph {et~al.}(2021)\citenamefont {Zhou},
  \citenamefont {Xie}, \citenamefont {Taniguchi}, \citenamefont {Watanabe},\
  and\ \citenamefont {Young}}]{Zhou2021RTG}%
  \BibitemOpen
  \bibfield  {author} {\bibinfo {author} {\bibfnamefont {H.}~\bibnamefont
  {Zhou}}, \bibinfo {author} {\bibfnamefont {T.}~\bibnamefont {Xie}}, \bibinfo
  {author} {\bibfnamefont {T.}~\bibnamefont {Taniguchi}}, \bibinfo {author}
  {\bibfnamefont {K.}~\bibnamefont {Watanabe}}, \ and\ \bibinfo {author}
  {\bibfnamefont {A.~F.}\ \bibnamefont {Young}},\ }\bibfield  {title} {\enquote
  {\bibinfo {title} {Superconductivity in rhombohedral trilayer graphene},}\
  }\href {\doibase 10.1038/s41586-021-03926-0} {\bibfield  {journal} {\bibinfo
  {journal} {Nature}\ }\textbf {\bibinfo {volume} {598}},\ \bibinfo {pages}
  {434} (\bibinfo {year} {2021})}\BibitemShut {NoStop}%
\end{thebibliography}%

\onecolumngrid

\begin{appendix}

\section{Continuum model}\label{a:continuum}
The continuum Hamiltonian is described in detail in Section \ref{ContinuumModel}. For clarity, here we additionally provide the explicit momentum space representation. 

To establish the Hamiltonian in momentum space, we denote by $c_{\vec{k};\rho,l,\eta,s,\vec{G}}$ the electron annihilation operator with the following quantum numbers: crystalline momentum $\vec{k}$ within the moir\'e Brillouin zone (MBZ); spin $s=\uparrow,\downarrow$; sublattice $\rho=A,B$; valley $\eta = \pm$ of the graphene layer $\ell=1,2,3$;  and reciprocal lattice (RL) vector $\vec{G} = \sum_{j=1,2} n_j \vec{G}_j$, $n_j \in \mathbbm{Z}$ of the effective moir\'e lattice. 

The unitary transformation in layer space,
\begin{equation}
    c_{\vec{k};\rho,l,\eta,s,\vec{G}} = V_{l,\ell} \psi_{\vec{k};\rho,\ell,\eta,s,\vec{G}}, \qquad V=\frac{1}{\sqrt{2}} \begin{pmatrix} 1 & 0 & -1 \\ 0 & \sqrt{2} & 0 \\ 1 & 0 & 1 \end{pmatrix}, \label{TrafoToMirrorEigenbasis}
\end{equation}
conveniently decomposes the system into mirror-even ($\ell = 1,2$) and mirror-odd ($\ell = 3$) subspaces, which become mixed at nonzero $D_0$ or SOC. Working in this mirror basis, the full continuum model described in the main text is
\begin{align}
   \notag H_0 &= \sum_{\vec{k} \in \text{MBZ}} \sum_{\rho,\rho'=A,B} \sum_{\ell,\ell'=1,2,3} \sum_{\eta=\pm} \sum_{s=\uparrow,\downarrow}  \sum_{\vec{G},\vec{G}' \in \text{RL}} \psi^\dagger_{\vec{k};\rho,\ell,\eta,s,\vec{G}} \left(h_{\vec{k},\eta}\right)_{\rho,\ell,\vec{G};\rho',\ell',\vec{G}'} \psi^\pdagger_{\vec{k};\rho',\ell',\eta,s,\vec{G}'}, \\
    h_{\vec{k},\eta} &= h^{(g)}_{\vec{k},\eta} + h^{(t)}_{\vec{k},\eta} + h^{(D)}_{\vec{k}} + h^{(\text{SOC})}_{\vec{k},\eta}. \label{TheBandHamiltonian}
\end{align}
The corresponding components are given by, 
\begin{subequations}\begin{align}
    \left(h^{(g)}_{\vec{k},+}\right)_{\rho,\ell,\vec{G};\rho',\ell',\vec{G}'} &= \delta_{\ell,\ell'} \delta_{\vec{G},\vec{G}'} v_F (\vec{\rho}_{\theta_\ell})_{\rho,\rho'} \left(\vec{k} + \vec{G} - (-1)^\ell \vec{q}_{1}/2 \right), \label{DiracCones} \\ \left(h^{(g)}_{\vec{k},-}\right)_{\rho,\ell,\vec{G};\rho',\ell',\vec{G}'} &= \left(h^{(g)}_{-\vec{k},+}\right)^*_{\rho,\ell,-\vec{G};\rho',\ell',-\vec{G}'},\\
    \left(h^{(t)}_{\vec{k},+}\right)_{\rho,\ell,\vec{G};\rho',\ell',\vec{G}'} &= \sqrt{2} \begin{pmatrix} 0 & (T_{\vec{G}-\vec{G}'})_{\rho,\rho'} & 0 \\  (T_{\vec{G}'-\vec{G}}^*)_{\rho',\rho} & 0 & 0 \\ 0 & 0 & 0 \end{pmatrix}_{\ell,\ell'}, \\  \left(h^{(t)}_{\vec{k},-}\right)_{\rho,\ell,\vec{G};\rho',\ell',\vec{G}'} &= \left(h^{(t)}_{-\vec{k},+}\right)^*_{\rho,\ell,-\vec{G};\rho',\ell',-\vec{G}'},\\
     \left(h^{(D)}_{\vec{k}}\right)_{\rho,\ell,\vec{G};\rho',\ell',\vec{G}'} &= -D_0 \delta_{\rho,\rho'}\delta_{\vec{G},\vec{G}'} \begin{pmatrix} 0 & 0 & 1 \\ 0 & 0 & 0 \\ 1 & 0 & 0 \end{pmatrix}_{\ell,\ell'},\\
    \left(h^{(\text{SOC})}_{\vec{k},\eta}\right)_{\rho,\ell,\vec{G};\rho',\ell',\vec{G}'} &= \frac{\delta_{\vec{G},\vec{G}'}}{2} \begin{pmatrix} (h^{\text{SOC},l=1}_{\eta})_{\rho,\rho'} & 0 & (h^{\text{SOC},l=1}_{\eta})_{\rho,\rho'} \\ 0 & 0 & 0 \\ (h^{\text{SOC},l=1}_{\eta})_{\rho,\rho'} & 0 & (h^{\text{SOC},l=1}_{\eta})_{\rho,\rho'} \end{pmatrix}.
\end{align}\end{subequations}
Here  $\vec{\rho}_{\theta} = e^{i \theta \rho_3/2} \vec{\rho} e^{-i \theta \rho_3/2}$, and $\vec{q}_1$ connects the K and K' points in the MBZ; the tunneling matrices are
\begin{align}
   \notag T_{\delta\vec{G}} = \sum_{j=-1,0,1}\delta_{\delta\vec{G}+\vec{A}_j,0} \left[w_0 \rho_0 + w_1 \begin{pmatrix} 0 & \omega^j \label{FormOfT} \\  \omega^{-j} & 0 \end{pmatrix} \right], \\ \omega = e^{i \frac{2\pi}{3}}, \quad \vec{A}_0 =0, \quad \vec{A}_1 = \vec{G}_1, \quad \vec{A}_2 = \vec{G}_1 + \vec{G}_2,
\end{align}
with the property $T_{\delta\vec{G}}^\dagger = T_{\delta\vec{G}}^\pdagger$ and $\rho_x T_{\delta\vec{G}} \rho_x = T^*_{\delta\vec{G}}$; and the spin-orbit components are,
\begin{align}
 h^{\text{SOC},l=1}_{\eta} &=\lambda_{\text{I}} s_z \eta + \lambda_{\text{R}} \left(\eta \rho_x s_y - \rho_y s_x \right) + \lambda_{\text{KM}} \eta \rho_z s_z + m \rho_z.
\end{align}
This completes the continuum model in momentum space.

\section{Pairing states in the opposite limit}\label{FirstLambdaI}
In this appendix, we complement the discussion of \secref{PossiblePairingStates} of the main text and present the details of the evolution of pairing states when first turning on $\lambda_{\text{I}}$, before $\lambda_{\text{R}}$, which is illustrated schematically in \figref{fig:states}(b). This provides important insights into the form of the pairing state in the limit $\lambda_{\text{R}} \ll \lambda_{\text{I}}$.

As follows readily by inspection of \tableref{ActionOfSymmetries}, the set of point symmetries (apart from time-reversal) for $\lambda_{\text{R}}=0$, $\lambda_{\text{I}}\neq0$ is generated by $C_{3z}$, SO(2)$_s$, and $C_{2z}^{s'}$. Upon noting that the first of these symmetry operations commutes with the latter two, it is readily seen that they form the point group $C_3 \times D_{\infty}$, where $D_{\infty}$ is generated by SO(2)$_s$ and $C_{2z}^{s'}$ and, thus, can be thought of as $D_n$ in the limit $n\rightarrow \infty$.

As in the main text, we start from the $A_g^1$ singlet and the $B_u^3$ triplet defined in \equref{SCStartingPoint} as the two parent pairing states in the limit without SOC, $\lambda_{\text{R}}=\lambda_{\text{I}}=0$. Transforming trivially under $C_{3z}$, SO(2)$_s$, and $C_{2z}^{s'}$, the $A_g^1$ singlet transitions into the pairing state of IR $\Sigma_+^A$ of $C_3 \times D_{\infty}$ (here the superscript indicates the IR $A$ of $C_3$ and $\Sigma_+$ labels the trivial IR $D_{\infty}$); since SO(3)$_s$ and $C_{2z}$ are broken, an additional triplet component (from $B_u^3$) along the $z$ direction is admixed and the order parameter can be written as
\begin{equation}
        \psi_{\vec{k},\eta} = \chi_{\vec{k},\eta}, \qquad \vec{d}_{\vec{k},\eta} = \alpha_1 \eta \chi_{\vec{k},\eta}  \vec{e}_z, \label{ASigmaPlus}
\end{equation}
where $\alpha_1$ is proportional to $\lambda_{\text{I}}$ for small $\lambda_{\text{I}}$. No triplet component in the $xy$-plane can be admixed due to SO(2)$_s$. 
This includes the remaining two triplet components of $B_u^3$, which belong to the IR $\Pi^A$ of $C_3 \times D_{\infty}$ (as above, the superscript $A$ indicates the fact that the state transforms trivially under $C_{3z}$ and $\Pi$ is defined as the two-dimensional IR of $D_{\infty}$ with components transforming as $(x,y)$ under it). These symmetries do not allow for singlet-triplet admixture in this state and the two basis functions of the order parameter simply read as
\begin{equation}
    \begin{pmatrix} \psi^1_{\vec{k},\eta} \\ \psi^2_{\vec{k},\eta}  \end{pmatrix} = 0, \qquad \begin{pmatrix} \vec{d}^1_{\vec{k},\eta} \\ \vec{d}^2_{\vec{k},\eta} \end{pmatrix} = \eta \chi_{\vec{k},\eta} \begin{pmatrix}
    \vec{e}_x \\ \vec{e}_y
    \end{pmatrix}. \label{PiAState}
\end{equation}
If we eventually also turn on $\lambda_{\text{R}}$, we have to arrive at the same IRs and associated pairing states as discussed in \secref{PossiblePairingStates}. However, as long as $\lambda_{\text{R}} \ll \lambda_{\text{I}}$, the relative strength of the different admixed components can differ. For the $\Sigma_+^A$ state in \equref{ASigmaPlus}, breaking SO(2)$_s$ and $C_{2z}^{s'}$ will add in-plane triplet components,
\begin{equation}
    \psi_{\vec{k},\eta} = \chi_{\vec{k},\eta}, \qquad
    \vec{d}_{\vec{k},\eta} = \alpha_1 \eta\, \chi_{\vec{k},\eta} \vec{e}_z + \alpha_2 \begin{pmatrix} X_{\vec{k}} \\ Y_{\vec{k}} \\ 0 \end{pmatrix}  + \alpha_3\, \eta \begin{pmatrix} 2 X_{\vec{k}} Y_{\vec{k}} \\  X^2_{\vec{k}}-Y^2_{\vec{k}} \\ 0\end{pmatrix}. \label{FinalFormOfSinglet}
\end{equation}
As required, it is of the same form as \equref{ABAdmixed} [or \equref{ABAdmixed2} for that matter], but the relative weights are different since $\alpha_{2,3} \ll \alpha_1$ in \equref{FinalFormOfSinglet} for $\lambda_{\text{R}} \ll \lambda_{\text{I}}$ [as compared to $\alpha_{1,2}\gg \alpha_3$ in \equref{ABAdmixed} for $\lambda_{\text{I}} \ll \lambda_{\text{R}}$].

Finally, the $\Pi^A$ state in \equref{PiAState} becomes the $E$ state of $\widetilde{C}_3$. Its order parameter is again of the form of \equref{FormOfTheEState}, but with the crucial difference that, this time, all three coefficients $\alpha_{1,2,3}$ are small in $\lambda_{\text{R}}$.

\section{Diode effect of the $E$ state}\label{AppendixOnEPairingDiode}
In this appendix, we discuss the free energy expansion and associated critical current for pairing in the $E$ representation, introduced in \secref{PossiblePairingStates} of the main text.

\subsection{Free energy}
As opposed to \equref{EExpansionInBasisFunc}, we here use the ``chiral basis'', $c_\pm = (c_1 \mp ic_2)/2$, as it will be more convenient. Generalizing to finite momentum pairing, $c_\pm \rightarrow c_\pm(\vec{q})$, the symmetries of the system act as
\begin{subequations}\begin{align}
    C_{3z}^s:\quad \left(c_+(\vec{q}),c_-(\vec{q})\right) \quad &\longrightarrow \quad  \left(\omega c_+(C_{3z}\vec{q}),\omega^*c_-(C_{3z}\vec{q})\right), \quad \omega = e^{i\frac{2\pi}{3}}, \label{C3Symmetry} \\
    \Theta_s: \quad \left(c_+(\vec{q}),c_-(\vec{q})\right) \quad &\longrightarrow \quad \left(c^*_-(-\vec{q}),c^*_+(-\vec{q})\right). \label{TRSSymmetry}
\end{align}\label{SymmetryRepsOnEState}\end{subequations}
Neglecting the momentum dependence of the quartic terms, $b_1$ and $b_2$, the free energy reads 
\begin{align}\begin{split}
    \mathcal{F}[\{c_{\pm}(\vec{q})\}] &\sim \sum_{\vec{q}} \left[ a_{\vec{q}-2e\vec{A}} \left( |c_+(\vec{q})|^2 + |c_-(\vec{q})|^2 \right) + \delta a_{\vec{q}-2e\vec{A}} \left( |c_+(\vec{q})|^2 - |c_-(\vec{q})|^2 \right) + \left(\alpha_{\vec{q}-2e\vec{A}} c_+^*(\vec{q})c_-(\vec{q})  + \text{c.c.}\right) \right] \\
    & \quad + \sum_{\vec{q}_1,\vec{q}_2,\vec{q}_3,\vec{q}_4} \delta_{\vec{q}_1+\vec{q}_3,\vec{q}_2+\vec{q}_4}  \left[ b_1 \left(\sum_{\mu=\pm} c^*_\mu(\vec{q}_1) c_\mu(\vec{q}_2) \right)\left(\sum_{\mu=\pm} c^*_\mu(\vec{q}_3) c_\mu(\vec{q}_4) \right) + b_2 \, c^*_+(\vec{q}_1)c_+(\vec{q}_2)c^*_-(\vec{q}_3)c_-(\vec{q}_4) \right] \label{FreeEnergyExpansionERep}
\end{split}\end{align}
where $a_{\vec{q}}, \delta a_{\vec{q}} \in \mathbbm{R}$,  $\alpha(\vec{q}) \in \mathbb{C}$  obeying 
\begin{equation}
    a_{\vec{q}} = a_{C_{3z}\vec{q}}, \, \delta a_{\vec{q}} = \delta a_{C_{3z}\vec{q}}, \, \alpha_{\vec{q}}=\omega^* \alpha_{C_{3z}\vec{q}} \quad \text{ and } \quad  a_{\vec{q}} = a_{-\vec{q}}, \, \delta a_{\vec{q}} = -\delta a_{-\vec{q}}, \, \alpha_{\vec{q}}=\alpha_{-\vec{q}} \label{PropertiesOfThePrefactors}
\end{equation}
as follows from \equsref{C3Symmetry}{TRSSymmetry}, respectively. Consequently, both $\delta a_{\vec{q}}$ and $\alpha_{\vec{q}}$ have to vanish at $\vec{q}=0$, reproducing the form of the free energy in \cite{PhysRevResearch.2.033062}. Note that $\alpha$ will also have to vanish at finite $\vec{q}$, if we only have Ising SOC ($\lambda_{\text{R}}=0$): in this case, the order parameter will be the $\Pi^A$ state in \equref{PiAState} [see also \figref{fig:states}(b)] and the SO(2)$_s$ rotation symmetry in \tableref{ActionOfSymmetries} prohibits finite $\alpha_{\vec{q}}$ in \equref{FreeEnergyExpansionERep}.

To illustrate these statements and the microscopic origin of the terms in the free energy (\ref{FreeEnergyExpansionERep}), consider the following minimal low-energy, mean-field model for pairing in the $E$ representation, 
\begin{equation}
    H^{E} = \sum_{\vec{k},\eta,s,s'} f^\dagger_{\vec{k},\eta,s} \left(h^{\text{LE}}_{\vec{k};\eta}\right)_{s,s'} f^\pdagger_{\vec{k},\eta,s'} + \sum_{\vec{k},\vec{q},s,s'} \left[ f^\dagger_{\vec{k}+\vec{q},+,s}  \sum_{\mu=\pm}c_\mu(\vec{q}) \left(s^\mu  \right)_{s,s'} f^\dagger_{-\vec{k},-,s'} +\text{H.c.} \right] + \frac{1}{g_E} \sum_{\mu=\pm} |c_\mu(\vec{q})|^2,  \label{MeanFieldHamForRepE}
\end{equation}
using the low-energy fermions $f_{\vec{k},\eta,s}$ introduced in \secref{LowEnergyModel} and $s^{\pm} = s_x \pm i s_y$. This is equivalent to setting $\alpha_j=0$ and $\chi_{\vec{k},\eta}=1$ in \equref{FormOfTheEState}.
It is easy to check that applying the representations of the symmetries $C_{3z}^s$ and $\Theta_s$ in \tableref{ActionOfSymmetries} to the fermions in \equref{MeanFieldHamForRepE} reproduces \equref{SymmetryRepsOnEState}. We parametrize the normal-state Hamiltonian in \equref{MeanFieldHamForRepE} as
\begin{equation}
    h^{\text{LE}}_{\vec{k};+} = \epsilon_{\vec{k}} s_0 + \vec{g}_{\vec{k}}\cdot \vec{s}, \qquad h^{\text{LE}}_{\vec{k};-} = s_y(h^{\text{LE}}_{-\vec{k};+})^*s_y,
\end{equation}
i.e., explicitly restrict the analysis to time-reversal symmetric \text{normal states}. As such, any diode effect must come from the pairing-induced time-reversal-symmetry breaking.
Unless further symmetries are imposed (if $\lambda_{\text{R}}$ or $\lambda_{\text{I}}$ vanishes), the only additional constraints on the normal-state Hamiltonian come from $C_{3z}$ and read as
\begin{equation}
    \epsilon_{\vec{k}} = \epsilon_{C_{3z}\vec{k}}, \qquad \vec{g}_{\vec{k}} = C_{3z}\vec{g}_{C_{3z}\vec{k}}. \label{ConstraintOnEpsAndg}
\end{equation}
Integrating out the fermions in an action description of \equref{MeanFieldHamForRepE} and expanding the resultant effective action in terms of $c_{\mu}$, one can readily derive all prefactors in the free-energy expansion (\ref{FreeEnergyExpansionERep}). Defining the functional
\begin{equation}
    \mathcal{I}[X(i\omega_n,\vec{k},\vec{q})] := T \sum_{\omega_n} \frac{4}{N}\sum_{\vec{k}\in\text{MBZ}} \frac{X(i\omega_n,\vec{k},\vec{q})}{\left[(i\omega_n-\epsilon_{\vec{k}+\vec{q}/2})^2-\vec{g}^2_{\vec{k}+\vec{q}/2}\right]\left[(i\omega_n+\epsilon_{\vec{k}-\vec{q}/2})^2-\vec{g}^2_{\vec{k}-\vec{q}/2}\right]},
\end{equation}
where $\omega_n$ denote fermionic Matsubara frequencies, we have
\begin{align}
    a_{\vec{q}} &= \frac{1}{g_E} + \mathcal{I}\left[(i\omega_n+\epsilon_{\vec{k}-\frac{\vec{q}}{2}})(i\omega_n-\epsilon_{\vec{k}+\frac{\vec{q}}{2}})+ g^z_{\vec{k}+\frac{\vec{q}}{2}} g^z_{\vec{k}-\frac{\vec{q}}{2}} \right], \\
    \delta a_{\vec{q}} &=  \mathcal{I}\left[ i\omega_n (g^z_{\vec{k}+\frac{\vec{q}}{2}}+g^z_{\vec{k}-\frac{\vec{q}}{2}}) + g^z_{\vec{k}+\frac{\vec{q}}{2}}\epsilon_{\vec{k}+\frac{\vec{q}}{2}} - g^z_{\vec{k}-\frac{\vec{q}}{2}}\epsilon_{\vec{k}-\frac{\vec{q}}{2}} \right], \\
    \alpha_{\vec{q}} &= - \mathcal{I}\left[ (g^x_{\vec{k}+\frac{\vec{q}}{2}}- i g^y_{\vec{k}+\frac{\vec{q}}{2}}) (g^x_{\vec{k}-\frac{\vec{q}}{2}}- i g^y_{\vec{k}-\frac{\vec{q}}{2}})  \right]. \label{FormOfAlpha}
\end{align}
While it is straightforward to evaluate all involved Matsubara sums, we can already see immediately, using \equref{ConstraintOnEpsAndg}, that all properties in \equref{PropertiesOfThePrefactors} are obeyed---as required by symmetry. Furthermore, in accordance with \equref{DispersionForm}, it holds  $g^{x,y}=0$ if we only have Ising SOC, $\lambda_{\text{R}}=0$, leading to $\alpha_{\vec{q}}=0$ in \equref{FormOfAlpha}; this, in turn, agrees with our statement above that the SO(2)$_s$ symmetry forces $\alpha$ to vanish in the free energy (\ref{FreeEnergyExpansionERep}).

\subsection{Critical current}
We next discuss the resulting critical current and diode effect. To simplify the presentation, let us first set $\alpha=0$. As discussed in \secref{GeneralExpression}, the critical current is computed by setting $c_\mu(\vec{q}) = \Psi_\mu(\vec{q}_0) \delta_{\vec{q},\vec{q}_0}$, with $\Psi_\mu(\vec{q}_0)$ chosen to minimize $\mathcal{F}$ (at $\vec{A}=0$), and studying the maximum value of 
\begin{equation}
    \vec{J}_{\vec{q}_0} = -\partial_{\vec{A}} \mathcal{F}[\{\Psi_\mu(\vec{q}_0) \delta_{\vec{q},\vec{q}_0}\}]|_{\vec{A}=0} = 2 e \left[  (|\Psi_+(\vec{q}_0)|^2 + |\Psi_-(\vec{q}_0)|^2)\partial_{\vec{q}_0}a_{\vec{q}_0} + (|\Psi_+(\vec{q}_0)|^2 - |\Psi_-(\vec{q}_0)|^2) \partial_{\vec{q}_0}\delta a_{\vec{q}_0} \right] \label{ERepExprForCurrent}
\end{equation}
along a given direction $\hat{n}$.

If $b_2 > 0$, which is the sign that one obtains within mean-field theory \cite{SelectionRulePaper}, and we use the \textit{global minimum} of the free energy for $\Psi_\mu(\vec{q}_0)$, we will have $\Psi_{\sign(\delta a_{\vec{q}_0})}(\vec{q}_0) = 0$ and $\Psi_{-\sign(\delta a_{\vec{q}_0})}(\vec{q}_0) = \sqrt{(|\delta a_{\vec{q}_0}| - a_{\vec{q}_0})/(2 b_1)}$. Inserting this into the expression for the current in \equref{ERepExprForCurrent} yields
\begin{equation}
    \vec{J}_{\vec{q}} = e (|\delta a_{\vec{q}}|-a_{\vec{q}})(\partial_{\vec{q}} a_{\vec{q}} - \sign(\delta a_{\vec{q}}) \partial_{\vec{q}} \delta a_{\vec{q}}  )/b_1. \label{ChiralCurrentMomentumRelation}
\end{equation}
Recalling \equref{PropertiesOfThePrefactors}, we see that it holds $\vec{J}_{\vec{q}} = -\vec{J}_{-\vec{q}}$, implying $J_c(\hat{n}) = J_c(-\hat{n})$ and, hence, ruling out a diode effect.

It can be shown by explicit calculation that this is also the case when finite $\alpha$ are taken into account (or when $b_2<0$). In fact, this can be understood more generally, without having to neglect the momentum dependence of higher-order terms in $\mathcal{F}$ or even without resorting to an expansion of the free energy---it is just a manifestation of the fact that the normal state above the superconducting phase has time-reversal symmetry. To see this formally in our current description, let us define
\begin{equation}
    f(\psi_\mu,\vec{q}_0) := \mathcal{F}[\{c_{\mu}(\vec{q})=\psi_\mu\delta_{\vec{q},\vec{q}_0}\}]. \label{DefinitonOfFunctionf}
\end{equation}
Time-reversal symmetry implies that the free-energy $\mathcal{F}$ is invariant under \equref{TRSSymmetry} which leads to the constraint
\begin{equation}
    f(\psi_\mu,\vec{q}) = f(\psi^*_{-\mu},-\vec{q}) \label{TRSConstraintOnf}
\end{equation}
on the function defined in \equref{DefinitonOfFunctionf}. Denoting the minimum of $f(\psi_\mu,\vec{q})$ at fixed $\vec{q}$ by $\Psi_\mu(\vec{q})$ (note that for generic $\vec{q}$ this is expected to be unique), \equref{TRSConstraintOnf} then implies
\begin{equation}
    \Psi_\mu(-\vec{q}) = \Psi^*_{-\mu}(\vec{q}). \label{ConstraintOnSolution}
\end{equation}
With this, we are in position to relate $\vec{J}_{\vec{q}} = 2e(\partial_{\vec{q}} f(\psi_{\mu},\vec{q}))|_{\psi_{\mu} = \Psi_\mu(\vec{q})}$ and $\vec{J}_{-\vec{q}}$: 
\begin{align}
    \vec{J}_{-\vec{q}} & = 2 e \left(\partial_{\vec{q}'} f (\psi_{\mu},\vec{q}') \right)|_{\vec{q}'=-\vec{q}; \,\psi_{\mu} = \Psi_\mu(-\vec{q})} \\
    & = -2 e \left(\partial_{\vec{q}'} f (\psi_{\mu},-\vec{q}') \right)|_{\vec{q}'=\vec{q}; \,\psi_{\mu} = \Psi_\mu(-\vec{q})} \\
    & \stackrel{(\ref{TRSConstraintOnf})}{=} -2 e \left(\partial_{\vec{q}'} f (\psi^*_{-\mu},\vec{q}') \right)|_{\vec{q}'=\vec{q}; \,\psi_{\mu} = \Psi_\mu(-\vec{q})} \\
    & \stackrel{(\ref{ConstraintOnSolution})}{=} -2 e \left(\partial_{\vec{q}'} f (\psi_{\mu},\vec{q}') \right)|_{\vec{q}'=\vec{q}; \,\psi_{\mu} = \Psi_\mu(\vec{q})} \\
    & = -\vec{J}_{\vec{q}}.
\end{align}
Form this immediately follows that there will be no diode effect.

However, as mentioned in \secref{ExoticPairingCurrent}, the situation is different if the system remains in the \textit{local minimum} of $\mathcal{F}$ which is smoothly connected to the configuration of the superconductor without external current. This is expected to be particularly relevant when the time-reversal-symmetry-breaking order parameter in \equref{VestigialEOP} orders at temperatures well above the resistive superconducting transition. 

As before, we illustrate this situation in the particularly simple case where $\alpha=0$ in \equref{FreeEnergyExpansionERep}. Let us assume that $b_1,b_2>0$, $a<0$ and the superconductor, without current, has $c_+\neq 0$, $c_-=0$. Then the Ising-like order parameter in \equref{VestigialEOP} is positive, $\mathcal{C}>0$, and
\begin{equation}
    \Psi_{+}(\vec{q}) = \sqrt{(|\delta a_{\vec{q}}| - a_{\vec{q}})/(2 b_1)}, \quad  \Psi_{-}(\vec{q}) = 0, \quad \text{as long as }\, \delta a_{\vec{q}} > \frac{a_{\vec{q}}b_2}{2b_1 + b_2} \label{LocalMinimumOfPsiPlusMinus}
\end{equation}
at the associated local minimum in the presence of a current-induced finite-momentum order parameter (if $\delta a_{\vec{q}}$ is smaller than the indicated lower bound, the minimum becomes locally unstable). 
We see that, irrespective of the direction of $\vec{q}$, the Ising-like order parameter $\mathcal{C}$ now varies smoothly and remains positive when turning on $\vec{q}$. Assuming that the inequality in \equref{LocalMinimumOfPsiPlusMinus} is valid for all $\vec{q}$ relevant to determine the critical current, we can determine the latter by inserting this into \equref{ChiralCurrentMomentumRelation}, yielding
\begin{equation}
    \vec{J}_{\vec{q}} = e (|\delta a_{\vec{q}}|-a_{\vec{q}})(\partial_{\vec{q}} a_{\vec{q}} +  \partial_{\vec{q}} \delta a_{\vec{q}}  )/b_1.
\end{equation}
Since $\delta a_{\vec{q}}$ is an odd function of $\vec{q}$, see \equref{PropertiesOfThePrefactors}, we generally have $\vec{J}_{\vec{q}} \neq -\vec{J}_{-\vec{q}}$ and a diode effect becomes possible. 

Note that this is different for the nematic state: if $b_2<0$, we obtain a nematic superconducting state, which---in analogy to the time-reversal-odd composite order parameter in \equref{VestigialEOP}---can be characterized by the nematic composite order parameter
\begin{equation}
    \mathcal{N}^E_j := \sum_{\vec{q}}\begin{pmatrix} \text{Re}\left[ c_+^*(\vec{q}) c_-^{\phantom{*}}(\vec{q}) \right] \\  \text{Im}\left[ c_+^*(\vec{q}) c_-^{\phantom{*}}(\vec{q}) \right]  \end{pmatrix}_j.
\end{equation}
Note that $c_+^* c^{\phantom{*}}_- = \mathcal{N}^E_1 + i\, \mathcal{N}^E_2 \rightarrow \omega c_+^* c^{\phantom{*}}_-$ under $C_{3z}$ while remaining invariant under $\Theta_s$. Breaking a discrete symmetry, $\mathcal{N}^E_j$ can order at a non-zero temperature $T^*$. Just as in case of the chiral state, let us assume that $T^*$ is significantly larger than the resistive superconducting transition and that, when applying a current, the superconducting order parameter adiabatically follows the \textit{local minimum} that has already been chosen spontaneously at $T^*$ (note that sixth order terms need to be added to \equref{FreeEnergyExpansionERep} to break the artificial continuous rotational symmetry of the quartic free-energy expansion without current, see \cite{PhysRevResearch.2.033062}). Writing as before $c_\mu(\vec{q}) = \Psi_\mu(\vec{q}_0) \delta_{\vec{q},\vec{q}_0}$, it holds under these assumptions
\begin{equation}
    \phi(\vec{q}) = \phi(-\vec{q}), \quad |\Psi_+(\vec{q})| = |\Psi_-(-\vec{q})|, \label{ConstraintsForNematicE}
\end{equation}
where $\phi$ is defined via $\Psi^*_+(\vec{q})\Psi_-(\vec{q}) = |\Psi^*_+(\vec{q})||\Psi_-(\vec{q})| e^{i \phi(\vec{q})}$.
It is a smooth function of momentum such that $\mathcal{N}^E_j$ varies smoothly when applying a current. The properties in \equref{ConstraintsForNematicE} follow from $a_{\vec{q}} = a_{-\vec{q}}$, $\delta a_{\vec{q}} = -\delta a_{-\vec{q}}$,  $\alpha_{\vec{q}}=\alpha_{-\vec{q}}$ in \equref{PropertiesOfThePrefactors} and that $|\Psi_+(\vec{q}=0)| = |\Psi_-(\vec{q}=0)|$. From this we get
\begin{equation}
    \vec{J}_{\vec{q}} = 2e\left[ \partial_{\vec{q}}a_{\vec{q}} \sum_{\mu}|\Psi_\mu(\vec{q})|^2 + \partial_{\vec{q}}\delta a_{\vec{q}} \sum_{\mu}\mu|\Psi_\mu(\vec{q})|^2 + 2 |\Psi_+(\vec{q})||\Psi_-(\vec{q})|\text{Re} \left(\partial_{\vec{q}}\alpha_{\vec{q}} e^{i\phi(\vec{q})}\right) \right].
\end{equation}
We, thus, see that $\vec{J}_{\vec{q}} = - \vec{J}_{-\vec{q}}$ and no diode effect is possible as expected since neither the normal nor the superconducting state breaks time-reversal symmetry (in the absence of an applied current).

\section{Patch theory expansion}\label{A:patch}
To obtain analytic expressions from \eqref{GammapExpression}, we consider $T\sim T_c$, with $T_c\sim v\Lambda_x e^{-1/(\nu g)}$ set by the energy cut-off of the patch, which is approximately set by the energy cut-off in the direction perpendicular to the patched Fermi surface, i.e. $k_x$ direction. Assuming large  density of states times the Cooper channel interaction strength, $\nu g$, we arrive at $T\sim v \Lambda_x$. With this we may expand the hyperbolic tangents in the kernel of $\Gamma(\bm q)$, $\tanh(x)\approx x -x^3/3 + 2x^5/15 + O(x^7)$, i.e.
\begin{align}
\label{ExpandedGammapExpression} 
    \notag \Gamma^p(\vec{q}) &\approx \int_{-\Lambda_y}^{\Lambda_y} \frac{\diff  k_y}{2 \pi} \int_{-\Lambda_x}^{\Lambda_x} \frac{\diff \delta k_x}{4 \pi} \frac{1}{\xi_1+\xi_2}\left[\frac{ \xi_1}{2 T}+\frac{ \xi_2}{2 T} -\frac{1}{3}\left(\frac{ \xi_1}{2 T}\right)^3-\frac{1}{3}\left(\frac{ \xi_2}{2 T}\right)^3+ \frac{2}{15}\left(\frac{ \xi_1}{2 T}\right)^5+ \frac{2}{15}\left(\frac{ \xi_2}{2 T}\right)^5\right], \\
  \notag  \xi_1&= v(\delta k_x + \delta q_x/2) - \alpha_+(k_y+q_y/2)^2,\\
   \xi_2&= v(\delta k_x - \delta q_x/2)- \alpha_-(k_y-q_y/2)^2.
\end{align}
Based on the above reasoning, $\xi_i/(2T)\lesssim1/2$, therefore justifying the order of the expansion. The expression \eqref{ExpandedGammapExpression} affords an analytic integration; performing such integration and extracting the leading coefficients, as per the expansion given in \equref{GammaExpand}, gives
\begin{align}
\label{ExpressionsForAB}
a_1&=\frac{\left(\alpha _+-\alpha _-\right) v \Lambda _x \Lambda _y^3
   \left(-126 T^2+14 v^2 \Lambda _x^2+9 \left(\alpha _-^2+\alpha
   _+^2\right) \Lambda _y^4\right)}{3024 T^5},\\
a_2&=\frac{v^2 \Lambda _x \Lambda _y \left(45 T^2-5 v^2 \Lambda _x^2-3
   \left(\alpha_+^2-\alpha_- \alpha_++\alpha_-^2\right) \Lambda
   _y^4\right)}{360 T^5 (8\pi^2)},\\
\notag   a_3&= \frac{\left(\alpha _+-\alpha _-\right) v^3 \Lambda _x \Lambda _y^3}{144
   T^5  (8\pi^2)},\\
\notag   a_4&= -\frac{v^4 \Lambda _x \Lambda _y}{192 T^5  (8\pi^2)},\\
  \notag c&= \frac{\Lambda _x \Lambda _y^3 \left(105 \left(3 \alpha _+^2+\alpha _-
   \alpha _++3 \alpha _-^2\right) T^2-7 \left(6 \alpha _+^2+\alpha _-
   \alpha _++6 \alpha _-^2\right) v^2 \Lambda _x^2+9 \left(-7 \alpha
   _1^4+\alpha _- \alpha _+^3+\alpha _-^2 \alpha _+^2+\alpha _-^3 \alpha
   _1-7 \alpha _-^4\right) \Lambda _y^4\right)}{3780 T^5  (8\pi^2)}.
\end{align}
In particular, we see that $a_1$ and $a_2$ vanish as $\alpha_+-\alpha_-\rightarrow 0$, which we stated and used in \secref{PatchSection}. We emphasize that this also holds when no additional approximations are made in the evaluation of \equref{GammapExpression}. Since $\Gamma$ in \equref{GammapExpression} is invariant under $\delta q_x \rightarrow -\delta q_x$ and simultaneous exchange $\alpha_+ \leftrightarrow \alpha_-$, we immediately conclude that
\begin{equation}
    a_{1,3}(\alpha_+,\alpha_-) = -a_{1,3}(\alpha_-,\alpha_+), \qquad a_{2,4}(\alpha_+,\alpha_-) = a_{2,4}(\alpha_-,\alpha_+)
\end{equation}
and, hence, $a_{1,3}(\alpha,\alpha)=0$.

\section{LG theory in the presence of strong spin-orbit coupling or intravalley pairing} \label{A:DerivationOfGamma}

\subsection{Strong spin-orbit coupling}
We look now at strong SOC, such that bands are non-degenerate. Explicitly, we work with the low energy bands that cross the Fermi level in the vicinity of $\vec{k}$; the band energies in valley $\eta$ are denoted by $\xi_{\vec{k},\eta}$. These states are obtain by diagonalising the full tTLG noninteracting Hamiltonian \eqref{DifferentPartsofContHam}, in the presence of either valley polarization or applied magnetic field, which are captured by adding the perturbations
\begin{align}
\delta h_1 & = V_z \eta_z,\quad
\delta h_2 = \bm B\cdot \bm s
\end{align} 
to $h$.
In the absence of these perturbations, time-reversal symmetry is preserved, and the band energies satisfy $\xi_{\vec{k},+} = \xi_{-\vec{k},-}$. 

To describe pairing between states at opposite valleys, we consider the effective interacting Hamiltonian
\begin{align}
    \notag &H = \sum_{\vec{k},\eta} \tilde{f}^\dagger_{\vec{k},\eta} \xi_{\vec{k},\eta} \tilde{f}^\pdagger_{\vec{k},\eta} - \frac{\tilde{g}}{2} \sum_{\vec{k},\vec{k}',\vec{q}} \tilde{f}^\dagger_{\vec{k}+\vec{q},\eta} \tilde{f}^\dagger_{\vec{k}'-\vec{q},\eta'} \tilde{f}^\pdagger_{\vec{k}',\eta'}  \tilde{f}^\pdagger_{\vec{k},\eta}, \label{SOCHamiltonian}
\end{align}
where $\tilde{f}_{\vec{k},\eta}$ and $\tilde{f}^\dagger_{\vec{k},\eta}$ are annihilation and creation operators of electrons of valley $\eta$, in the low-energy bands crossing the Fermi level in the vicinity of $\vec{k}$. Due to strong SOC, this Hamiltonian is effectively spinless. It is important to keep in mind that, as a consequence of the underlying SOC, the electronic states created by $\tilde{f}^\dagger_{\vec{k},\eta}$ exhibit a momentum-dependent spin quantum number.  

Performing a mean-field decoupling in the intervalley channel, we obtain
\begin{align}
     &H = \sum_{\vec{k},\eta} \tilde{f}^\dagger_{\vec{k},\eta} \xi_{\vec{k},\eta}  \tilde{f}^\pdagger_{\vec{k},\eta}  + \frac{1}{2\tilde{g}} \sum_{\vec{q}} \Delta_{\vec{q}}^\dagger \Delta_{\vec{q}}^\pdagger + \frac{1}{2}\sum_{\vec{k},\vec{q}} \left[ \tilde{f}^\dagger_{\vec{k}+\vec{q},+} \Delta_{\vec{q}} \tilde{f}^\dagger_{-\vec{k},-} + \text{H.c.} \right]
\end{align}
where the complex scalar $\Delta_{\vec{q}}$ is the superconducting order parameter. As a result of the aforementioned momentum-dependent spin quantum number of the low-energy fermions, this single complex number describes a, in general, singlet-triplet mixed state; for $\vec{q}=0$ pairing, the structure of the order parameter in terms of the physical spin basis is discussed in \secref{PossiblePairingStates} of the main text. 
The associated Ginzburg-Landau expansion is analogous to the expression presented in the main text \eqref{ExpressionForGamma}, 
\begin{align}
\label{ExpressionForGammaSOC}
 \notag   &\mathcal{F} \sim \frac{1}{2}\sum_{\vec{q}} a_{\vec{q}} \Delta_{\vec{q}}^\dagger \Delta_{\vec{q}}^\pdagger +\mathcal{O}(\Delta^4), \quad  a_{\vec{q}} = \frac{1}{\tilde{g}} - \Gamma(\vec{q}), \\
   & \Gamma(\vec{q}) = \frac{1}{2N} \sum_{\vec{k}\in\text{MBZ}} \frac{\tanh\left(\frac{\xi_{\vec{k}+\frac{\vec{q}}{2},+}}{2T}\right) + \tanh\left(\frac{\xi_{-\vec{k}+\frac{\vec{q}}{2},-}}{2T}\right)}{\xi_{\vec{k}+\frac{\vec{q}}{2},+}+\xi_{-\vec{k}+\frac{\vec{q}}{2},-}}. 
\end{align}
From which the expression for the current follows via \equref{ExpressionForCurrent1}.

\subsection{Intravalley pairing}
Finally, we state, without derivation, that for the case of intravalley, spin-singlet pairing with spin-degenerate bands---considered in the context of the toy-models of \secref{MBZmodels}---the corresponding expression for $\Gamma(\bm q)$ is 
\begin{align}
\label{ExpressionForGammaIV}
& \Gamma(\vec{q}) = \frac{1}{2N} \sum_{\vec{k}\in\text{MBZ}} \frac{\tanh\left(\frac{\xi_{\vec{k}+\frac{\vec{q}}{2},+,\uparrow}}{2T}\right) + \tanh\left(\frac{\xi_{-\vec{k}+\frac{\vec{q}}{2},+,\downarrow}}{2T}\right)}{\xi_{\vec{k}+\frac{\vec{q}}{2},+,\uparrow}+\xi_{-\vec{k}+\frac{\vec{q}}{2},+,\downarrow}}. 
\end{align}

\end{appendix}

\end{document}